\documentclass[fleqn,usenatbib]{mnras}
\usepackage[T1]{fontenc}
\usepackage{ae,aecompl}
\usepackage{graphicx}   % Including figure files
\usepackage{amsmath}    % Advanced maths commands
\usepackage{amssymb}    % Extra maths symbols
\usepackage{siunitx}    % for numbers with units
\DeclareSIUnit{\degree}{deg} % we want degrees written with 'deg' and
\DeclareSIUnit{\arcmin}{arcmin} % we want degrees written  'arcmin' and
\DeclareSIUnit{\arcsec}{arcsec} % we want arcsec written  'arcsec' and
\usepackage{color,verbatim,url}
\usepackage{tabularx}
\usepackage[table,usenames,dvipsnames]{xcolor}
\usepackage{booktabs}
\usepackage{pdflscape}
\usepackage{lastpage}

\usepackage{minitoc}

\definecolor{pink}{rgb}{0.858, 0.188, 0.478}
\definecolor{purple}{RGB}{76, 0,153}

\newcommand{\x}{\vec{x}}
\renewcommand{\d}[0]{{\rm d}}

\newcommand{\be}{\begin{equation}}  \newcommand{\ee}{\end{equation}}
  
\newcommand{\ba}{\begin{eqnarray}}\newcommand{\ea}{\end{eqnarray}}
\newcommand{\mat}[1]{\mathbfss{#1}}

\title[KiDS: Cosmological Parameters]{KiDS-450:  Cosmological parameter constraints from tomographic weak gravitational lensing}

\author[Hildebrandt, Viola, Heymans, Joudaki, Kuijken \& the KiDS collaboration]
{H. Hildebrandt$^1$\thanks{Email: hendrik@astro.uni-bonn.de}, 
M. Viola$^2$\thanks{Email: viola@strw.leidenuniv.nl}, 
C. Heymans$^3$, 
S. Joudaki$^4$, 
K. Kuijken$^2$,
C. Blake$^4$,
\newauthor
T. Erben$^1$,
B. Joachimi$^5$,
D. Klaes$^1$,
L. Miller$^{6}$,
C.B. Morrison$^1$,
R. Nakajima$^1$,
\newauthor
G. Verdoes Kleijn$^7$,
A. Amon$^3$,
A. Choi$^3$,
G. Covone$^8$,
J.T.A. de Jong$^2$,
\newauthor
A. Dvornik$^2$,
I. Fenech Conti$^{9,10}$,
A. Grado$^{11}$,
J. Harnois-D\'eraps$^{3,12}$,
\newauthor
R. Herbonnet$^2$,
H. Hoekstra$^2$,
F. K\"ohlinger$^2$,
J. McFarland$^7$,
A. Mead$^{12}$,
\newauthor
J. Merten$^6$,
N. Napolitano$^{11}$,
J.A. Peacock$^3$,
M. Radovich$^{13}$,
P. Schneider$^1$,
\newauthor
P. Simon$^1$,
E.A. Valentijn$^7$,
J.L. van den Busch$^1$,
E. van Uitert$^5$
\newauthor
and
L. Van Waerbeke$^{12}$
\\ 
$^1$Argelander-Institut f\"ur Astronomie, Auf dem H\"ugel 71, 53121 Bonn, Germany\\
$^2$Leiden Observatory, Leiden University, Niels Bohrweg 2, 2333 CA Leiden, the Netherlands\\
$^3$Institute for Astronomy, University of Edinburgh, Royal Observatory, Blackford Hill, Edinburgh EH9 3HJ, UK\\
$^4$Centre for Astrophysics \& Supercomputing, Swinburne University of Technology, PO Box 218, Hawthorn, VIC 3122, Australia\\
$^5$University College London, Gower Street, London WC1E 6BT, UK\\
$^6$Department of Physics, University of Oxford, Denys Wilkinson Building, Keble Road, Oxford OX1 3RH, U.K.	\\
$^7$Kapteyn Astronomical Institute, University of Groningen, 9700AD Groningen, the Netherlands\\
$^8$Department of Physics, University of Napile Federico II, via Cintia, 80126, Napoli, Italy\\
$^9$Institute of Space Sciences and Astronomy (ISSA), University of Malta, Msida MSD 2080, Malta\\
$^{10}$Department of Physics, University of Malta, Msida, MSD 2080, Malta\\
$^{11}$INAF -- Osservatorio Astronomico di Capodimonte, Via Moiariello 16, 80131 Napoli, Italy\\
$^{12}$Department of Physics and Astronomy, University of British Columbia, 6224 Agricultural Road, Vancouver, BC V6T 1Z1, Canada\\
$^{13}$INAF -- Osservatorio Astronomico di Padova, via dell'Osservatorio 5, 35122 Padova, Italy\\
\vspace{-0.5cm}  %necessary so a LaTeX crash doesn't occur...
}

\date{Released 16/6/2016}

\pagerange{\pageref{firstpage}--\pageref{LastPage}} \pubyear{2016}

\begin{document}
\setlength{\voffset}{-12mm}

\doparttoc % Tell to minitoc to generate a toc for the parts
\faketableofcontents % Run a fake tableofcontents command for the partocs
\noptcrule  %No horizontal rules in TOC

\label{firstpage}

\maketitle

\begin{abstract}
We present cosmological parameter constraints from a tomographic weak gravitational lensing analysis of $\sim$\SI{450}{\square\degree} of imaging data from the Kilo Degree Survey (KiDS).  For a flat $\Lambda$CDM cosmology with a prior on $H_0$ that encompasses the most recent direct measurements, we find $S_8\equiv\sigma_8\sqrt{\Omega_{\rm m}/0.3}=0.745\pm0.039$.  This result is in good agreement with other low redshift probes of large scale structure, including recent cosmic shear results, along with pre-Planck cosmic microwave background constraints.  A $2.3$-$\sigma$ tension in $S_8$ and `substantial discordance' in the full parameter space is found with respect to the Planck 2015 results.
We use shear measurements for nearly 15 million galaxies, determined with a new improved `self-calibrating' version of \emph{lens}fit validated using an extensive suite of image simulations. Four-band $ugri$ photometric redshifts are calibrated directly with deep  spectroscopic surveys. The redshift calibration is confirmed using two independent techniques based on angular cross-correlations and the properties of the photometric redshift probability distributions. Our covariance matrix is determined using an analytical approach, verified numerically with large mock galaxy catalogues.  We account for uncertainties in the modelling of intrinsic galaxy alignments and the impact of baryon feedback on the shape of the non-linear matter power spectrum, in addition to the small residual uncertainties in the shear and redshift calibration.
The cosmology analysis was performed blind.
Our high-level data products, including shear correlation functions, covariance matrices, redshift distributions, and Monte Carlo Markov Chains are available at \url{http://kids.strw.leidenuniv.nl}.
\end{abstract}

\begin{keywords}
cosmology: observations -- gravitational lensing: weak -- galaxies: photometry -- surveys
\vspace{-1.9cm}
\end{keywords}

\clearpage
\section{INTRODUCTION}
\label{sec:introduction}

The current `standard cosmological model' ties together a diverse set of properties of the observable Universe. Most importantly, it describes the statistics of anisotropies in the cosmic microwave background radiation \citep[CMB; e.g.,][]{hinshaw/etal:2013,planck/cosmo:2015}, the Hubble diagram of supernovae of type Ia \citep[SNIa; e.g.,][]{betoule/etal:2014}, big bang nucleosynthesis \citep[e.g.,][]{fields/olive:2006}, and galaxy clustering. It successfully predicts key aspects of the observed large-scale structure, from baryonic acoustic oscillations \citep[e.g.,][]{ross/etal:2015,kazin/etal:2014,anderson/etal:2014} on the largest scales down to Mpc-scale galaxy clustering and associated inflow velocities \citep[e.g.,][]{peacock/etal:2001}. It is also proving to be a successful paradigm for (predominantly hierarchical) galaxy formation and evolution theories.

This model, based on general relativity, is characterised by a flat geometry, a non-zero cosmological constant $\Lambda$ that is responsible for the late-time acceleration in the expansion of the Universe, and cold dark matter (CDM) which drives cosmological structure formation. Increasingly detailed observations can further stress-test this model, search for anomalies that are not well described by flat $\Lambda$CDM, and potentially yield some guidance for a deeper theoretical understanding. Multiple cosmological probes are being studied, and their concordance will be further challenged by the next generation of cosmological experiments.

The two main ways in which to test the cosmological model are observations of the large-scale geometry and the expansion rate of the Universe, and of the formation of structures (inhomogeneities) in the Universe. Both aspects are exploited by modern imaging surveys using the weak gravitational lensing effect of the large-scale structure \citep[cosmic shear; for a review see][]{kilbinger:2015}. Measuring the coherent distortions of millions of galaxy images as a function of angular separation on the sky and also as a function of their redshifts provides a great amount of cosmological information complementary to other probes. The main benefits of this tomographic cosmic shear technique are its relative insensitivity to galaxy biasing, its clean theoretical description \citep[though there are complications due to baryon physics; see e.g.][]{semboloni/etal:2011}, and its immense potential statistical power compared to other probes \citep{albrecht/etal:2006}. 

In terms of precision, currently cosmic shear measurements do not yet yield cosmological parameter constraints that are competitive with other probes, due to the limited cosmological volumes covered by contemporary imaging surveys \citep[see][table 1 and fig.~7]{kilbinger:2015}. The volumes surveyed by cosmic shear experiments will, however, increase tremendously with the advent of very large surveys such as LSST\footnote{\url{http://www.lsst.org/}} \citep[see for example][]{chang/etal:2013}, Euclid\footnote{\url{http://sci.esa.int/euclid/}} \citep{laureijs/etal:2011}, and WFIRST\footnote{\url{http://wfirst.gsfc.nasa.gov/}} over the next decade. In order to harvest the full statistical power of these surveys, our ability to correct for several systematic effects inherent to tomographic cosmic shear measurements will have to keep pace. Each enhancement in statistical precision comes at the price of requiring increasing control on low-level systematic errors. Conversely, only this statistical precision gives us the opportunity to identify, understand, and correct for new systematic effects. It is therefore of utmost importance to develop the cosmic shear technique further and understand systematic errors at the highest level of precision offered by the best data today.

%\defcitealias{abbott/etal:2015}{The Dark Energy Survey 2015}

Confidence in the treatment of systematic errors becomes particularly important when tension between different cosmological probes is found.  Recent tomographic cosmic shear results from the Canada France Hawaii Telescope Lensing Survey \citep[CFHTLenS\footnote{\url{http://www.cfhtlens.org/}};][]{heymans/etal:2012, heymans/etal:2013} are in tension with the CMB results from Planck \citep{planck/cosmo:2015} as described in \citet{maccrann/etal:2015}, yielding a lower amplitude of density fluctuations (usually parametrised by the root mean square fluctuations in spheres with a radius of 8\,Mpc, $\sigma_8$) at a given matter density ($\Omega_{\rm m}$). A careful re-analysis of the data \citep{joudaki/etal:2016} incorporating new knowledge about systematic errors in the photometric redshift (photo-$z$) distributions \citep{choi/etal:2015} was not found to alleviate the tension.  
Only conservative analyses, measuring the lensing power-spectrum \citep{kitching/etal:2014,kohlinger/etal:2016} or limiting the real-space measurements to large angular-scales \citep{joudaki/etal:2016}, reduce the tension primarily as a result of the weaker cosmological constraints.

The first results from the Dark Energy Survey \citep[DES; ][]{abbott/etal:2015}
%(DES\footnote{\url{http://www.darkenergysurvey.org/}}; 
%\citetalias{abbott/etal:2015},
%\defcitealias{abbott/etal:2015}{DES2015}
%henceforth \citetalias{abbott/etal:2015})
do not show such tension, but their uncertainties on cosmological parameters are roughly twice as large as the corresponding constraints from CFHTLenS.   In addition to rigorous re-analyses of CFHTLenS with new tests for weak lensing systematics \citep{asgari/etal:2016}, there have also been claims in the literature of possible residual systematic errors or internal tension in the Planck analysis \citep{spergel/etal:2015,addison/etal:2016,riess/etal:2016}. It is hence timely to re-visit the question of inconsistencies between CMB and weak lensing measurements with the best data available.

The ongoing Kilo Degree Survey \citep[KiDS\footnote{\url{http://kids.strw.leidenuniv.nl/}};][]{dejong/etal:2015} was designed specifically to measure cosmic shear with the best possible image quality attainable from the ground. In this paper we present intermediate results from \SI{450}{\square\degree} (about one third of the full target area) of the KiDS dataset, with the aim to investigate the agreement or disagreement between CMB and cosmic shear observations with new data of comparable statistical power to CFHTLenS but from a different telescope and camera. In addition the analysis includes an advanced treatment of several potential systematic errors. 
This paper is organised as follows.
We present the KiDS data and their reduction in Section~\ref{sec:data}, and describe how we calibrate the photometric redshifts in Section~\ref{sec:photoz_calibration}. Section~\ref{sec:theory} summarises the theoretical basis of cosmic shear measurements. Different estimates of the covariance between the elements of the cosmic shear data vector are described in Section~\ref{sec:covariance}. We present the shear correlation functions and the results of fitting cosmological models to them in Section~\ref{sec:results}, followed by a discussion in Section~\ref{sec:discussion}. A summary of the findings of this study and an outlook (Section~\ref{sec:summary}) conclude the main body of the paper. The more technical aspects of this work are available in an extensive Appendix, which covers requirements on shear and photo-$z$ calibration (Appendix~\ref{sec:app_A}), the absolute photometric calibration with stellar locus regression (SLR, Appendix~\ref{sec:app_SLR}), systematic errors in the photo-$z$ calibration (Appendix~\ref{sec:app_z_tests}), galaxy selection, shear calibration and E/B-mode analyses (Appendix~\ref{sec:app_shear_tests}), a list of the independent parallel analyses that provide redundancy and validation, right from the initial pixel reduction all the way through to the cosmological parameter constraints (Appendix~\ref{sec:redundancy}), and an exploration of the full multi-dimensional likelihood chain (Appendix~\ref{sec:detailed_results}).

Readers who are primarily interested in the cosmology findings of this study may wish to skip straight to Section~\ref{sec:results}, referring back to the earlier sections for details of the data and covariance estimate, and of the fitted models.
%1

\section{DATASET AND REDUCTION}
\label{sec:data}
In this section we briefly describe the KiDS-450 dataset, highlighting significant updates to our analysis pipeline since it was first documented in the context of the earlier KiDS-DR1/2 data release \citep{dejong/etal:2015,kuijken/etal:2015}.  These major changes include incorporating a global astrometric solution in the data reduction, improved photometric calibration, using spectroscopic training sets to increase the accuracy of our photometric redshift estimates, and analysing the data using an upgraded `self-calibrating' version of the shear measurement method \emph{lens}fit \citep{fenechconti/etal:2016}. 

\subsection{KiDS-450 data}
\begin{figure*}
\includegraphics[clip=true, trim=3cm 5cm 0cm 6cm, width=\hsize]{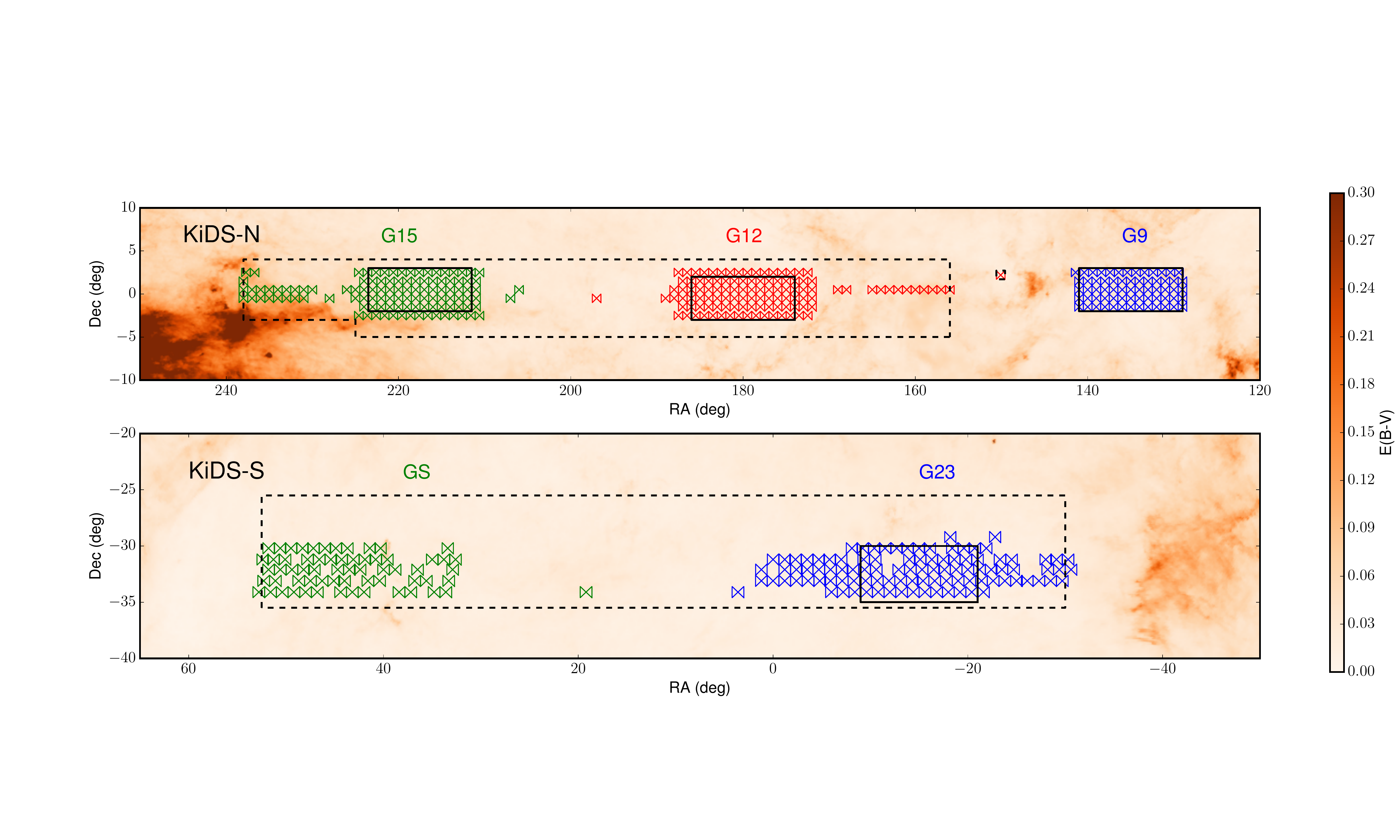}
\caption{\label{fig:footprint}Footprint of the KiDS-450 dataset. The dashed contours outline the full KiDS area (observations ongoing) and the {\large $\Join$} symbols represent the pointings included in KiDS-450 and used in this study correponding to \SI{449.7}{\square\degree}. The different colours indicate which pointing belongs to which of the five patches (G9, G12, G15, G23, GS). The solid rectangles indicate the areas observed by the GAMA spectroscopic survey. The background shows the reddening $E(B-V)$ from the Schlegel et al. (1998) maps.}
\end{figure*}

KiDS is a four-band imaging survey conducted with the OmegaCAM CCD mosaic camera mounted at the Cassegrain focus of the VLT Survey Telescope (VST). This telescope-camera combination, with its small camera shear and its well-behaved and nearly round point spread function (PSF), was specifically designed with weak lensing measurements in mind. Observations are carried out in the SDSS-like $u$-, $g$-, $r$-, and $i$-bands with total exposure times of 17, 15, 30 and 20 minutes, respectively. This yields limiting magnitudes of  24.3, 25.1, 24.9, 23.8 (5$\sigma$ in a \SI{2}{\arcsec} aperture) in $ugri$, respectively. The observations are queue-scheduled such that the best-seeing dark time is reserved for the $r$-band images, which are used to measure the shapes of galaxies (see Section~\ref{sec:lensfit}). KiDS targets two $\sim$\SI{10}{\degree}$\times$\SI{75}{\degree} strips, one on the celestial equator (KiDS-N) and one around the South Galactic Pole (KiDS-S). The survey is constructed from individual dithered exposures that each cover a `tile' of roughly \SI{1}{\square\degree} at a time. 

The basis for our dataset are the 472 KiDS tiles which had been observed in four bands on July 31st, 2015.  These data had also survived initial quality control, but after further checks some $i$-band and $u$-band images were rejected and placed back in the observing queue.  Those that were re-observed before October 4th, 2015 were incorporated into the analysis where possible such that the final dataset consists of 454 tiles covering a total area of \SI{449.7}{\square\degree} on the sky. The median seeing of the $r$-band data is \SI{0.66}{\arcsec} with no $r$-band image having a seeing larger than \SI{0.96}{\arcsec}. The sky distribution of our dataset, dubbed `KiDS-450', is shown in Fig.~\ref{fig:footprint}.  It consists of 2.5\,TB of coadded $ugri$ images (for the photometry, see Section~\ref{sec:AW}), 3\,TB of individual $r$-band exposures for shear measurements (Section~\ref{sec:theli}), and similar amounts of calibration, masks and weight map data.

Initial KiDS observations prioritised the parts of the sky covered by the spectroscopic GAMA survey \citep{driver/etal:2011}, and these were the basis of the first set of lensing analyses \citep{viola/etal:2015,sifon/etal:2015,vanuitert/etal:2016,brouwer/etal:2016}. Even though KiDS currently extends beyond the GAMA regions, we continue to group the tiles in five `patches' that we call G9, G12, G15, G23, and GS following the convention of the GAMA survey, with each patch indicated by the letter `G' and a rough RA (hour) value. Note that GS does not have GAMA observations, however we decided to maintain the naming scheme nevertheless. GS should not be confused with the G2 GAMA patch, which does not overlap with KiDS. Each KiDS patch consists of a central core region as well as nearby survey tiles observed outside the GAMA boundaries. As the survey progresses these areas will continue to be filled. 

\subsection{Multi-colour processing with Astro-WISE}
\label{sec:AW}
The multi-colour KiDS data, from which we estimate photometric redshifts, are reduced and calibrated with the \textsc{Astro-WISE} system \citep{valentijn/etal:2007,begeman/etal:2013}. The reduction closely follows the procedures described in \cite{dejong/etal:2015} for the previous KiDS data release (DR1/2), and we refer the reader to that paper for more in-depth information.

The first phase of data reduction involves de-trending the raw data, consisting of the following steps: correction for cross-talk, de-biasing, flat-fielding, illumination correction, de-fringing (only in the $i$-band), masking of hot and cold pixels as well as cosmic rays, satellite track removal, and background subtraction.

Next the data are photometrically calibrated. This is a three stage process. First the 32 individual CCDs are assigned photometric zeropoints based on nightly observations of standard star fields. Second, all CCDs entering a coadd are relatively calibrated with respect to each other using sources in overlap areas. The third step, which was not applied in DR1/2 and is only described as a quality test in \cite{dejong/etal:2015}, involves a tile-by-tile stellar locus regression (SLR) with the recipe of \cite{ivezic/etal:2004}. This alignment of the colours of the stars in the images (keeping the $r$-band magnitudes fixed) further homogenises the data and ensures that the photometric redshifts are based on accurate colours. In the SLR procedure, which is described in detail in Appendix~\ref{sec:app_SLR}, we use the \cite{schlegel/etal:1998} maps to correct for Galactic extinction for each individual star. 

Astrometric calibration is performed with 2MASS \citep{skrutskie/etal:2006} as an absolute reference. After that the calibrated images are coadded and further defects (reflections, bright stellar light halos, previously unrecognised satellite tracks) are masked out.

\subsection{Lensing reduction with \textsc{THELI}}
\label{sec:theli}
Given the stringent requirements of weak gravitational lensing observations on the quality of the data reduction we employ a second pipeline, \textsc{theli} \citep{erben/etal:2005,schirmer:2013}, to reduce the KiDS-450 $r$-band data. The handling of the KiDS data with this pipeline evolved from the CARS \citep{erben/etal:2009} and CFHTLenS \citep{erben/etal:2013} projects, and is described in more detail in \citet{kuijken/etal:2015}; the key difference to the multi-colour data reduction described in Section~\ref{sec:AW} is the preservation of the individual exposures, without the re-gridding or interpolation of pixels, which allows for a more accurate measurement of the sheared galaxy shapes. The major refinement for the KiDS-450 analysis over KiDS-DR1/2 concerns the astrometric calibration of the data. A cosmic shear analysis is particularly sensitive to optical camera distortions, and it is therefore essential to aim for the best possible astrometric alignment of the images. The specific improvements in the KiDS-450 data reduction are as follows.
\begin{enumerate}
  \item We simultaneously astrometrically calibrate \emph{all} data from a given patch, i.e., we perform a patch-wide global astrometric calibration of the data. This allows us to take into account information from overlap areas of individual KiDS tiles\footnote{The global astrometric solution is not calculated for the nine isolated tiles that do not currently overlap with other tiles (see Fig.~\ref{fig:footprint}).}.
  \item For the northern KiDS-450 patches G9, G12, and G15 we use accurate astrometric reference sources from the SDSS-Data Release 12 \citep{alam/etal:2015} for the absolute astrometric reference frame.
  \item The southern patches G23 and GS do not overlap with the SDSS, and we have to use the less accurate 2MASS catalogue (see \citealt{skrutskie/etal:2006}) for the absolute astrometric reference frame. However, the area of these patches is covered by the public VST ATLAS Survey \citep{shanks/etal:2015}. ATLAS is significantly shallower than KiDS (each ATLAS pointing consists of two \SI{45}{\second} OmegaCAM exposures) but it covers the area with a different pointing footprint than KiDS. This allows us to constrain optical distortions better, and to compensate for the less accurate astrometric 2MASS catalogue. Our global patch-wide astrometric calibration includes \emph{all} KiDS and ATLAS $r$-band images covering the corresponding area.
\end{enumerate}
We obtain a master detection catalogue for each tile by running  \textsc{SExtractor} \citep{bertin/arnouts:1996} on the corresponding co-added \textsc{theli} $r$-band image. These catalogues are the input for both the shape measurements and the multi-colour photometry.

Masks that cover image defects, reflections and ghosts, are also created for the \textsc{theli} reduction. Those are combined with the masks for the multi-colour catalogues described above and applied to the galaxy catalogues. After masking and accounting for overlap between the tiles, we have a unique effective survey area of \SI{360.3}{\degree\squared}.

\subsection{Galaxy photometry and photo-\textit{z}}
\label{sec:gal_photometry}
The KiDS-450 galaxy photometry is based on the same algorithms as were used in KiDS-DR1/2. We extract multi-colour photometry for all objects in the $r$-band master catalogue from PSF-homogenised \textsc{Astro-WISE} images in the $ugri$-bands.

We model the PSFs of the calibrated images in the four bands with shapelets \citep{refregier:2003}, and calculate convolution kernels that transform the PSFs into circular Gaussians. After convolving the images, we extract the photometry using elliptical Gaussian-weighted apertures designed to maximise the precision of colour measurements while properly accounting for seeing differences. The only significant difference in the photometric analysis procedures of the KiDS-450 data with respect to those used for KiDS-DR1/2 is the adjustment of the zero points using SLR as mentioned in Section~\ref{sec:AW}. The resulting improved photometric homogeneity is particularly important for the calibration of the photometric redshifts, which relies on a small number of calibration fields with deep spectroscopy (see Section~\ref{sec:photoz_calibration} below).

For photometric redshift estimation we use the \textsc{bpz} code \citep{benitez:2000} as described in \cite{hildebrandt/etal:2012}. The quality of the Bayesian point estimates of the photo-$z$, $z_{\rm B}$, is presented in detail in \citet[see figs.~10-12 of that paper]{kuijken/etal:2015}. Based on those findings we restrict the photo-$z$ range for the cosmic shear analysis to $0.1<z_{\rm B}\le0.9$ to limit the outlier\footnote{Outliers are defined as objects with $\left|\frac{z_{\rm spec}-z_{\rm B}}{z_{\rm spec}}\right|>0.15$} rates to values below 10 per cent. In order to achieve a sufficient resolution in the radial direction for the tomographic weak lensing measurement, we subdivide this range into four equally spaced tomographic bins of width $\Delta z_{\rm B} =0.2$. A finer binning is not useful given our photo-$z$ uncertainty, and would compromise our ability to calibrate for additive shear (see Section~\ref{sec:lensfit} and Appendix~\ref{sec:c_term}). Table~\ref{tab:tomo_bins} summarises the properties of the source samples in those bins.

\begin{table*}
\caption{\label{tab:tomo_bins}Properties of the galaxy source samples in the four tomographic bins used in the cosmic shear measurement as well as the full KiDS-450 shear catalogue. The effective number density in column 4 is determined with the method by \citet{heymans/etal:2012} whereas the one in column 5 is determined with the method by \citet{chang/etal:2013}. The ellipticity dispersion in column 6 includes the effect of the \emph{lens}fit weight. Columns 7 and 8 are obtained with the DIR calibration, see Section~\ref{sec:DIR}.}
\begin{center}
 \begin{tabular}{ccrcccccc}
   \hline
   bin & $z_{\rm B}$ range & no. of objects & $n_{\rm eff}$ H12 & $n_{\rm eff}$ C13 & $\sigma_e$ & median$(z_{\rm DIR})_{\rm weighted}$ & $\left<z_{\rm DIR}\right>_{\rm weighted}$ & \textsc{bpz} mean $P(z)$ \\
   &&&  [arcmin$^{-2}$] &  [arcmin$^{-2}$] &&&&\\
   \hline
   1 & $0.1<z_{\rm B}\le0.3$ & 3\,879\,823 & 2.35 & 1.94 & 0.293 & 0.418$\pm$0.041 & 0.736$\pm$0.036 & 0.495\\
   2 & $0.3<z_{\rm B}\le0.5$ & 2\,990\,099 & 1.86 & 1.59 & 0.287 & 0.451$\pm$0.012 & 0.574$\pm$0.016 & 0.493\\
   3 & $0.5<z_{\rm B}\le0.7$ & 2\,970\,570 & 1.83 & 1.52 & 0.279 & 0.659$\pm$0.003 & 0.728$\pm$0.010 & 0.675\\
   4 & $0.7<z_{\rm B}\le0.9$ & 2\,687\,130 & 1.49 & 1.09 & 0.288 & 0.829$\pm$0.004 & 0.867$\pm$0.006 & 0.849\\
   \hline
   total & no $z_{\rm B}$ cuts & 14\,640\,774 & 8.53 & 6.85 & 0.290 & & \\
   \hline
 \end{tabular}
\end{center}
\end{table*}

It should be noted that the photo-$z$ code is merely used to provide a convenient quantity (the Bayesian redshift estimate $z_{\rm B}$) to bin the source sample, and that in this analysis we do not rely on the posterior redshift probability distribution functions $P(z)$ estimated by \textsc{bpz}. Instead of stacking the $P(z)$ to obtain an estimate of the underlying true redshift distribution, i.e., the strategy adopted by CFHTLenS \citep[see for example][]{heymans/etal:2013,kitching/etal:2014} and the KiDS early-science papers \citep{viola/etal:2015,sifon/etal:2015,vanuitert/etal:2016,brouwer/etal:2016}, we now employ spectroscopic training data to estimate the redshift distribution in the tomographic bins directly (see Section~\ref{sec:photoz_calibration}). The reason for this approach is that the output of \textsc{bpz} (and essentially every photo-$z$ code; see e.g. \citealt{hildebrandt/etal:2010}) is biased at a level that cannot be tolerated by contemporary and especially future cosmic shear measurements \citep[for a discussion see][]{newman/etal:2015}.

\subsection{Shear measurements with \emph{lens}fit}
\label{sec:lensfit}
Gravitational lensing manifests itself as small coherent distortions of background galaxies. Accurate measurements of galaxy shapes are hence fundamental to mapping the matter distribution across cosmic time and to constraining cosmological parameters.  In this work we use the \emph{lens}fit likelihood based model-fitting method to estimate the shear from the shape of a galaxy \citep{miller/etal:2007, miller/etal:2013, kitching/etal:2008, fenechconti/etal:2016}. 

We refer the reader to the companion paper \cite{fenechconti/etal:2016} for a detailed description of the most recent improvements to the \emph{lens}fit algorithm, shown to successfully `self-calibrate' against noise bias effects as determined through the analysis of an extensive suite of image simulations.   This development is a significant advance on the version of the algorithm used in previous analyses of CFHTLenS, the KiDS-DR1/2 data, and the Red Cluster Sequence Lensing Survey \citep[][RCSLenS]{hildebrandt/etal:2016}.   The main improvements to the \emph{lens}fit algorithm and to our shape measurement analysis since \citet{kuijken/etal:2015} are summarised as follows:
\begin{enumerate}
\item All measurements of galaxy ellipticities are biased by pixel noise in the images.  Measuring ellipticity involves a non-linear transformation of the pixel values which causes a skewness of the likelihood surface and hence a bias in any single point ellipticity estimate \citep{refregier/etal:2012, melchior/viola:2012, miller/etal:2013,viola/etal:2014}. In order to mitigate this problem for \emph{lens}fit we apply a correction for noise bias, based on the actual measurements, which we refer to as `self-calibration'. When a galaxy is measured, a nominal model is obtained for that galaxy, whose parameters are obtained from a maximum likelihood estimate. The idea of `self-calibration' is to create a  simulated noise-free test galaxy with those parameters, re-measure its shape using the same measurement pipeline, and measure the difference between the re-measured ellipticity and the known test model ellipticity. We do not add multiple noise realisations to the noise-free galaxies, as this is computationally too expensive, but we calculate the likelihood as if noise were present. It is assumed that the measured difference is an estimate of the true bias in ellipticity for that galaxy, which is then subtracted from the data measurement. This method approximately corrects for noise bias only, not for other effects such as model bias. It leaves a small residual noise bias, of significantly reduced amplitude, that we parameterise and correct for using image simulations (see Appendix~\ref{sec:imsim}).
\item The shear for a population of galaxies is computed as a weighted average of the measured ellipticities. The weight accounts both for shape-noise variance and ellipticity measurement-noise variance, as described in \cite{miller/etal:2013}. As the measurement noise depends to some extent on the degree of correlation between the intrinsic galaxy ellipticity and the PSF distortion, the weighting introduces biases in the shear measurements. We empirically correct for this effect \citep[see][for further details]{fenechconti/etal:2016} by quantifying how the variance of the measured mean galaxy ellipticity depends on galaxy ellipticity, signal-to-noise ratio and isophotal area.  We then require that the distribution of the re-calibrated weights is neither a strong function of observed ellipticity nor of the relative PSF-galaxy position angle.   The correction is determined from the full survey split into 125 subsamples.  The sample selection is based on the local PSF model ellipticity ($\epsilon_1^*$, $\epsilon_2^*$) and PSF model size in order to accommodate variation in the PSF across the survey using 5 bins for each PSF observable.
\item The sampling of the likelihood surface is improved in both speed and accuracy, by first identifying the location of the maximum likelihood and only then applying the adaptive sampling strategy described by \cite{miller/etal:2013}. More accurate marginalisation over the galaxy size parameter is also implemented.
\item In surveys at the depth of CFHTLenS or KiDS, it is essential to deal with contamination from closely neighbouring galaxies (or stars). The \emph{lens}fit algorithm fits only individual galaxies, masking contaminating stars or galaxies in the same postage stamp during the fitting process. The masks are generated from an image segmentation and masking algorithm, similar to that employed in \textsc{SExtractor}.   We find that the CFHTLenS and KiDS-DR1/2 version of \emph{lens}fit rejected too many target galaxies that were close to a neighbour. For this analysis, a revised de-blending algorithm is adopted that results in fewer rejections and thus a higher density of measured galaxies. The distance to the nearest neighbour, known as the `contamination radius',  is recorded in the catalogue output so that any bias as a function of neighbour distance can be identified and potentially rectified by selecting on that measure (see Fig.~\ref{fig:set_contam_radius} in Appendix~\ref{sec:app_shear_tests}).
\item A large set of realistic, end-to-end image simulations (including chip layout, gaps, dithers, coaddition using \textsc{swarp}, and object detection using \textsc{sextractor}) is created to test for and calibrate a possible residual multiplicative shear measurement bias in \emph{lens}fit. These simulations are briefly described in Appendix~\ref{sec:imsim} with the full details presented in \citet{fenechconti/etal:2016}. We estimate the multiplicative shear measurement bias $m$ to be less than about 1 per cent with a statistical uncertainty, set by the volume of the simulation, of $\sim0.3$ per cent. We further quantify the additional systematic uncertainty coming from differences between the data and the simulations and choices in the bias estimation to be 1 per cent. Such a low bias represents a factor of four improvement over previous \emph{lens}fit measurements (e.g. CFHTLenS) that did not benefit from the `self-calibration'. As shown in Fig.~\ref{fig:PS_marginal} of Appendix~\ref{sec:PS_marginal} this level of precision on the estimate of $m$ is necessary not to compromise the statistical power of the shear catalogue for cosmology.
\item We implement a blinding scheme designed to prevent or at least suppress confirmation bias in the cosmology analysis, along similar lines to what was done in KiDS-DR1/2. The catalogues used for the analysis contain three sets of shear and weight values: the actual measurements, as well as two fake versions. The fake data contain perturbed shear and weight values that are derived from the true measurements through parameterized smooth functions designed to prevent easy identification of the true data set. The parameters of these functions as well as the labelling of the three sets are determined randomly using a secret key that is known only to an external `blinder', Matthias Bartelmann. The amplitude of the changes is tuned to ensure that the best-fit $S_8$ values for the three data sets differ by at least the 1-$\sigma$ error on the Planck measurement. All computations are run on the three sets of shears and weights and the lead authors add a second layer of blinding (i.e. randomly shuffling the three columns again for each particular science project) to allow for phased unblinding within the consortium. In this way co-authors can remain blind because only the second layer is unblinded for them. Which one of the three shear datasets in the catalogues is the truth is only revealed to the lead authors once the analysis is complete. 
\end{enumerate} 

In Appendix~\ref{sec:galaxy_selection} we detail the object selection criteria that are applied to clean the resulting \emph{lens}fit shear catalogue.  The final catalogue provides shear measurements for close to 15 million galaxies, with an effective number density of $n_{\rm eff} = 8.53$ galaxies arcmin$^{-2}$ over a total effective area of \SI{360.3}{\degree\squared}. The inverse shear variance per unit area of the KiDS-450 data, $\hat{w}=\sum w_i/A$, is \SI{105}{\arcmin^{-2}}. We use the effective number density $n_{\rm eff}$ as defined in \cite{heymans/etal:2012} as this estimate can be used to directly populate numerical simulations to create an unweighted mock galaxy catalogue, and it is also used in the creation of the analytical covariance (Section~\ref{sec:analyticalcov}). We note that this value represents a $\sim 30$ per cent increase in the effective number density over the previous KiDS DR1/2 shear catalogue.  This increase is primarily due to the improved \emph{lens}fit masking algorithm.  Table~\ref{tab:tomo_bins} lists the effective number density for each of the four tomographic bins used in this analysis and the corresponding weighted ellipticity variance. For completeness we also quote the number densities according to the definition by \citet{chang/etal:2013}.
%2

\section{CALIBRATION OF PHOTOMETRIC REDSHIFTS}
\label{sec:photoz_calibration}
The cosmic shear signal depends sensitively on the redshifts of all sources used in a measurement. Any cosmological interpretation requires a very accurate calibration of the photometric redshifts that are used for calculating the model predictions \citep{huterer/etal:2006,vanwaerbeke/etal:2006}. The requirements for a survey like KiDS are already quite demanding if the systematic error in the photo-$z$ is not to dominate over the statistical errors. For example, as detailed in Appendix~\ref{sec:app_A}, even a Gaussian 1-$\sigma$ uncertainty on the measured mean redshift of each tomographic bin of $0.05(1+z)$ can degrade the statistical errors on relevant cosmological parameters by $\sim25$ per cent. While such analytic estimates based on Gaussian redshift errors are a useful guideline,  photometric redshift distributions of galaxy samples typically have highly non-Gaussian tails, further complicating the error analysis.

In order to obtain an accurate calibration and error analysis of our redshift distribution we compare three different methods that rely on spectroscopic redshift (spec-$z$) training samples.
\begin{description}
  \item DIR: A weighted direct calibration obtained by a magnitude-space re-weighting \citep{lima/etal:2008} of spectroscopic redshift catalogues that overlap with KiDS.
  \item CC: An angular cross-correlation based calibration \citep{newman:2008} with some of the same spectroscopic catalogues.
  \item BOR: A re-calibration of the $P(z)$ of individual galaxies estimated by \textsc{bpz} in probability space as suggested by \citet{bordoloi/etal:2010}. 
\end{description}

An important aspect of our KiDS-450 cosmological analysis is an investigation into the impact of these different photometric redshift calibration schemes on the resulting cosmological parameter constraints, as presented in Section~\ref{sec:MCMC_photoz}.

\subsection{Overlap with spectroscopic catalogues}
\label{sec:overlap_specz}
KiDS overlaps with several spectroscopic surveys that can be exploited to calibrate the photo-$z$: in particular GAMA \citep{driver/etal:2011}, SDSS \citep{alam/etal:2015}, 2dFLenS (Blake et al., in preparation), and various spectroscopic surveys in the COSMOS field \citep{scoville/etal:2007}. Additionally there are KiDS-like data obtained with the VST in the Chandra Deep Field South (CDFS) from the VOICE project \citep{vaccari/etal:2012} and in two DEEP2 \citep{newman/etal:2013} fields, as detailed in Appendix~\ref{app:deepzdata}.

The different calibration techniques we apply require different properties of the spec-$z$ catalogues. The weighted direct calibration as well as the re-calibration of the $P(z)$ require a spec-$z$ catalogue that covers the same volume in colour and magnitude space as the photometric catalogue that is being calibrated. This strongly limits the use of GAMA, 2dFLenS, and SDSS for these methods since our shear catalogue is limited at $r>20$ whereas all three of these spectroscopic projects target only objects at brighter magnitudes. 

The cross-correlation technique does not have this requirement. In principle one can calibrate a faint photometric sample with a bright spectroscopic sample, as long as both cluster with each other. Being able to use brighter galaxies as calibrators represents one of the major advantages of the cross-correlation technique. However, for this method to work it is still necessary for the spec-$z$ sample to cover the full redshift range that objects in the photometric sample could potentially span given their apparent magnitude. For our shear catalogue with $r\la25$ this means that one needs to cover redshifts all the way out to $z\sim4$. While GAMA and SDSS could still yield cross-correlation information at low $z$ over a wide area those two surveys do not cover the crucial high $z$ range where most of the uncertainty in our redshift calibration lies. Hence, we limit ourselves to the deeper surveys in order to reduce processing time and data handling. The SDSS QSO redshift catalogue can be used out to very high-$z$ for cross-correlation techniques, but due to its low surface density the statistical errors when cross-correlated to KiDS-450 are too large for our purposes.

In the COSMOS field we use a non-public catalogue that was kindly provided by the zCOSMOS \citep{lilly/etal:2009} team and goes deeper than the latest public data release. It also includes spec-$z$ measurements from a variety of other spectroscopic surveys in the COSMOS field which are all used in the weighted direct calibration and the re-calibration of the $P(z)$ but are not used for the calibration with cross-correlations (for the reasons behind this choice see Section~\ref{sec:CC}). In the CDFS we use a compilation of spec-$z$ released by ESO\footnote{\url{http://www.eso.org/sci/activities/garching/projects/goods/MasterSpectroscopy.html}}. This inhomogeneous sample cannot be used for cross-correlation studies but is well suited for the other two approaches. The DEEP2 catalogue is based on the fourth data release \citep{newman/etal:2013}. While DEEP2 is restricted in terms of redshift range, in comparison to zCOSMOS and CDFS it is more complete at $z\ga1$. Thus, it adds crucial information for all three calibration techniques. Table~\ref{tab:specz_surveys} summarises the different spec-$z$ samples used for photo-$z$ calibration.   The number of objects listed refers to the number of galaxies in the spec-$z$ catalogues for which we have photometry from KiDS-450 or the auxilliary VST imaging data described in Appendix~\ref{app:deepzdata}. For details about the completeness of DEEP2 see \citet{newman/etal:2013}. COSMOS and CDFS, however, lack detailed information on the survey completeness.

\begin{table}
\caption{\label{tab:specz_surveys}Spectroscopic samples used for KiDS photo-$z$ calibration. The COSMOS catalogue is dominated by objects from zCOSMOS--bright and zCOSMOS--deep but also includes spec-$z$ from several other projects. While the DIR and BOR approaches make use of the full sample, the CC approach is limited to the DEEP2 sample and the original zCOSMOS sample.}
\begin{center}
\begin{tabular}{lrll}
\hline
sample  & no. of objects & $r_{\lim}$  & $z_{\rm spec}$ range \\
\hline
COSMOS  & 13\,397 & $r\la24.5$ & $0.0<z<3.5$ \\
CDFS    &  2\,290 & $r\la25$   & $0.0<z<4$   \\
DEEP2   &  7\,401 & $r\la24.5$ & $0.6<z<1.5$ \\
\hline
\end{tabular}
\end{center}
\end{table}

\subsection{Weighted direct calibration (DIR)}
\label{sec:DIR}
The most direct way to calibrate photo-$z$ distributions is simply to use the distribution of spec-$z$ for a sample of objects selected in the same way as the photometric sample of interest (e.g. a tomographic photo-$z$ bin). While this technique requires very few assumptions, in practice spec-$z$ catalogues are almost never a complete, representative sub-sample of contemporary shear catalogues. The other main disadvantage of this method is that typical deep spec-$z$ surveys cover less area than the photometric surveys they are supposed to calibrate, such that sample variance becomes a concern.

A way to alleviate both problems has been suggested by \citet{lima/etal:2008}. Using a $k$-nearest-neighbour search, the volume density of objects in multi-dimensional magnitude space is estimated in both the photometric and spectroscopic catalogues. These estimates can then be used to up-weight spec-$z$ objects in regions of magnitude space where the spec-$z$ are under-represented and down-weight them where they are over-represented. It is clear that this method will only be successful if the spec-$z$ catalogue spans the whole volume in magnitude space that is occupied by the photo-$z$ catalogue and samples this colour space densely enough. Another requirement is that the dimensionality of the magnitude space is high enough to allow a unique matching between colour and redshift. These two requirements certainly also imply that the spec-$z$ sample covers the whole redshift range of the photometric sample. A first application of this method to a cosmic shear measurement is presented in \citet{bonnett/etal:2015}.

Since the spectroscopic selection function is essentially removed by the re-weighting process, we can use any object with good magnitude estimates as well as a secure redshift measurement. Thus, we employ the full spec-$z$ sample described in Section~\ref{sec:overlap_specz} for this method. 

When estimating the volume density in magnitude space of the photometric sample we incorporate the \emph{lens}fit weight into the estimate. Note that we use the full distribution of \emph{lens}fit weights in the \emph{unblinded} photometric catalogue for this. Weights are different for the different blindings but we separate the data flows for calibration and further catalogue processing to prevent accidental unblinding. By incorporating the \emph{lens}fit weight we naturally account for the weighting of the shear catalogue without analysing the VST imaging of the spec-$z$ fields with the \emph{lens}fit shear measurement algorithm. This yields a more representative and robust estimate of the weighted redshift distribution. 

Special care has to be taken for objects that are not detected in all four bands. Those occur in the photometric as well as in the spectroscopic sample, but in different relative abundances. We treat these objects as separate classes essentially reducing the dimensionality of the magnitude space for each class and re-weighting those separately. After re-weighting, the classes are properly combined taking their relative abundances in the photometric and spectroscopic catalogue into account. Errors are estimated from 1000 bootstrap samples drawn from the full spec-$z$ training catalogue. These bootstrap errors include shot noise but do not correct for residual effects of sample variance, which can still play a role because of the discrete sampling of magnitude space by the spec-$z$ sample. Note though that sample variance is strongly suppressed by the re-weighting scheme compared to an unweighted spec-$z$ calibration since the density in magnitude space is adjusted to the cosmic (or rather KiDS-450) average.
A discussion of the influence of sample variance in the DIR redshift calibration can be found in Appendix~\ref{sec:sysDIR}.

\begin{figure*}
\includegraphics[width=\textwidth,trim=0 0.45cm 0 1cm,clip]{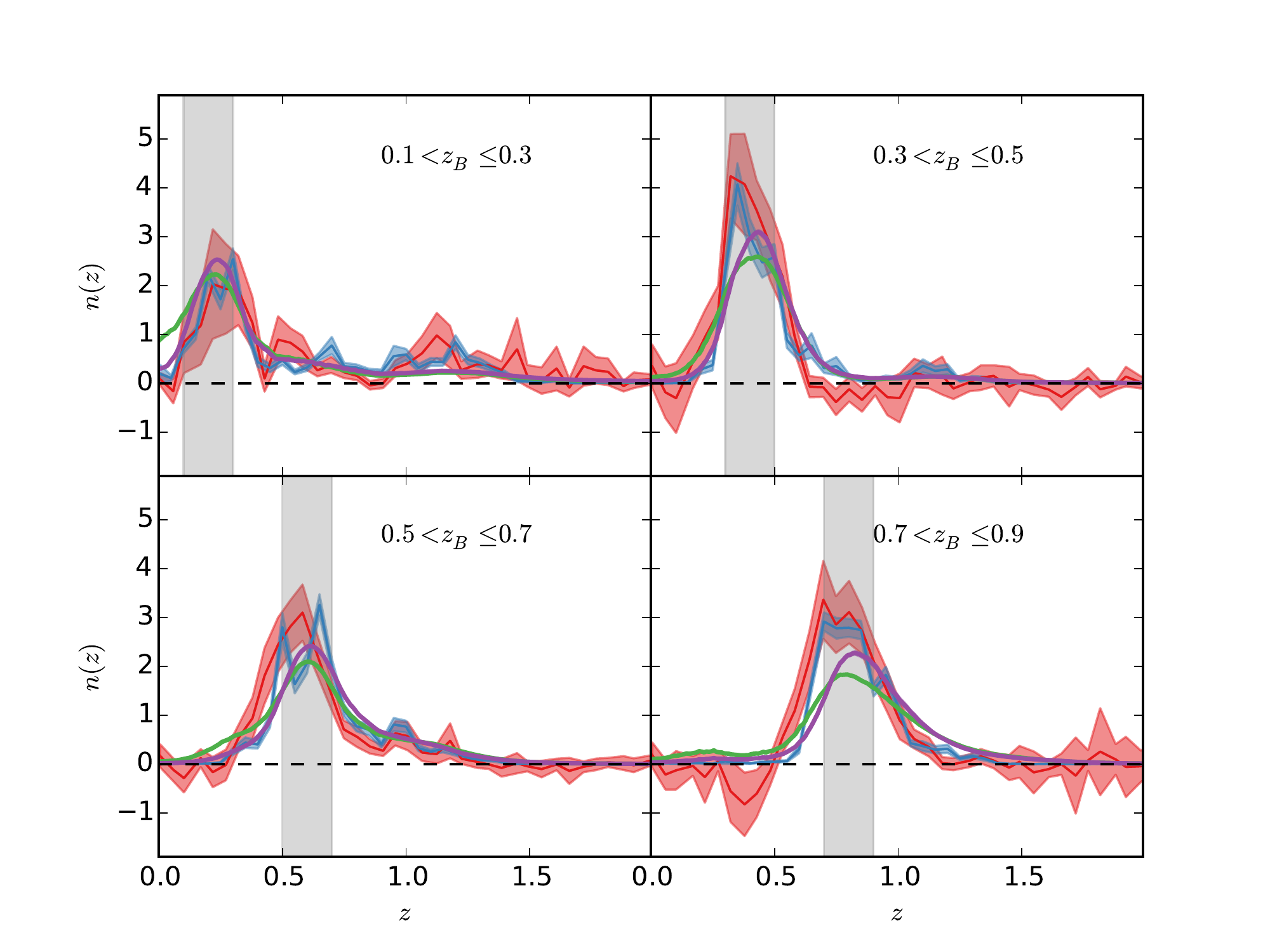}
\caption{\label{fig:z_dist_comp}Comparison of the normalised redshift distributions for the four tomographic bins as estimated from the weighted direct calibration (DIR, blue with errors), the calibration with cross-correlations (CC, red with errors), the re-calibrated stacked $P_{\rm recal}(z)$ (BOR, purple with errors that are barely visible), and the original stacked $P(z)$ from \textsc{bpz} (green). The gray-shaded regions indicate the target redshift range selected by cuts on the Bayesian photo-$z$ $z_{\rm B}$. Errors shown here do not include the effects of sample variance in the spec-$z$ calibration sample.}
\end{figure*}

A comparison of the resulting redshift distributions of the weighted direct calibration and the stacked $P(z)$ from \textsc{bpz} (see Section~\ref{sec:gal_photometry}) for the four tomographic bins is shown in Fig.~\ref{fig:z_dist_comp} (blue line with confidence regions). Note that especially the $n(z)$ in the first tomographic bin is strongly affected by the $r>20$ cut introduced by \emph{lens}fit which skews the distribution to higher redshifts and increases the relative amplitude of the high-$z$ tail compared to the low-$z$ bump. This is also reflected in the large difference between the mean and median redshift of this bin given in Table~\ref{tab:tomo_bins}. In Appendix~\ref{sec:sysDIR} we discuss and test the assumptions and parameter choices made for this method. Note that we determine the redshift distributions up to the highest spectroscopic reshifts of $z\sim4$ but only plot the range $0<z<2$ in Fig.~\ref{fig:z_dist_comp}. There are no significant $z>2$ bumps in the DIR redshift distribution for these four tomographic bins.

\subsection{Calibration with cross-correlations (CC)}
\label{sec:CC}
The use of angular cross-correlation functions between photometric and spectroscopic galaxy sample for re-constructing photometric redshift distributions was described in detail by \citet{newman:2008}. This approach has the great advantage of being rather insensitive to the spectroscopic selection function in terms of magnitude, galaxy type, etc., as long as it spans the full redshift range of interest. However, angular auto-correlation function measurements of the spectroscopic as well as the photometric samples are needed, to measure and correct for the -- typically unknown -- galaxy bias. In order to estimate these auto-correlations, precise knowledge of the angular selection function (i.e., the weighted footprint) of the samples is required. 

For the photometric catalogues, the angular selection functions can be estimated from the masks mentioned in Section~\ref{sec:AW}. We do not correct for depth and seeing variations as described in \citet{morrison/etal:2015} since those are relatively unimportant on the small spec-$z$ fields used here. Regarding the spectroscopic datasets, DEEP2 provide maps of the angular selection function, allowing us to calculate all correlation functions over the full \SI{0.8}{\square\degree} overlap area with KiDS-like VST imaging.  We do not have a similar spectroscopic selection function for COSMOS or CDFS. Given the small size and heterogeneity of the CDFS catalogue we cannot use it for the cross-correlation calibration; for COSMOS we restrict ourselves to the central \SI{0.7}{\square\degree} region covered very homogeneously by zCOSMOS, and we assume a constant selection function outside the masks of the KiDS data\footnote{Using the KiDS masks here makes sense since photometric as well as spectroscopic surveys are affected by e.g. bright stars and typical footprints often look quite similar.}. We do not use spec-$z$ measurements from other surveys in the COSMOS field for the cross-correlations. Both samples, DEEP2 and zCOSMOS, are analysed independently, and only at the very end of the analysis the redshift distributions are averaged with inverse variance weighting.

We employ an advanced version of the original technique proposed by \citet{newman:2008} and \citet{matthews/etal:2010} that is described in \citet{menard/etal:2013} and \citet{schmidt/etal:2013}. Unlike \citet{newman:2008}, who proposed using only linear scales, \citet{menard/etal:2013} and \citet{schmidt/etal:2013} advocate exploiting the much higher signal-to-noise ratio available on smaller non-linear scales, even though this comes at the cost of more complicated galaxy bias modelling. Additionally they describe how pre-selection of the photometric sample by photometric quantities can narrow down the underlying redshift distribution and make the technique less susceptible to the galaxy bias correction \citep[see also][]{rahman/etal:2016}.

A description of the full details and tests of our implementation of this calibration method can be found in Appendix~\ref{sec:sysCC}. We summarize the steps here.

All correlation functions are estimated over a fixed range of proper separation of 30--300\,kpc. The conversion of angular to proper scales requires a cosmological model. Here we assume a WMAP5 cosmology \citep{komatsu/etal:2009}, noting that the redshift recovery is insensitive to this choice and therefore does not bias the constraints given in Section~\ref{sec:results}. The auto-correlation functions of the spec-$z$ samples are estimated with a coarse redshift binning to allow for reliable power-law fits with small errors. We assume a linear relation between redshift and the power-law parameters $r_0$ and $\gamma$ and fit it to the results of all the redshift bins with $0<z_{\rm spec}<1.2$. For $z_{\rm spec}>1.2$ we fit a constant $r_0$ and $\gamma$.

The cross-correlation functions are estimated with a finer binning in spec-$z$ in order to obtain redshift distributions for the tomographic bins with high resolution. The raw cross-correlations are corrected for evolving galaxy bias with the recipe by \citet{newman:2008} and \citet{matthews/etal:2010}. We estimate statistical uncertainties from a bootstrap re-sampling of the spectroscopic training set (1000 bootstrap samples). The whole re-calibration procedure, including correlation function estimates and bias correction, is run for each bootstrap sample.

Note that the cross-correlation function can attain negative values that would lead to unphysical negative amplitudes in the $n(z)$. Nevertheless, it is important to allow for these negative values in the estimation of the cross-correlation functions so as not to introduce any bias. Such negative amplitudes can for example be caused by local over- or underdensities in the spec-$z$ catalogue as explained by \citet{rahman/etal:2015}. Only after the full redshift recovery process do we re-bin the distributions with a coarser redshift resolution to attain positive values for $n(z)$ throughout.

The redshift distributions from this method, based on the combination of the DEEP2 and zCOSMOS results, are displayed in Fig.~\ref{fig:z_dist_comp} (red line with confidence regions). Note that the uncertainties on the redshift distributions from the cross-correlation technique are larger than the uncertainties on the weighted direct calibration, owing to the relatively small area of sky covered by the spec-$z$ catalogues. As will be shown in Section~\ref{sec:results}, propagating the $n(z)$ and associated errors from the CC method into the cosmological analysis yields cosmological parameters that are consistent with the ones that are obtained when using the DIR redshift distributions, despite some differences in the details of the redshift distributions.

\subsection{Re-calibration of the photometric \textit{P}(\textit{z}) (BOR)} 
\label{sec:BOR}
Many photo-$z$ codes estimate a full redshift likelihood, $\mathcal{L}(z)$, for each galaxy or a posterior probability distribution, $P(z)$, in case of a Bayesian code like \textsc{bpz}. \citet{bordoloi/etal:2010} suggested to use a representative spectroscopic training sample and analyse the properties of the photometric redshift likelihoods of those galaxies. 

For each spectroscopic training object the photometric $P(z)$ is integrated from zero to $z_{\rm spec}$ yielding the cumulative quantity:
\begin{equation}
P_{\Sigma}(z_{\rm spec}) = \int_0^{z_{\rm spec}} P(z') \, {\rm d}z'\,.
\label{eqn:PBOR}
\end{equation}
If the $P(z)$ are a fair representation of the underlying probability density, the $P_{\Sigma}$ for the full training sample should be uniformly distributed between zero and one. If this distribution $N(P_{\Sigma})$ is not flat, its shape can be used to re-calibrate the original $P(z)$ as explained in \citet{bordoloi/etal:2010}.

One requirement for this approach to work is that the training sample is completely representative of the photometric sample to be calibrated. Since this is not the case for KiDS-450 we employ this re-calibration technique in combination with the re-weighting procedure in magnitude space described in Section~\ref{sec:DIR}. Some tests on the performance of this method are described in Appendix~\ref{sec:sysBOR}.

We make use of the full spec-$z$ sample, similar to the weighted direct calibration mentioned above. The resulting re-calibrated, stacked $P_{\rm recal}(z)$ are also included in Fig.~\ref{fig:z_dist_comp} (purple lines). Errors are estimated from 1000 bootstrap samples. The re-calibration changes very little between the bootstrap samples which is reflected in the comparably small errors on the purple lines. This is due to the fact that the BOR method uses the $P(z)$ output from \textsc{bpz} directly whereas the DIR and CC methods are completely ignorant about this information.

\subsection{Discussion}
The four sets of redshift distributions from the different techniques displayed in Fig.~\ref{fig:z_dist_comp} show some differences, most prominently in the first and fourth tomographic bin. While most of these differences are not very significant within the errors\footnote{Note that errors at different redshifts are correlated.} it is clear that the resulting theoretical model will differ depending on which set is chosen. This is particularly true for the first redshift bin where the redshift distribution obtained with the stacked $P(z)$ from \textsc{BPZ} is quite different from the re-calibrated distributions obtained by DIR and CC. This is also reflected in the different mean redshift in this bin for DIR and \textsc{BPZ} reported in Table~\ref{tab:tomo_bins}. Due to the more pronounced high-$z$ tail in the DIR (and CC) distributions the mean redshift in this first bin is actually higher than the mean redshift in the 2nd and 3rd bin in contrast to what is found for \textsc{BPZ}. The fact that both, DIR and CC, independently recover this high-$z$ tail with similar amplitude makes us confident that it is real. As discussed in Section~\ref{sec:results} this has profound consequences for the best-fit intrinsic alignment amplitude, $A_{\rm IA}$. Apart from these differences it is encouraging to see that some of the features that are not present in the stacked \textsc{BPZ} $P(z)$ are recovered by all three re-calibration techniques, e.g. the much lower amplitude for DIR, CC, and BOR compared to BPZ at very low redshift in the first tomographic bin.

Applying the calibrations determined from a few deep spectroscopic fields to the full survey requires a consistent photometric calibration. As briefly mentioned above (Section~\ref{sec:AW}) and described in more detail in Appendix~\ref{sec:app_SLR} we rely on stellar locus regression to achieve homogeneous photometry over the full survey area.
%3

\section{COSMOLOGICAL ANALYSIS}
\label{sec:theory}
\subsection{Shear two-point correlation functions}
In this analysis we measure the tomographic angular two-point shear correlation function $\hat{\xi}_{\pm}^{ij}$ which can be estimated from two tomographic redshift bins $i$ and $j$ as:
\be
\hat{\xi}_{\pm}^{ij}(\theta) = \frac{\sum_{ab} w_a w_b \left[ \epsilon_\rmn{t}^i (\x_a) \epsilon_\rmn{t}^j (\x_b) \, \pm \, \epsilon_\times^i (\x_a) \epsilon_\times ^j(\x_b)
\right]}{
\sum_{ab} w_a w_b } \, .
\label{eqn:xipm_est}
\ee
Galaxy weights $w$ are included when the sum is taken over pairs of galaxies with angular separation $|\x_a - \x_b|$ within an interval $\Delta \theta$ around $\theta$. The tangential and cross components of the ellipticities $\epsilon_{\rmn{t},\times}$ are measured with respect to the vector $\x_a - \x_b$ joining each pair of objects \citep{bartelmann/schneider:2001}.     This estimator $\hat{\xi}_{\pm}$ can be related to the underlying matter power spectrum $P_\delta$, via
\be
\xi_\pm^{ij}(\theta) = \frac{1}{2\pi}\int \d\ell \,\ell \,P^{ij}_\kappa(\ell) \, J_{0,4}(\ell \theta) \, , 
\label{eqn:xiGG}
\ee
where $J_{0,4} (\ell \theta)$ is the zeroth (for $\xi_+$) or fourth (for $\xi_- $) order Bessel function of the first kind. $P_\kappa(\ell)$ is the convergence power spectrum at angular wave number $\ell$.  Using the Limber approximation one finds
\be 
P^{ij}_\kappa(\ell) = \int_0^{\chi_{\rm H}} \d \chi \, \frac{q_i(\chi)q_j(\chi)}{[f_K(\chi)]^2} \, P_\delta \left( \frac{\ell}{f_K(\chi)},\chi \right),
\label{eqn:Pkappa} 
\ee
where $\chi$ is the comoving radial distance, and $\chi_{\rm H}$ is the horizon distance.  The lensing efficiency function $q(\chi)$ is given by
\be
q_i(\chi) = \frac{3 H_0^2 \Omega_{\rm m}}{2c^2} \frac{f_K(\chi)}{a(\chi)}\int_\chi^{\chi_{\rm H}}\, \d \chi'\ n_i(\chi') 
\frac{f_K(\chi'-\chi)}{f_K(\chi')}, 
\label{eqn:qk} 
\ee
where $a(\chi)$ is the dimensionless scale factor corresponding to the comoving radial distance $\chi$, $n_i(\chi)\,\d \chi$ is the effective number of galaxies in $\d \chi$ in redshift bin $i$, normalised so that $\int_0^{\chi_{\rm H}} n(\chi)\,\d \chi = 1$. $f_K(\chi)$ is the comoving angular diameter distance out to comoving radial distance $\chi$, $H_0$ is the Hubble constant and $\Omega_\rmn{m}$ the matter density parameter at $z=0$. Note that in this derivation we ignore the difference between shear and reduced shear as it is completely negligible for our analysis. For more details see \citet{bartelmann/schneider:2001} and references therein.

Cosmological parameters are directly constrained from KiDS-450 measurements of the observed angular two-point shear correlation function $\xi_{\pm}$ in Section~\ref{sec:results}.  This base measurement could also be used to derive a wide range of alternative statistics.   \citet{SchvWKM02} and \citet{COSEBIS} discuss the relationship between a number of different real-space two-point statistics. Especially the COSEBIs \citep[Complete Orthogonal Sets of E-/B-Integrals]{COSEBIS} statistic yields a very useful separation of E and B modes as well as an optimal data compression. We choose not to explore these alternatives in this analysis, however, as \citet{kilbinger/etal:2013} showed that they provide no significant additional cosmological information over the base $\xi_{\pm}$ measurement.    The real-space measurements of $\xi_{\pm}$ are also input data for the two Fourier-mode conversion methods to extract the power spectrum presented in \citet{becker/etal:2016}. This conversion does not result in additional cosmological information over the base $\xi_{\pm}$ measurement, however, if the observed shear field is B-mode free.

Direct power spectrum measurements that are not based on $\xi_{\pm}$ with CFHTLenS were made by \citet{kohlinger/etal:2016} who present a measurement of the tomographic lensing power spectra using a quadratic estimator, and  \citet{kitching/etal:2014, kitching/etal:2016} present a full 3-D power spectrum analysis.  The benefit of using these direct power spectrum estimators is a cleaner separation of Fourier modes which are blended in the $\xi_{\pm}$ measurement. Uncertainty in modelling the high-$k$ non-linear power spectrum can therefore be optimally resolved by directly removing these $k$-scales \citep[see for example][]{kitching/etal:2014, alsing/etal:2016}.  The alternative for real-space estimators is to remove small $\theta$ scales.   The conclusions reached by these alternative and more conservative analyses however still broadly agree with those from the base $\xi_{\pm}$ statistical analysis \citep{heymans/etal:2013, joudaki/etal:2016}.   

Owing to these literature results we have chosen to limit this first cosmological analysis of KiDS-450 to the $\xi_{\pm}$ statistic, with a series of future papers to investigate alternative statistics.  In Appendix~\ref{sec:EB} we also present an E/B-mode decomposition and analysis of KiDS-450 using the $\xi_{E/B}$ statistic.
 
\subsection{Modelling intrinsic galaxy alignments} 
\label{sec:IAmodels}
The two-point shear correlation function estimator from Eq.~\ref{eqn:xipm_est} does not measure $\xi_\pm$ directly but is corrupted by the following terms:
\be
\left<\hat{\xi}_{\pm}\right> = \xi_\pm + \xi_\pm^{\rm II} + \xi_\pm^{\rm GI} \, ,
\label{eqn:xiobsIIGI}
\ee
where $\xi_\pm^{\rm II}$ measures correlations between the intrinsic ellipticities of neighbouring galaxies (known as `II'), and $\xi_\pm^{\rm GI}$ measures correlations between the intrinsic ellipticity of a foreground galaxy and the shear experienced by a background galaxy (known as `GI').  

We account for the bias introduced by the presence of intrinsic galaxy alignments by simultaneously modelling the cosmological and intrinsic alignment contributions to the observed correlation functions $\hat{\xi}_{\pm}$.  We adopt the `non-linear linear' intrinsic alignment model developed by \citet{hirata/seljak:2004, bridle/king:2007, joachimi/etal:2011}.  This model has been used in many cosmic shear analyses (\citealt{kirk/etal:2010}; \citealt{heymans/etal:2013}; \citealt{abbott/etal:2015};  %\citetalias{abbott/etal:2015}; 
\citealt{joudaki/etal:2016}) as it provides a reasonable fit to both observations and simulations of intrinsic galaxy alignments \citep[see][and references therein]{joachimi/etal:2015}.   In this model, the non-linear intrinsic alignment II and GI power spectra are related to the non-linear matter power spectrum as, 
\begin{equation}
\begin{split}
P_{\rm II}(k,z) = &\: F^2(z) P_\delta(k,z)\\
P_{\rm GI}(k,z) = &\: F(z) P_\delta(k,z) \, ,
\label{eqn:PIIGI} 
\end{split}
\end{equation}
where the redshift and cosmology-dependent modifications to the power spectrum are given by
\be 
F(z) = - A_{\rm IA} C_1 \rho_{\rm crit} \frac{\Omega_{\rm m}}{D_+(z)}  \left( \frac{1 + z}{1+z_0} \right)^\eta \left ( \frac{\bar{L}}{L_0} \right) ^\beta\, .
\label{eqn:Fz} 
\ee
Here $A_{\rm IA}$ is a free dimensionless amplitude parameter that multiplies the fixed normalisation constant $C_1 = 5 \times 10^{-14} \, h^{-2} M_\odot^{-1} {\rm Mpc}^3$, $\rho_{\rm crit}$ is the critical density at $z=0$, and $D_+(z)$ is the linear growth factor normalised to unity today.    The free parameters $\eta$ and $\beta$ allow for a redshift and luminosity dependence in the model around arbitrary pivot values $z_0$ and $L_0$, and $\bar{L}$ is the weighted average luminosity of the source sample.  The II and GI contributions to the observed two-point correlation function in Eq.~\ref{eqn:xiobsIIGI} are related to the II and GI power spectra as
\be
\xi_{\pm}^{ij}(\theta)_{\rm II, GI} = \frac{1}{2\pi}\int \d\ell \,\ell \,C_{\rm II,GI}^{ij}(\ell) \, J_{0,4}(\ell \theta) \, , 
\label{eqn:xiIIGI}
\ee
with 
\be 
C_{\rm II}^{ij}(\ell) = \int \d\chi \, 
\frac{n_i(\chi)n_j(\chi)}{[f_K(\chi)]^2} \, P_{\rm II} \left( \frac{\ell}{f_K(\chi)},\chi \right),
\label{eqn:CII} 
\ee
\be 
C_{\rm GI}^{ij}(\ell) = \int \d\chi \, 
\frac{q_i(\chi)n_j(\chi) + n_i(\chi)q_j(\chi) }{[f_K(\chi)]^2} \, P_{\rm GI} \left( \frac{\ell}{f_K(\chi)},\chi \right),
\label{eqn:CGI} 
\ee
where the projection takes into account the effective number of galaxies in redshift bin $i$, $n_i(\chi)$, and, in the case of GI correlations,  the lensing efficiency $q_{i}(\chi)$ (see Eq.~\ref{eqn:qk}).   

Late-type galaxies make up the majority of the KiDS-450 source sample, and no significant detection of intrinsic alignments for this type of galaxy exists.  A luminosity dependent alignment signal has, however, been measured in massive early-type galaxies with $\beta \simeq1.2 \pm 0.3$, with no evidence for redshift dependence \citep{joachimi/etal:2011,singh/etal:2015}.  
We therefore determine the level of luminosity evolution with redshift for a sample of galaxies similar to KiDS-450 using the `COSMOS2015' catalogue \citep{laigle/etal:2016}.  We select galaxies with $20<m_{\mathrm{r}}<24$ and compute the mean luminosity in the $r$-band for two redshift bins, $0.1<z<0.45$ and $0.45<z<0.9$.  We find the higher redshift bin to be only 3\% more luminous, on average, than the lower redshift bin.  Any luminosity dependence of the intrinsic alignment signal can therefore be safely ignored in this analysis given the very weak luminosity evolution across the galaxy sample and the statistical power of the current data.

\citet{joudaki/etal:2016} present cosmological constraints from CFHTLenS, which has similar statistical power as KiDS-450, using a range of priors for the model parameters $A_{\rm IA}$, $\eta$, and $\beta$ from Eq.~\ref{eqn:Fz}  (see also \citealt{abbott/etal:2015}  %\citetalias{abbott/etal:2015} 
who allow $A_{\rm IA}$ and $\eta$ to vary, keeping $\beta=0$).      Using the Deviance Information Criterion (DIC; see Section~\ref{sec:discussion}) to quantify the relative performance of different models, they find that a flexible two-parameter ($A_{\rm IA}, \beta$) or three-parameter ($A_{\rm IA}, \beta, \eta$) intrinsic alignment model, with or without informative priors, is disfavoured by the data, implying that the CFHTLenS data are insensitive to any redshift- or luminosity-dependence in the intrinsic alignment signal. 

Taking all this information into account, we fix $\eta=0$ and $\beta=0$ for our mixed population of early and late-type galaxies, and set a non-informative prior on the amplitude of the signal $A_{\rm IA}$, allowing it to vary between $-6 < A_{\rm IA} < 6$.

\subsection{Modelling the matter power spectrum including baryon physics}
\label{sec:feedback}

Cosmological parameter constraints are derived from the comparison of the measured shear correlation function with theoretical models for the cosmic shear and intrinsic alignment contributions (Eq.~\ref{eqn:xiobsIIGI}).    One drawback to working with the $\xi_\pm$ real-space statistic is that the theoretical models integrate the matter power spectrum $P_\delta$ over a wide range of $k$-scales (see for example Eq.~\ref{eqn:Pkappa}).  As such we require an accurate model for the matter power spectrum that retains its accuracy well into the non-linear regime.

The non-linear dark matter power spectrum model of \citet{takahashi/etal:2012} revised the `halofit' formalism of \citet{smith/etal:2003}.  The free parameters in the fit were constrained using a suite of N-body simulations spanning 16 different $\Lambda$CDM cosmological models.  This model has been shown to be accurate to $\sim 5$\% down to $k=10\, h\, {\rm Mpc^{-1}}$ when compared to the wide range of N-body cosmological simulations from the `Coyote Universe' \citep{heitmann/etal:2014}.    Where this model lacks flexibility, however, is when we consider the impact that baryon physics could have on the small-scale clustering of matter \citep{vandaalen/etal:2011}.

In \citet{semboloni/etal:2011}, matter power spectra from the `Overwhelmingly Large' (OWLS) cosmological hydrodynamical simulations were used to quantify the biases introduced in cosmic shear analyses that neglect baryon feedback.  The impact ranged from being insignificant to significant, where the most extreme case modelled the baryon feedback with a strong AGN component.  For the smallest angular scales ($\theta\ga\SI{0.5}{\arcmin}$) used in this KiDS-450 cosmic shear analysis, in the AGN case the amplitude of $\xi_\pm$ was found to decrease by up to 20\%, relative to a gravity-only model. This decrement is the result of changes in the total matter distribution by baryon physics, which can be captured by adjusting the parameters in the halo model. This provides a simple and sufficiently flexible parametrisation of this effect, and we therefore favour this approach over alternatives that include polynomial models and principal component analyses of the hydrodynamical simulations \citep{harnois-deraps/etal:2015, eifler/etal:2015}.

In order to model the non-linear power spectrum of dark matter and baryons we adopt the effective halo model from \citet{mead/etal:2015} with its accompanying software \textsc{HMcode} \citep{mead:2015}.   In comparison to cosmological simulations from the `Coyote Universe', the \textsc{HMcode} dark matter-only power spectrum has been shown to be as accurate as the \citet{takahashi/etal:2012} model.  As the model is built directly from the properties of haloes it has the flexibility to vary the amplitude of the halo mass-concentration relation $B$, and also includes a `halo bloating' parameter $\eta_0$ \citep[see eq.~14 and eq.~26 in][]{mead/etal:2015}.  Allowing these two parameters to vary when fitting data from the OWLS simulations results in a model that is accurate to $\sim 3\%$ down to $k=10\, h\, {\rm Mpc ^{-1}}$ for all the feedback scenarios presented in \citet{vandaalen/etal:2011}. \citet{mead/etal:2015} show that these two parameters are degenerate, recommending the use of a single free parameter $B$ to model the impact of baryon feedback on the matter power spectrum, fixing $\eta_0 = 1.03 - 0.11B$ in the likelihood analysis. For this reason we call $B$ the baryon feedback parameter in the following, noting that a pure dark matter model does not correspond to $B=0$ but to $B=3.13$. We choose to impose top-hat priors on the feedback parameter $2< B<4$ given by the range of plausible feedback scenarios from the OWLS  simulations.  Figure 9 of \citet{mead/etal:2015} illustrates how this range of $B$ broadens the theoretical expectation of $\xi_\pm(\theta)$ by less than a per cent for scales with $\theta >\SI{6}{\arcmin}$ for $\xi_+$ and $\theta >\SI{1}{\degree}$ for $\xi_-$.  We show in Section~\ref{sec:MCMC_Bmode} that taking a conservative approach by excluding small angular scales from our cosmological analysis does not significantly alter our conclusions.

We refer the reader to \citet{joudaki/etal:2016} who show that there is no strong preference for or against including this additional degree of freedom in the model of the matter power spectrum when analysing CFHTLenS.    They also show that when considering a dark matter-only power spectrum, the cosmological parameter constraints are insensitive to which power spectrum model is chosen; either \textsc{HMcode} with $B=3.13$, the best fitting value for a dark matter-only power spectrum, or \citet{takahashi/etal:2012}.  In the analysis that follows, whenever baryons are not included in the analysis, the faster (in terms of CPU time) \citet{takahashi/etal:2012} model is used.

\citet{mead/etal:2016} present an extension of the effective halo model to produce accurate non-linear matter power spectra for non-zero neutrino masses.  This allows for a consistent treatment of the impact of both baryon feedback and neutrinos, both of which affect the power spectrum on small scales.  We use this extension to verify that our cosmological parameter constraints are insensitive to a change in the neutrino mass from a fixed $\Sigma m_\nu = 0.00 \,{\rm eV}$ to a fixed $\Sigma m_\nu = 0.06\, {\rm eV}$,  the fiducial value used for example by \citet{planck/cosmo:2015}.   We therefore choose to fix $\Sigma m_\nu = 0.00\, {\rm eV}$ in order to minimise CPU time in the likelihood analysis.  Whilst we are insensitive to a small change of $0.06$\,eV in $\Sigma m_\nu$, KiDS-450 can set an upper limit on the sum of the neutrino masses, and a full cosmological parameter analysis where $\Sigma m_\nu$ varies as a free parameter will be presented in future work (Joudaki et al. in prep., K\"ohlinger et al. in prep.).
%4

\section{COVARIANCE MATRIX ESTIMATION}
\label{sec:covariance}

We fit the correlation functions $\xi_+$ and $\xi_-$ at seven and six  angular scales, respectively, and in four tomographic bins. With ten possible auto- and cross-correlation functions from the tomographic bins, our data vector therefore has 130 elements. We construct three different estimators of the covariance matrix to model the correlations that exist between these measurements: an analytical model, a numerical estimate from mock galaxy catalogues, and a direct measurement from the data using a Jackknife approach.    There are merits and drawbacks to each estimator which we discuss below.  In the cosmological analysis that follows in Section~\ref{sec:results} we use the analytical covariance matrix as the default.

We neglect the dependence of the covariance matrix on cosmological parameters. According to \citet{eifler/etal:2009,kilbinger/etal:2013} this is not expected to impact our conclusions as the cosmological parameter constraints from KiDS-450 data are consistent with the `WMAP9' cosmology adopted for both our numerical and analytical approaches, with $\Omega_{\rm m} = 0.2905$, $\Omega_\Lambda = 0.7095$, $\Omega_{\rm b} = 0.0473$, $h = 0.6898$, $\sigma_8 = 0.826$ and $n_{\rm s} = 0.969$ \citep{hinshaw/etal:2013}.

\subsection{Jackknife covariance matrix}
The Jackknife approach to determine a covariance matrix  is completely empirical and does not require any assumptions of a fiducial background cosmology \citep[see for example][]{heymans/etal:2005, friedrich/etal:2016}.  We measure $N_{\rm JK}=454$ Jackknife sample estimates of $\xi_{\pm}$ by removing a single KiDS-450 tile in turn.  We then construct a Jackknife covariance estimate from the variance between the partial estimates \citep{wall/jenkins:2012}. The main drawbacks of the Jackknife approach are the high levels of noise in the measurement of the covariance which results in a biased inversion of the matrix, the bias that results from measuring the covariance between correlated samples, and the fact that the Jackknife estimate is only valid when the removed sub-samples are representative of the data set \citep[see for example][]{zehavi/etal:2002}.   We therefore only trust our Jackknife estimate for angular scales less than half the extent of the excised Jackknife region, which in our analysis extends to \SI{1}{\degree}.  With the patchwork layout of KiDS-450 (see Fig.~\ref{fig:footprint}) larger Jackknife regions are currently impractical, such that we only use the Jackknife estimate to verify the numerical and analytical estimators on scales $\theta < \SI{30}{\arcmin}$.

\subsection{Numerical covariance matrix}
\label{sec:mockcov}
The standard approach to computing the covariance matrix employs a set of mock catalogues created from a large suite of $N$-body simulations.  With a sufficiently high number of independent simulations, the impact of noise on the measurement can be minimised and any bias in the inversion can be corrected to good accuracy \citep{hartlap/etal:2007,taylor/joachimi:2014,sellentin/heavens:2016}.  The main benefit of this approach is that small-scale masks and observational effects can readily be applied and accounted for with the mock catalogue.   The major drawback of this approach is that variations in the matter distribution that are larger than the simulation box are absent from the mock catalogues.  As small-scale modes couple to these large-scale modes (known as `super-sample covariance' or SSC), numerical methods tend to underestimate the covariance, particularly on large scales where sample variance dominates.  This could be compensated by simply using larger-box simulations, but for a fixed number of particles, the resulting lack of resolution then results in a reduction of power on small scales.  This dilemma accounts for the main drawback of using mocks, which we address by taking an alternative analytical approach that includes the SSC contribution to the total covariance in Section~\ref{sec:analyticalcov}.

Our methodology to construct a numerical covariance matrix follows that described in \citet{heymans/etal:2013}, which we briefly outline here.  We produce mock galaxy catalogues using 930 simulations from the SLICS (Scinet LIght Cone Simulation) project \citep{harnois-deraps/etal:2015}.  Each simulation follows the non-linear evolution of $1536^3$ particles within a box of size $505 \, h^{-1}$ Mpc.   The density field is output at 18 redshift snapshots in the range $0 < z < 3$.  The gravitational lensing shear and convergence are computed at these lens planes, and a survey cone spanning \SI{60}{\square\degree} with a pixel resolution of \SI{4.6}{\arcsec} is constructed.  In contrast to previous analyses, we have a sufficient number of simulations such that we do not need to divide boxes into sub-realizations to increase the number of mocks.

We construct mock catalogues for the four tomographic bins by Monte-Carlo sampling sources from the density field to match the mean DIR redshift distribution and effective number density in each  bin, from the values listed in Table~\ref{tab:tomo_bins}. Since this $n(z)$ already includes the {\it lens}fit weights, each mock source is assigned a weight $w_i= 1$. We assign two-component gravitational shears to each source by linearly interpolating the mock shear fields, and apply shape noise components drawn from a Gaussian distribution determined in each bin from the weighted ellipticity variance of the data (see Table~\ref{tab:tomo_bins}).  We apply representative small-scale masks to each realisation using a fixed mask pattern drawn from a section of the real data.  We hence produce 930 mock shear catalogues matching the properties of the KiDS-450 survey, each covering \SI{60}{\square\degree}.

We measure the cosmic shear statistics in the mock catalogue using an identical set-up to the measurement of the data. We derive the covariance through area-scaling of the effective area of the mock to match that of the effective area of the KiDS-450 dataset, accounting for regions lost through masking. Area-scaling correctly determines the total shape noise contribution to the covariance.  It is only approximate, however, when scaling the cosmological Gaussian and non-Gaussian terms.  We use a log-normal approximation \citep{hilbert/etal:2011} to estimate the error introduced by area-scaling the mock covariance.  We calculate that for the typical area of each KiDS patch ($\sim$\SI{100}{\square\degree}) relative to the area of each mock catalogue (\SI{60}{\square\degree}), area-scaling introduces less than a 10\% error on the amplitude of the cosmological contributions to the covariance.

\subsection{Analytical covariance matrix}
\label{sec:analyticalcov}
Our favoured approach to computing the correlation function covariance employs an analytical model.    The model is composed of three terms:\\
(i) a disconnected part that includes the Gaussian contribution to sample variance, shape noise, as well as a mixed noise-sample variance term, \\
(ii) a non-Gaussian contribution from in-survey modes that originates from the connected trispectrum of matter, and \\
(iii) a contribution due to the coupling of in-survey and super-survey modes. \\
This approach is an advance over the numerical or Jackknife approach as it does not suffer from the effects of noise, no area-scaling is required and the model readily accounts for the coupling with modes larger than the simulation box.  It does however require approximations to model higher-order correlations, survey geometry, and pixel-level effects. 

The first Gaussian term is calculated from the formula presented in \cite{joachimi08}, using the effective survey area (to account for the loss of area due to masking), the effective galaxy number density per redshift bin (to account for the impact of the \emph{lens}fit weights), and the weighted intrinsic ellipticity dispersion per redshift bin (see Table~\ref{tab:tomo_bins}). The underlying matter power spectrum is calculated assuming the same cosmology as the SLICS $N$-body simulations, using the transfer function by \cite{eisenstein98} and the non-linear corrections by \cite{takahashi/etal:2012}. Convergence power spectra are then derived by line-of-sight integration over the DIR redshift distribution from Section~\ref{sec:DIR}.

To calculate the second, non-Gaussian `in-survey' contribution, we closely follow the formalism of \cite{takada13}. The resulting convergence power spectrum covariance is transformed to that of the correlation functions via the relations laid out in \cite{kaiser92}. The connected trispectrum underlying this term is calculated via the halo model, using the halo mass function and halo bias of \cite{tinker10}.  We assume a \citet{navarro/etal:1996} halo profile with the concentration-mass relation by \cite{duffy08} and employ the analytical form of its Fourier transform by \cite{scoccimarro01}. The matter power spectra and line-of-sight integrations are performed in the same manner as for the Gaussian contribution. We do not account explicitly for the survey footprint in the in-survey covariance contributions. This will lead to a slight over-estimation of the covariance of $\xi_+$ on large scales \citep{sato/etal:2011}.

The final SSC term was modelled by \cite{takada13} as the response of the matter power spectrum to a background density composed of modes larger than the survey footprint. This response can again be expressed in terms of the halo model. It comprises contributions sometimes referred to as halo sample variance and beat coupling, plus a dilation term identified by \cite{li14}. The coupling of super-survey modes into the survey is caused by the finite survey footprint, which therefore needs to be modelled accurately. We account for this by creating a {\verb N_side=1024 }  pixel {\sc healpix} map \citep{gorski/etal:2005} of the current full KiDS survey footprint and convert the part of the formalism by \cite{takada13} pertaining to survey geometry into spherical harmonics.   We refer the reader to Joachimi et al (in prep.) for a detailed description of our analytical model.

\begin{figure}
\includegraphics[width=0.5\textwidth,trim=0 4cm 0 0.5cm,clip]{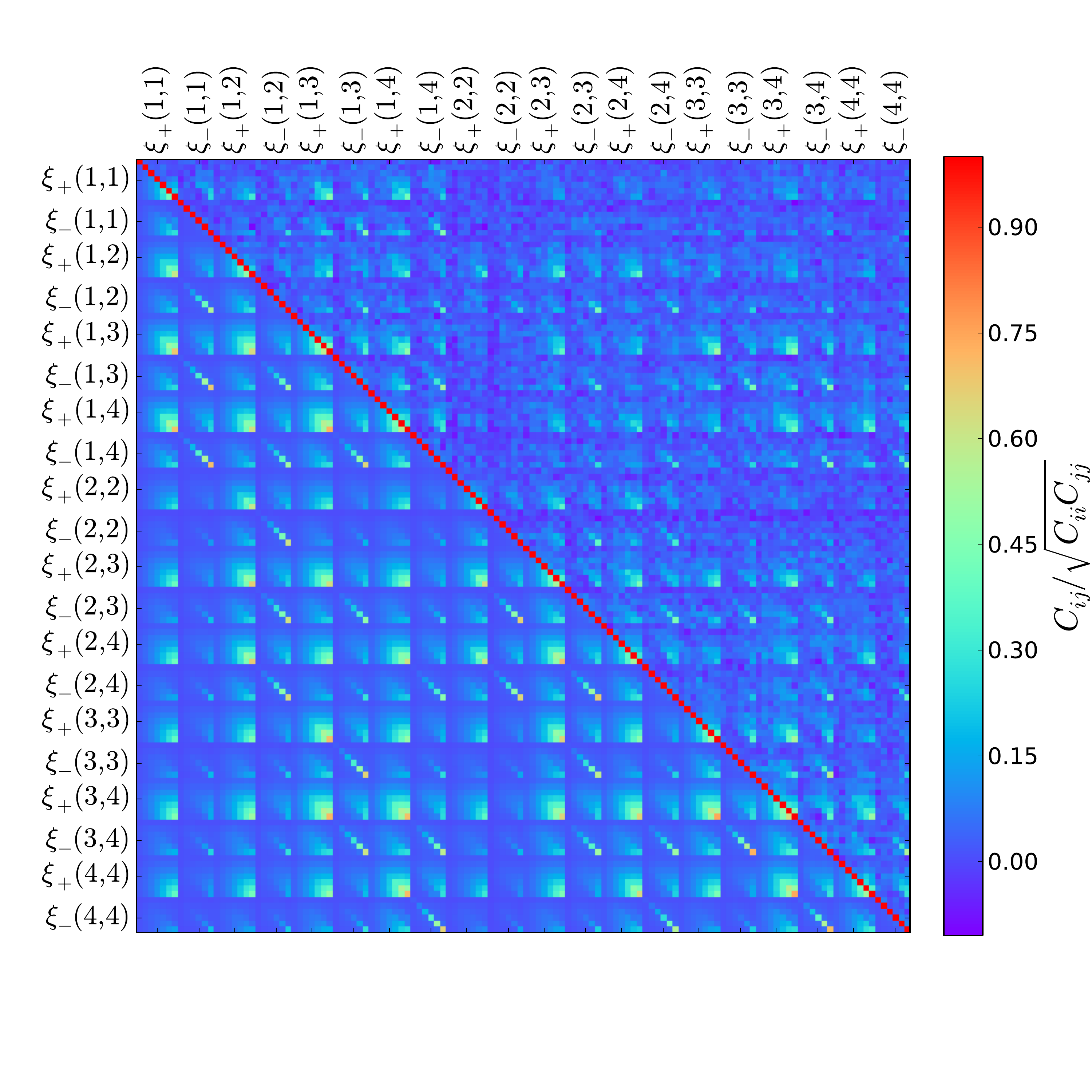}
\caption{\label{fig:cov_comp}Comparison between the analytical correlation matrix (lower triangle) and the numerical correlation matrix (upper triangle).  We order the $\xi_\pm$ values in the data vector by redshift bins $(m,n)$ as labelled, with the seven angular bins of $\xi_+$ followed by the six angular bins of $\xi_-$.  In this Figure the covariance $C_{ij}$ is normalised by the diagonal $\sqrt{C_{ii}C_{jj}}$ to display the correlation matrix.}
\end{figure}

\subsection{Comparison of covariance estimators}
\label{sec:cov_comp}
In Fig.~\ref{fig:cov_comp} we compare the correlation matrix of $\xi^{ij}_\pm$ estimated using the analytical approach (lower triangle) with the numerical approach (upper triangle) on the scales chosen for this analysis (see Section~\ref{sec:results}).  We see broad agreement between the two approaches that the $\xi_+$ statistic is highly correlated across angular scales and redshift bins, and that the correlation is less pronounced for the $\xi_-$ statistic.   The most striking result from this visual comparison, however, is that, even though we have 930 mock simulations, the noise on the numerical result is very pronounced.   As shown in section~\ref{sec:MCMC_cov} the differences highlighted by Fig.~\ref{fig:cov_comp}, on a point-by-point basis, do not, however, significantly change our cosmological parameter constraints.  These differences will be explored further in Joachimi et al (in prep).

\begin{figure}
\includegraphics[width=0.5\textwidth,trim=0 1cm 0 1cm,clip]{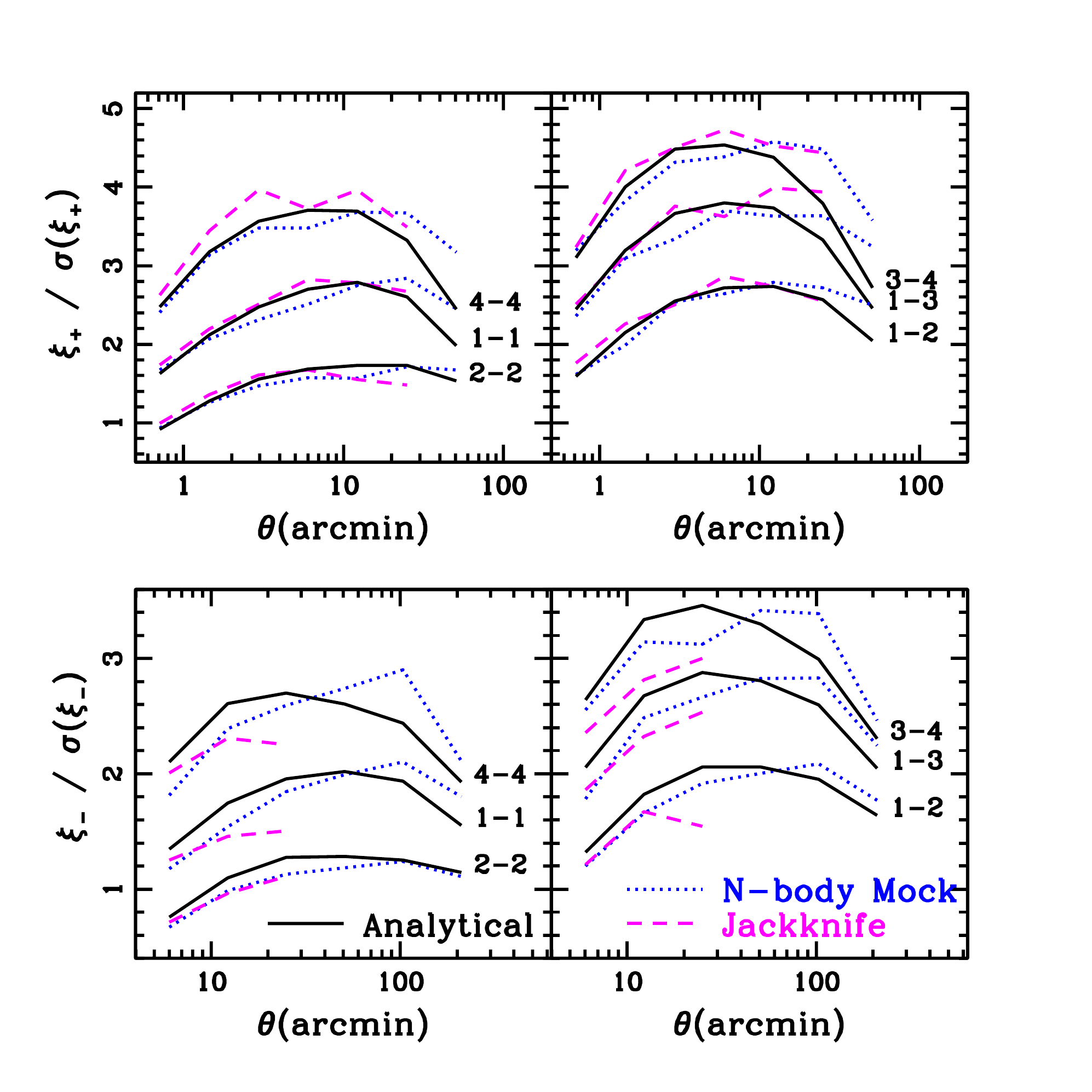}
\caption{\label{fig:covSN}Signal-to-noise estimates for the $\xi_+$ (upper) and $\xi_-$ (lower) statistics using three different error estimates; an analytical approach (solid), $N$-body mock simulations (dotted) and Jackknife (dashed).   A representative selection of six out of the ten different tomographic bin combinations are shown with auto-correlations on the left (bins 1-1, 2-2 and 4-4) and cross-correlations shown on the right (bins 1-2, 1-3 and 3-4).  There is good agreement between the three error estimates on scales $\theta<\SI{30}{\arcmin}$ with the highest signal-to-noise measurement coming from $\xi_+$ in the cross-correlated 3-4 bin.  Note the Jackknife errors are shown only for those scales for which the method is valid.}
\end{figure}

Fig.~\ref{fig:covSN} provides a quantitative comparison between the three covariance estimates, focusing on the `diagonal' components and showing the signal-to-noise ratio.  For a representative sample of six out of the ten different tomographic bin combinations we show the expected signal-to-noise across the angular scales used in our analysis.   The signal is taken from a theoretical model using the same cosmology as the SLICS simulations and the error is taken from the analytical (solid), numerical (dotted), and Jackknife (dashed) estimators.    We find good agreement between the three error estimates on scales $\theta<\SI{30}{\arcmin}$ with the highest signal-to-noise measurement coming from $\xi_+$ in the cross-correlated `3--4' tomographic bin.  On large scales we find that the numerical approach underestimates the variance of $\xi_\pm$, in comparison to the analytical approach.  This is expected as the mock galaxy catalogues do not include super-sample covariance and are subject to finite box effects which become significant on large scales \citep{harnois-deraps/vanwaerbeke:2015}.  Note that the Jackknife errors are invalid on these scales and hence are not shown.  

Based on this comparison, we conclude that the analytical method provides a reliable (and quick) recipe for obtaining a noise-free estimate of the covariance matrix that includes SSC. We therefore use it as the default in our analysis. In Section~\ref{sec:MCMC_cov} we run an additional analysis with the numerical covariance matrix.

\subsection{Propagation of shear calibration uncertainty}
As described in Section~\ref{sec:lensfit} and Appendix~\ref{sec:imsim} we apply a calibration correction factor of $(1+m)^{-1}$ to our shear measurements. The correction is at the per cent level: in the four tomographic bins we have $m=-0.0131$, $-0.0107$, $-0.0087$ and $-0.0217$.  In Appendix~\ref{sec:imsim} we estimate the systematic uncertainty in $m$ to be $\sigma_\rmn{m} = 0.01$. We therefore allow for an additional overall scaling of all shear values by a Gaussian random variable $f$ of mean 1 and standard deviation $\sigma_\rmn{m}\ll1$ by modifying the data covariance matrix $\mat{C}$ to 
\be
\label{eq:Cov_mcorr}
C^{\rm cal}_{ij}  =  4 \xi_i \xi_j \,\sigma_{\rmn m}^2 + C_{ij} \, .
\ee
(The factor 4 in the first term of Eq.~\ref{eq:Cov_mcorr} is due to the $\xi_\pm$ scaling with $f^2$, which has standard deviation $\sim2\sigma_\rmn{m}$.)

We use the data to determine the additive calibration term, as described in Appendix~\ref{sec:c_term} where the uncertainty on this correction is $\delta c \sim 2 \times 10^{-4}$ per tomographic bin.  On the angular scales used in this analysis, the error $(\delta c)^2$ on the additive correction to $\xi_+$ is negligible and is therefore not included in our error budget.  No additive correction is made to $\xi_-$.

%5

\section{RESULTS}
\label{sec:results}
We measure the two-point shear correlation functions $\xi_\pm$ with the public \textsc{athena} code\footnote{\url{http://www.cosmostat.org/software/athena/}} which implements the estimator from Eq.~\ref{eqn:xipm_est}. The measured ellipticities are corrected for the multiplicative and additive biases described in Appendix~\ref{sec:imsim} and~\ref{sec:c_term}.   In order to be insensitive to residual uncertainties in the additive shear bias calibration, we limit our analysis to scales $\theta<\SI{72}{\arcmin}$ for $\xi_+$. The angular range for $\xi_-$ is limited by the declination extent of the KiDS patches to $\theta<\SI{5}{\degree}$. At small angular separations, the uncertainties in the model at non-linear scales as well as the low signal-to-noise ratio lead us to impose lower limits of $\theta>\SI{0.5}{\arcmin}$ for $\xi_+$ and $\theta>\SI{4.2}{\arcmin}$ for $\xi_-$. Overall, we use nine logarithmically-spaced bins spanning $0.5<\theta<\SI{300}{\arcmin}$ of which the first seven are used for $\xi_+$ and the last six for $\xi_-$. 

\begin{figure*}
\includegraphics[width=\textwidth]{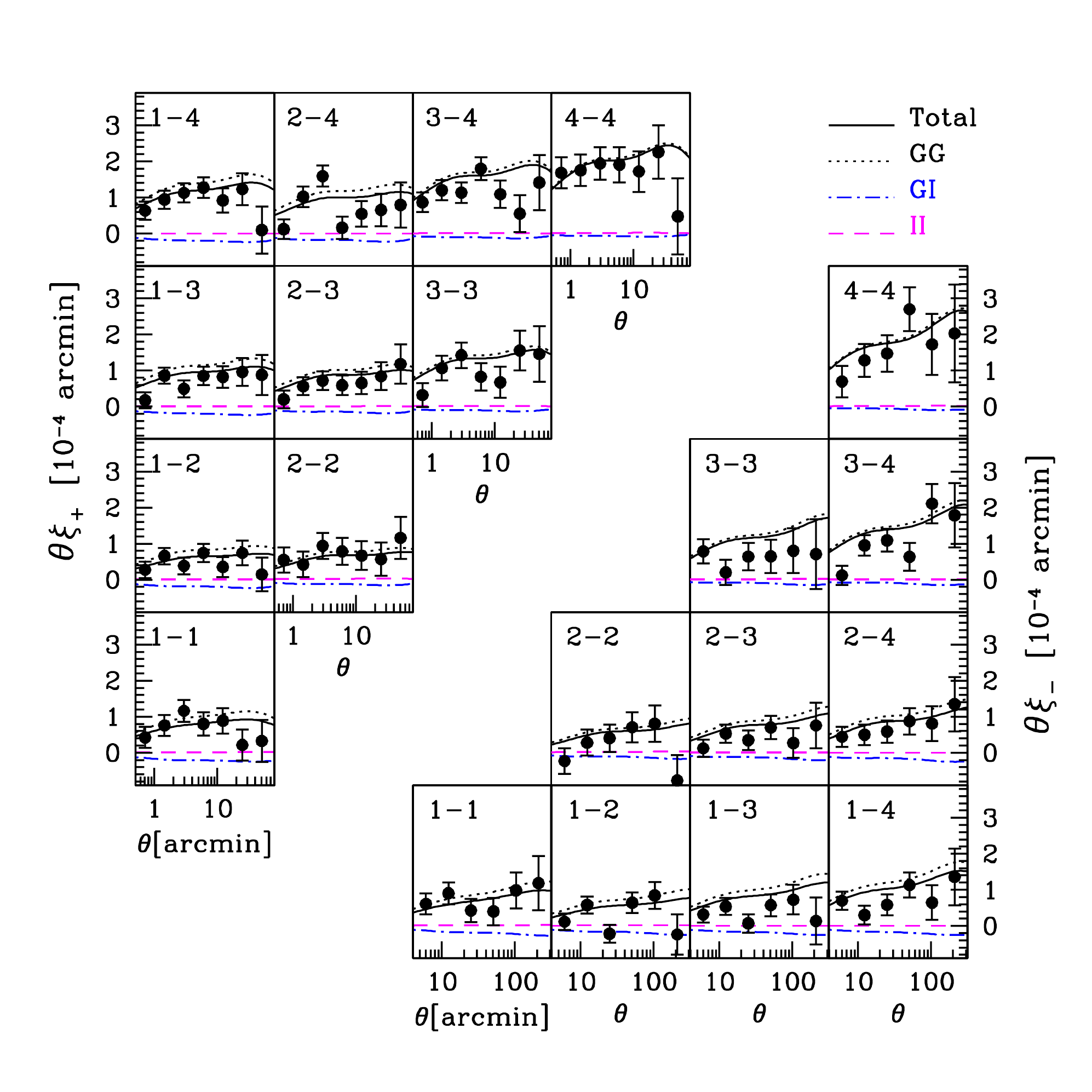}
\caption{\label{fig:xipm}Tomographic measurements of $\xi_+$ (upper-left panels) and $\xi_-$ (lower-right panels) from the full KiDS-450 dataset.  The errors shown here correspond to the diagonal of the analytical covariance matrix (Section~\ref{sec:analyticalcov}). The theoretical model using the best-fit cosmological parameters from Table~\ref{tab:primary_param} is shown (solid) which is composed of a cosmic shear term (GG, dotted), and two intrinsic alignment terms (GI, dot-dashed, and II, dashed).}
\end{figure*}

The resulting correlation functions for all possible combinations of the four tomographic bins are shown in Fig.~\ref{fig:xipm}. The errors correspond to the square root of the diagonal of the analytical covariance matrix (Section~\ref{sec:analyticalcov}) and are highly correlated as shown in Fig.~\ref{fig:cov_comp}. Overplotted is the best-fit cosmic shear and intrinsic galaxy alignment model, as obtained from our primary analysis described in Section~\ref{sec:fiducial} below.

Besides the $\xi_\pm$ correlation functions we also estimate the derived quantities $\xi_{\rm E}$ and $\xi_{\rm B}$ (where the theoretical background and measurements are presented in Appendix~\ref{sec:EB}).  These statistics represent an approximate way to separate gradient modes (E) from curl modes (B) in the shear field. The $\xi_{\rm B}$ correlation function is often used as a null test for systematic errors. We find a small, but significant $\xi_{\rm B}$ signal at the smallest angular scales with $\theta<\SI{4.2}{\arcmin}$. In Section~\ref{sec:MCMC_Bmode} we demonstrate that $\xi_{\rm B}$ has a negligible impact on the cosmological constraints. 

Cosmic shear is most sensitive to a degenerate combination of the cosmological parameters $\Omega_{\rm m}$ and $\sigma_8$ with the amplitude of $\xi_{\pm}$ roughly scaling with $S_8^{2.5}$ where $S_8\equiv\sigma_8\sqrt{\Omega_\rmn{m}/0.3}$ \citep{jain/seljak:1997}.  In the analysis presented in this section we therefore concentrate on these two parameters and in particular their combination $S_8$, by marginalising over all other parameters within the framework of a flat $\Lambda$CDM universe.  In Appendix~\ref{sec:detailed_results} we review our constraints within the full parameter space.

\subsection{Parameters, priors, and information criterion}
\label{sec:cosmo_priors}
In order to arrive at meaningful cosmological constraints and to avoid non-physical solutions we include top-hat priors on several cosmological parameters as well as the parameters that model the astrophysical systematic errors, the amplitude of the intrinsic alignment signal, $A_{\rm IA}$, and the baryon feedback parameter $B$.  
We summarize the priors in Table~\ref{tab:priors}. In several cases we expect these to be informative (and this is borne out by the analysis), but the choice is justified because 
the majority of these parameters are poorly constrained by current weak lensing surveys, with the notable exception of $S_8$. We refer the reader to \citet{joudaki/etal:2016} for a detailed analysis of how the choice of prior impacts upon the resulting parameter constraints, showing that using progressively more informative priors on $h$, $n_\rmn{s}$ and $A_\rmn{s}$ truncates the extremes of the $\sigma_8-\Omega_\rmn{m}$ degeneracy, but does not alter constraints on $S_8$. When comparing different weak lensing surveys, analysed using different priors, one should therefore be careful not to emphasise differences between the tails of the $\sigma_8$ and $\Omega_\rmn{m}$ distributions which could be artificially truncated by the choice of prior.

In this analysis we are interested in using KiDS-450 to explore the reported tension in cosmological parameter constraints between CFHTLenS and Planck.  We therefore ensure that any informative priors that we use are motivated by non-CMB cosmological probes.  For our prior on $h$ we use distance-ladder constraints from \citet{riess/etal:2016} who find $h = 0.730 \pm 0.018$.  We choose to adopt a top-hat prior with a conservative width $\pm 5 \sigma$ such that $0.64<h<0.82$.  This prior is also consistent with the values of $h$ preferred by \citet{planck/cosmo:2015} who find $h=0.673 \pm 0.007$. Note that $h$ is completely degenerate with $\theta_{\rm MC}$. The \textsc{COSMO}MC code used here and described in Sect.~\ref{sec:cosmo_param} samples in $\theta_{\rm MC}$ for technical reasons and hence $h$ is a derived parameter in our analysis. However, we choose the $\theta_{\rm MC}$ prior to be so wide as to be effectively irrelevant and add in any prior information through $h$. This is necessary as non-CMB analyses usually report constraints in terms of $h$ instead of $\theta_{\rm MC}$.

For our top-hat prior on $\Omega_\rmn{b} h^2$ we use big bang nucleosynthesis constraints from \citet{olive/etal:2014}, again adopting a conservative width $\pm 5 \sigma$ such that $0.019<\Omega_\rmn{b} h^2<0.026$.  Our other prior choices are broad.

The best-fit effective $\chi^2$ is defined as ${\chi^2_{\rm eff}(\hat{\theta})}  = -2 \ln {\mathcal{L}}_{{\rm max}}$, where $\hat\theta$ is the vector of the model parameters that yields the maximum likelihood ${\mathcal{L}}_{\rm max}$.
For purposes of model selection, we use the Deviance Information Criterion (DIC;~\citealt{spiegelhalter/etal:2002}, also see \citealt{joudaki/etal:2016} for further details):
\be
{\rm DIC}\equiv \chi^2_{\rm eff}(\hat\theta)+2p_{\rm D}\,,
\ee
where $p_{\rm D}=\overline{\chi^2_{\rm eff}(\theta)}-\chi^2_{\rm eff}(\hat\theta)$ is the Bayesian complexity, which acts to penalise more complex models. $\overline{\chi^2_{\rm eff}(\theta)}$ represents $\chi^2$ averaged over the posterior distribution. The difference in DIC values between two competing models is computationally less expensive to calculate than the Bayes factor (e.g.~\citealt{trotta08}), an alternative measure given by the evidence ratio of the two models. Furthermore, calculating the evidence is non-trivial due to our particular approach for propagating the photometric redshift uncertainties into the analysis. We take a DIC difference between two models in excess of 10 to constitute strong preference in favour of the model with the lower DIC (corresponding to odds of 1 in 148 for two models with the same complexity).

\begin{table}
\begin{center}
\caption{\label{tab:priors}Cosmological priors on \{$\Omega_\rmn{c} h^2$, $\Omega_\rmn{b} h^2$, $\theta_{\rm MC}$, $A_\rmn{s}$, $n_\rmn{s}$, $h$, $k_{\rm pivot}$, $\Sigma m_\nu$\} and astrophysical systematic priors on $\{A_{\rm IA}, B\}$.  $\theta_\rmn{s}$ denotes the angular size of the sound horizon at the redshift of last scattering and $k_{\rm pivot}$ corresponds to the scale where the scalar spectrum has the amplitude $A_{\rm S}$. Note that $h$ is closely tied to $\theta_{\rm MC}$ and we choose to add an informative prior on $h$ only.}
\begin{tabular}{p{3.75cm}p{1.7cm}p{1.9cm}}
\hline
Parameter & Symbol & Prior \\
\hline
Cold dark matter density & $\Omega_{\rm c}h^2$ & $[0.01 , 0.99]$ \\
Baryon density & $\Omega_{\rm b}h^2$ & $[0.019 , 0.026]$ \\
100 $\times$ approximation to $\theta_{\rm s}$ & $100 \, \theta_{\rm MC}$ & $[0.5 , 10]$ \\
Scalar spectrum amplitude & $\ln{(10^{10} A_{\rm s})}$ & $[1.7 , 5.0]$ \\
Scalar spectral index & $n_{\rm s}$ & $[0.7 , 1.3]$ \\
Hubble parameter & $h$ & $[0.64 , 0.82]$ \\
Pivot scale $[{\rm{Mpc}}^{-1}]$ & $k_{\rm pivot}$ & 0.05 \\
Neutrino Mass [eV] & $\Sigma m_\nu$ & 0.00 \\
\hline
IA amplitude & $A_{\rm IA}$ & $[-6 , 6]$ \\
Feedback amplitude & $B$ & $[2 , 4]$ \\
\hline
\end{tabular}
\end{center}
\end{table}

\begin{table*}
\caption{\label{tab:MCMC_setups}Setups for the different MCMC runs. The first column gives a short descriptive name to the setup and the second and third column refer the reader to the section and figure in which the setup is discussed. Columns 4--6 indicate which astrophysical systematics are marginalised over in each run.  Column 7 and column 8 report the choices for the redshift distribution and the covariance matrix, respectively. Column 8, 9, and 10 indicate whether the equation-of-state parameter $w$ is varied, the KiDS results are combined with Planck (TT + lowP), and $2\times\xi_{\rm B}$ is subtracted from $\xi_+$. The last column gives the angular scales used for $\xi_+$. For $\xi_-$ we use scales of $4.2$--$\SI{300}{\arcmin}$ for all setups. }
\begin{tabular}{lllcccllcccc}
\hline
Setup & Sect. & Fig. & baryons & IA & photo-$z$ & $n(z)$ & covariance & $w$ & comb. w. & B mode & scales\\
&&&&&error&&&&Planck&subtr.&$\xi_+$\\
\hline
KiDS-450     & \ref{sec:fiducial}    & \ref{fig:Om_s8_fiducial} & $\surd$ & $\surd$ & $\surd$ & DIR & analytical &  --      &  --      &  --     & $0\farcm5$ -- $72'$ \\
DIR                 & \ref{sec:MCMC_photoz} & \ref{fig:Om_s8_pzerr}    & --      & $\surd$ & $\surd$ & DIR & analytical &  --      &  --      &  --     & $0\farcm5$ -- $72'$ \\
CC                  & \ref{sec:MCMC_photoz} & \ref{fig:Om_s8_pzerr}    & --      & $\surd$ & $\surd$ & CC  & analytical &  --      &  --      &  --     & $0\farcm5$ -- $72'$ \\
BOR                 & \ref{sec:MCMC_photoz} & \ref{fig:Om_s8_pzerr}    & --      & $\surd$ & --      & BOR & analytical &  --      &  --      &  --     & $0\farcm5$ -- $72'$ \\
BPZ                 & \ref{sec:MCMC_photoz} & \ref{fig:Om_s8_pzerr}    & --      & $\surd$ & --      & BPZ & analytical &  --      &  --      &  --     & $0\farcm5$ -- $72'$ \\
no systematics      & \ref{sec:MCMC_cov}    &  --                      & --      & --      & --      & DIR & analytical &  --      &  --      &  --     & $0\farcm5$ -- $72'$ \\
$N$-body            & \ref{sec:MCMC_cov}    &  --                      & --      & --      & --      & DIR & $N$-body   &  --      &  --      &  --     & $0\farcm5$ -- $72'$ \\
DIR no error        & \ref{sec:MCMC_Bmode}  & \ref{fig:Om_s8_Bmode}    & --      & $\surd$ & --      & DIR & analytical &  --      &  --      & --      & $0\farcm5$ -- $72'$ \\
B mode              & \ref{sec:MCMC_Bmode}  & \ref{fig:Om_s8_Bmode}    & --      & $\surd$ & --      & DIR & analytical &  --      &  --      & $\surd$ & $0\farcm5$ -- $72'$ \\
$\xi_+$ large-scale & \ref{sec:MCMC_Bmode}  & \ref{fig:Om_s8_Bmode}    & --      & $\surd$ & --      & DIR & analytical &  --      &  --      &  --     & $4\farcm2$ -- $72'$ \\
wCDM                & \ref{sec:MCMC_wCDM}   & \ref{fig:Om_w}           & $\surd$ & $\surd$ & $\surd$ & DIR & analytical & $\surd$  &  --      &  --     & $0\farcm5$ -- $72'$ \\
+Planck             & \ref{sec:discussion}  &  --                      & $\surd$ & $\surd$ & $\surd$ & DIR & analytical &  --      & $\surd$  &  --     & $0\farcm5$ -- $72'$ \\
\hline
\end{tabular}
\end{table*}

\subsection{Cosmological parameter constraints}
\label{sec:cosmo_param}
\label{sec:fiducial}
\begin{figure*}
\begin{center}
\resizebox{8.50cm}{!}{{\includegraphics{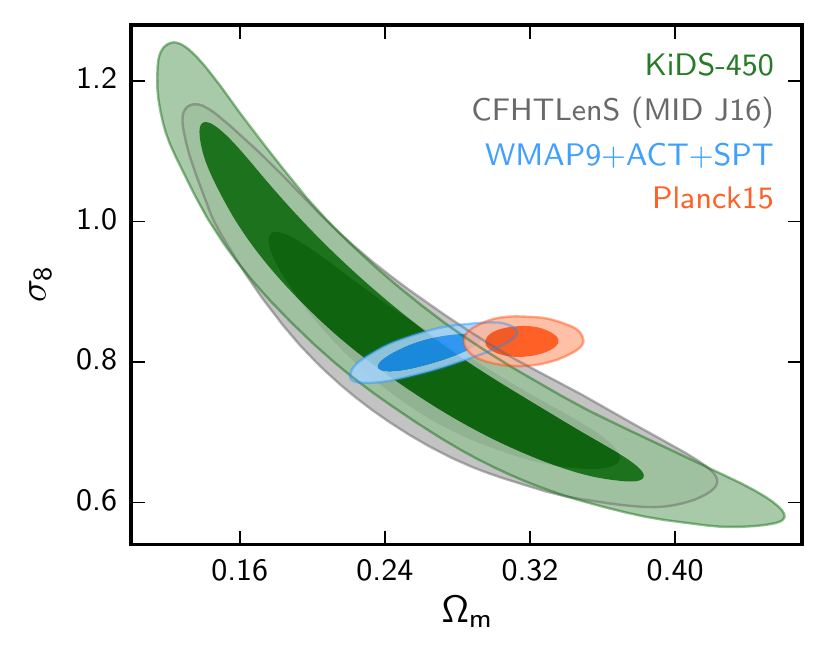}}}
\resizebox{8.63cm}{!}{{\includegraphics{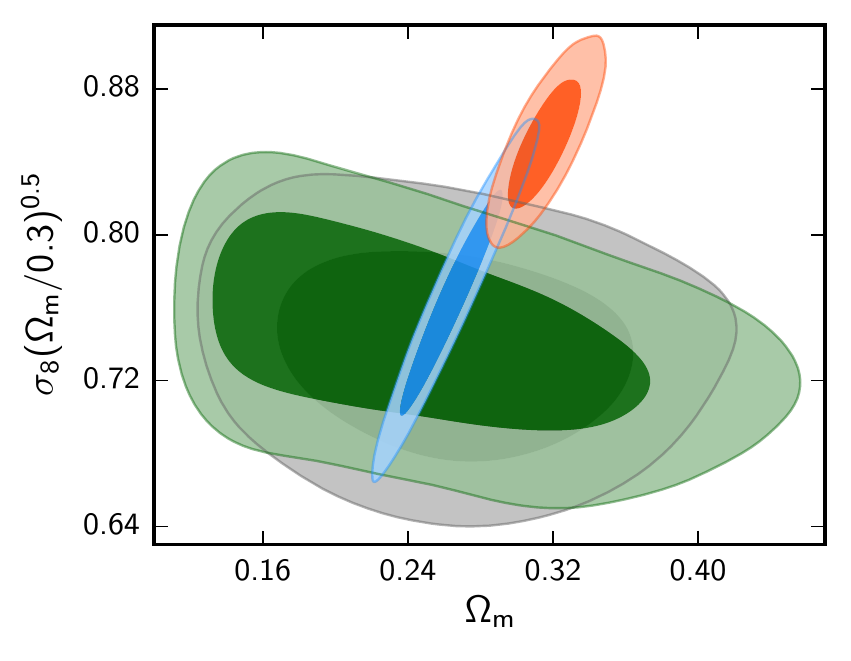}}}
\end{center}
\vspace{-2.3em}
\caption{\label{fig:Om_s8_fiducial}Marginalized posterior contours (inner 68\%~CL, outer 95\%~CL) in the $\Omega_{\rm m}$-$\sigma_8$ plane (left) and $\Omega_{\rm m}$-$S_8$ plane (right) from the present work (green), CFHTLenS (grey), pre-Planck CMB measurements (blue), and Planck 2015  (orange). Note that the horizontal extent of the confidence contours of the lensing measurements is sensitive to the choice of the prior on the scalar spectrum amplitude $A_{\rm s}$. The CFHTLenS results are based on a more informative prior on $A_{\rm s}$ artificially shortening the contour along the degeneracy direction.}
\end{figure*}

We obtain cosmological parameter estimates from a Bayesian likelihood analysis using the \textsc{CosmoMC} software including \textsc{camb} \citep{lewis/bridle:2002, lewis/etal:1999}.  Our extended version uses a halo model recipe based on \textsc{HMcode} \citep{mead/etal:2015} to calculate the effect of baryons on the total matter power spectrum and closely follows the \citet{joudaki/etal:2016} re-analysis of the CFHTLenS data, with the exception of the handling of photo-$z$ errors. Our primary KiDS-450 analysis includes the full modelling for intrinsic galaxy alignments (see Section~\ref{sec:IAmodels}) and baryon feedback (see Section~\ref{sec:feedback}), the weighted direct calibration (DIR) of the photometric redshift distribution with error estimate (see Section~\ref{sec:DIR}), and the analytic estimate of the covariance matrix (see Section~\ref{sec:analyticalcov}).   Fig.~\ref{fig:Om_s8_fiducial} shows the confidence contours of the cosmologically most relevant parameters constrained, $\Omega_{\rm m}$ and $\sigma_8$ (and their combination $S_8$), in comparison to the CFHTLenS results, as well as pre-Planck CMB measurements \citep{calabrese/etal:2013}, and Planck \citep{planck/cosmo:2015}.
%\footnote{
Note that the CFHTLenS constraints use a somewhat more informative prior on $A_{\rm s}$ which artificially decreases the extent of the confidence contours along the degeneracy direction in comparison to the KiDS-450 constraints.  The measurements for $S_8$ and the comparison to CMB measurements is however unaffected by this informative prior.
%} 
The confidence contours for all pair-wise combinations of the model parameters are presented in Fig.~\ref{fig:triangle}.

While the two lensing analyses (KiDS-450, CFHTLenS) and the pre-Planck CMB results are consistent with each other, with overlapping 1-$\sigma$ contours, there is tension between the KiDS-450 and Planck results, similar to that found for CFHTLenS. The tension with respect to Planck is significant at the $2.3$-$\sigma$ level.   We explore concordance in the full parameter space in Section~\ref{sec:discordance}. Note that a recent re-analysis of the Planck data \citep{planck/etal:2016} finds slightly different values for $\sigma_8$ and $\Omega_{\rm m}$ but essentially the same $S_8$. Hence the tension with respect to this KiDS-450 study is not affected.

We find that the KiDS-450 cosmic shear analysis is not particularly sensitive to the Hubble parameter so that constraints on this parameter are relatively loose and dominated by the prior employed in the analysis.  The choice of the prior on $h$ does not change the results for $S_8$: a change in $h$ moves the constraints along the curved degeneracy direction in the $\Omega_{\rm m}$ vs. $\sigma_8$ plane, effectively keeping $S_8$ and its error constant (see Fig~\ref{fig:triangle_derived}).

We chose to adopt the DIR method as our primary calibration of the redshift distributions for the four tomographic bins because arguably it gives the smallest systematic uncertainties (see Appendix~\ref{sec:app_z_tests} for a detailed discussion).
We use bootstrap realisations to model the uncertainties and to capture the correlations between the different tomographic bins (we build the bootstrap sample from the spec-$z$ catalogue and run the whole DIR process for each sample). We run $N=750$ MCMCs, varying the input set of tomographic redshift distributions each time by picking one bootstrap realisation at a time.  By combining all $N$ chains we accurately marginalise over our full uncertainty on the photometric redshift distribution without having to resort to modelling the photometric redshift error as an uncorrelated shift in the mean as in \citet{joudaki/etal:2016} and \citet{abbott/etal:2015}. %\citetalias{abbott/etal:2015}. 
The value of $N=750$ was determined through convergence tests on the final combined chain.  We use a conservative criterion of $(R - 1) < 2 \times 10^{-2}$ where $R$ is defined as the variance of chain means divided by the mean of chain variances \citep{Gelman92}.  We have verified that our results are stable to further exploration in the tails of the distribution.  

In the following sections we explore a series of restricted models that allow us to test the impact of different effects on the resulting cosmological parameters. The setups for the different analyses are summarised in Table~\ref{tab:MCMC_setups} and the results are described in Section~\ref{sec:MCMC_photoz} to \ref{sec:MCMC_wCDM}.  A one-dimensional comparison of the constraints on the combined cosmological parameter $S_8$ for different setups of our KiDS-450 analysis and different external datasets can be found in Section~\ref{sec:S_8}.

\subsection{Impact of photometric redshift uncertainty}
\label{sec:MCMC_photoz}
In this section we explore the sensitivity of the cosmological parameter constraints to the method with which the photometric redshift distributions are determined.  We consider the three cases discussed in Section~\ref{sec:photoz_calibration}: the weighted direct calibration (DIR, Section~\ref{sec:DIR}), the cross-correlation analysis (CC, Section~\ref{sec:CC}), and the re-calibration of the photometric $P(z)$ (BOR, Section~\ref{sec:BOR}). We compare those three re-calibrations to the uncalibrated redshift distributions that are based directly on the stacked $P(z)$ from \textsc{bpz}.

We use the same model and priors as for the primary analysis in Section~\ref{sec:fiducial}, with the exception of the baryon feedback amplitude, which we set to zero. As discussed in Sect.~\ref{sec:S_8}, this astrophysical systematic has only a small impact on the overall result, and since for a sensitivity test we are more interested in parameter changes than in actual values, we revert to a dark-matter only power spectrum in this comparison. This choice also enables us to switch from \textsc{HMcode} to the faster \citet{takahashi/etal:2012} model for the non-linear power spectrum.

For each of the three calibration methods (DIR, CC, BOR) we estimate statistical errors from a bootstrap re-sampling of the spectroscopic calibration sample (see Section~\ref{sec:fiducial} for details of the implementation). Including those uncertainties will broaden the contours. As can be seen in Fig.~\ref{fig:z_dist_comp} these bootstrap errors are very small for the BOR method. This is due to the fact that a lot of information in that technique is based on the photometric $P(z)$ and the re-calibration is more stable under bootstrap re-sampling of the spectroscopic calibration sample than for the other two methods. Hence to further speed up the MCMC runs we neglect the BOR errors in the following with no visible impact on the results. The uncertainties on the DIR method -- while larger than the BOR errors -- are also negligible compared to the shot noise in the shear correlation function (see Appendix~\ref{sec:DIR_stat_err}). We nevertheless include these errors here (as before) since DIR is our primary calibration method. The statistical errors on the CC method are larger than for the two other methods, owing to the as yet small area covered by the spectroscopic surveys that we can cross-correlate with. More importantly, we estimate that the limited available area also gives rise to a larger systematic uncertainty on the CC method compared to the DIR technique. All major requirements for the DIR technique are met in this analysis whereas the CC method will only realise its full potential when larger deep spec-$z$ surveys become available.

\begin{figure}
\includegraphics[width=\hsize]{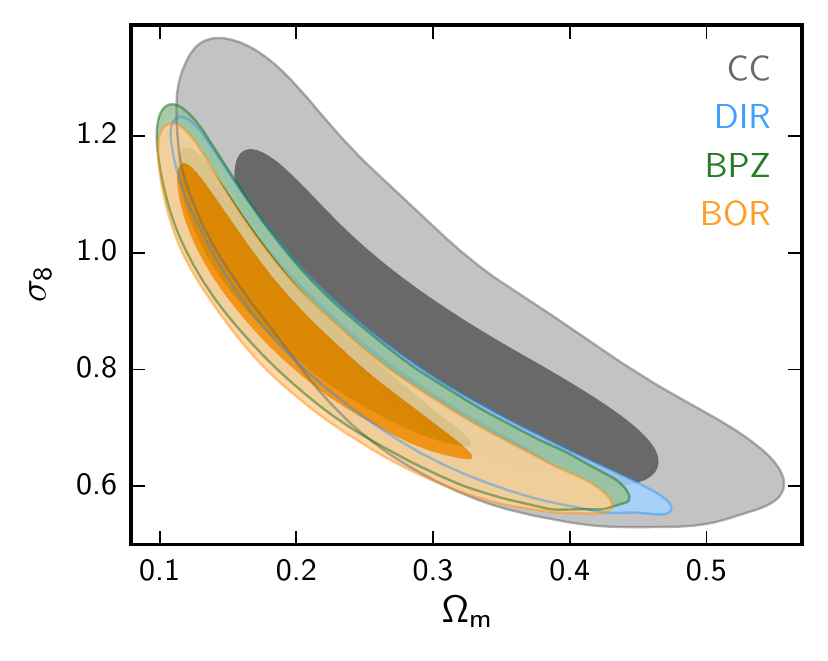}
\vspace{-2.3em}
\caption{\label{fig:Om_s8_pzerr}Marginalized posterior contours (inner 68\%~CL, outer 95\%~CL) in the $\Omega_{\rm m}$-$\sigma_8$ plane, examining the impact of photometric redshift uncertainty and calibration methods. Shown are the constraints in the $\Omega_{\rm m}$-$\sigma_8$ plane for the weighted direct calibration with errors (DIR, blue), the calibration with cross-correlations with errors (CC, grey), the original stacked $P(z)$ from \textsc{bpz} (green), and their re-calibrated version (BOR, yellow).
}
\end{figure}

The resulting confidence contours in the $\Omega_{\rm m}$-$\sigma_8$ plane for the four cases are shown in Fig.~\ref{fig:Om_s8_pzerr}. All four cases give fully consistent results although there are some shifts in the contours with respect to each other. 
However, with $\Delta{\chi^2_{\rm eff}} \simeq -10$, we find that the DIR and CC methods provide a better fit to the data as compared to the BPZ and BOR methods. For future cosmic shear surveys, with considerably larger datasets, it will be essential to reduce the statistical uncertainty in the redshift calibration in order to not compromise the statistical power of the shear measurement. For KiDS-450 the uncertainty for our favoured DIR calibration scheme is still subdominant.

In summary, we find that the four possible choices for the photometric redshift calibration technique yield consistent cosmological parameters.

\subsection{Impact of analytical and numerical covariance matrices}
\label{sec:MCMC_cov}
For our primary analysis we choose to adopt the analytical estimate of the covariance matrix described in Section~\ref{sec:analyticalcov}, as it yields the most reliable estimate of large-scale sample variance (including super-sample contributions), is free from noise, and is broadly consistent with the $N$-body covariance (see Section~\ref{sec:cov_comp}).  In this section we compare the cosmological parameter constraints obtained with the analytical covariance matrix to the alternative numerical estimate as described in Section~\ref{sec:mockcov}. For this test, we set all astrophysical and data-related systematics to zero: this applies to the intrinsic alignment amplitude, the baryon feedback amplitude, the errors on the shear calibration, and the errors on the redshift distributions. Fixing these parameters allows us to focus on the effect of the different covariance matrices on the cosmological parameters. 

 We correct for noise bias in the inverse of the numerical covariance matrix estimate using the method proposed by \citet{sellentin/heavens:2016}.  As we have a significant number of N-body simulations, however,  we note that the constraints derived using our numerical covariance matrix are unchanged if we use the less precise but alternative \citet{hartlap/etal:2007} bias correction scheme.

We find consistency between the results for the different covariance matrices given the statistical errors of KiDS-450.  
There is however a shift in the central values of the best-fit parameters; the $S_8$ constraint for the numerical covariance is 0.04 lower than the constraint for the analytical covariance.  This shift is equivalent to the size of the $1\sigma$ error on $S_8$ when all systematic effects are included in the analysis. We attribute these shifts to super-sample-covariance terms that are correctly included only in the analytical estimate (which is also the reason why we adopt it as our preferred covariance).  The SSC reduces the significance of the large angular $\xi_\pm$ measurements (see Fig.~\ref{fig:covSN}) where our measured signal is rather low in comparison to the best-fit model (see Fig.~\ref{fig:xipm}).  The numerical covariance incorrectly gives too much weight to the large-scale results, resulting in a shift to lower $S_8$ values when the numerical covariance is used.

In this case, where we have neglected all systematic uncertainties, the reduced $\chi^2$ when using the numerical covariance ($\chi^2_{\rm red} = 1.2$) is lower than the analytic covariance analysis ($\chi^2_{\rm red} = 1.5$). This difference can be understood from Fig.~\ref{fig:covSN} where the numerical covariance predicts slightly larger errors for the angular scales which carry the most information.  This is particularly true for the $\xi_-$ statistic.

\subsection{Impact of B modes}
\label{sec:MCMC_Bmode}
As detailed in Appendix~\ref{sec:EB}, we find small but significant B modes in the KiDS-450 data on angular scales $\theta<\SI{4.2}{\arcmin}$.  In order to assess their importance we tested two mitigation strategies; excluding the small-scale measurements, and subtracting $2\times\xi_{\rm B}$ from our $\xi_+$ measurements.  The latter correction is valid if the origin of the systematic creates E modes with the same amplitude as the B mode. Note that $\xi_-$ is not modified under this assumption, as explained in Appendix~\ref{sec:EB}.  Fig.~\ref{fig:Om_s8_Bmode} shows the effect of these two B-mode correction schemes on the constraints in the $\Omega_{\rm m}$-$\sigma_8$ plane.  The contours shift somewhat when the correction is applied, and grow when only large scales are used, but the changes are well within the 1-$\sigma$ confidence region. It therefore appears that our analysis is not significantly affected by B modes: in particular the B modes are not sufficient to explain the tension with respect to the Planck results. If anything, the B-mode correction increases the tension.  Applying the B mode correction does however result in an improvement in the goodness-of fit, with the $\chi^2_{\rm red}$ reducing from 1.3 to 1.1.

\begin{figure}
\includegraphics[width=\hsize]{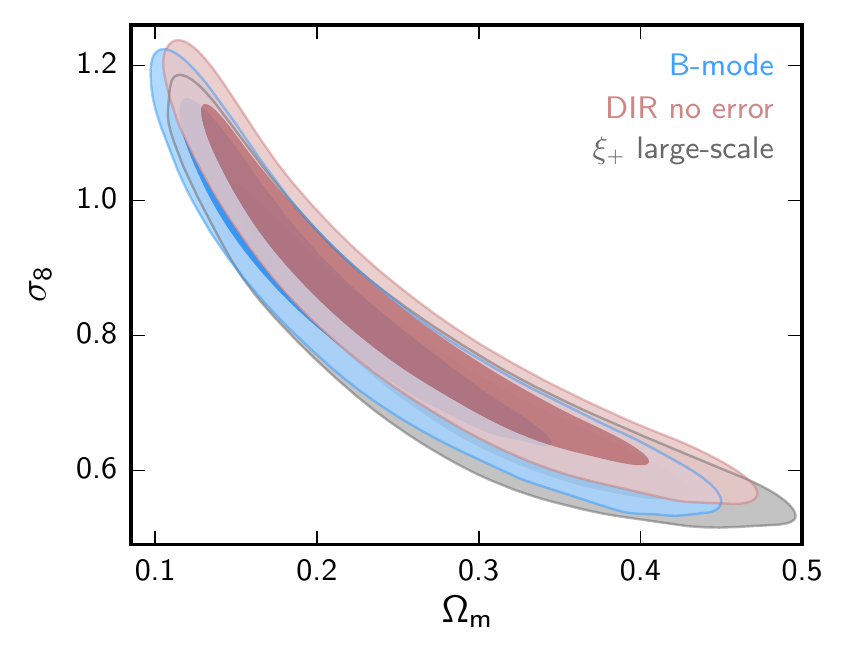}
\vspace{-2.3em}
\caption{\label{fig:Om_s8_Bmode}Marginalized posterior contours (inner 68\%~CL, outer 95\%~CL) in the $\Omega_{\rm m}$-$\sigma_8$ plane, examining the impact of different ways of handling the B modes. Shown are the primary constraints in the $\Omega_{\rm m}$-$\sigma_8$ plane but neglecting baryon feedback and photo-$z$ errors (red), an analysis that subtracts $2\times\xi_{\rm B}$ from $\xi_+$ (B mode; blue), and an analysis that only uses large scales in $\xi_+$ (grey). 
}
\end{figure}

\subsection{Impact of intrinsic galaxy alignment and baryon feedback modelling}
In our primary analysis, we constrain the amplitude of the intrinsic alignments to $A_{\rm IA}=1.10\pm0.64$. This is in contrast to the different CFHTLenS analyses: from a combined analysis with WMAP7 \citet{heymans/etal:2013} find an overall negative amplitude with $A_{\rm IA}=-1.18^{+0.96}_{-1.17}$, and \citet{joudaki/etal:2016} find $A_{\rm IA}=-3.6\pm1.6$ from lensing alone. Interestingly, if we switch from our preferred $n(z)$ (DIR, determined from the weighted direct calibration) to the stacked $P(z)$ estimated by the photo-$z$ code \textsc{bpz} (see Section~\ref{sec:MCMC_photoz}), i.e. the redshift distribution methodology used for CFHTLenS, we also find a negative $A_{\rm IA}$ for KiDS and a considerably worse $\chi^2$ (for details see Appendix~\ref{sec:detailed_results} and in particular Table~\ref{tab:primary_param}). Since the $n(z)$ for these two different cases differ significantly in the first tomographic bin where the relative influence of intrinsic alignments is greatest, we conclude that the \textsc{bpz} distributions are particularly biased in this bin which is properly calibrated by our now favoured DIR approach from Section~\ref{sec:DIR}. The inclusion of the IA parameter gives $\Delta{\rm DIC} = -2.7$, such that it is slightly preferred by the data.

The KiDS-450 data do not strongly constrain the baryon feedback amplitude $B$, reflecting that this astrophysical effect is relatively unimportant for our study. Only future cosmic shear surveys with higher signal-to-noise measurements and finer binning in angle and redshift or cross-correlations between lensing and baryonic probes will allow $B$ to be constrained to reasonable levels. Moreover, the inclusion of baryon feedback only improves the DIC by $1.0$, such that it is neither favoured nor disfavoured by the data.

\subsection{\textit{w}CDM cosmology}
\label{sec:MCMC_wCDM}
\begin{figure}
\includegraphics[width=\hsize]{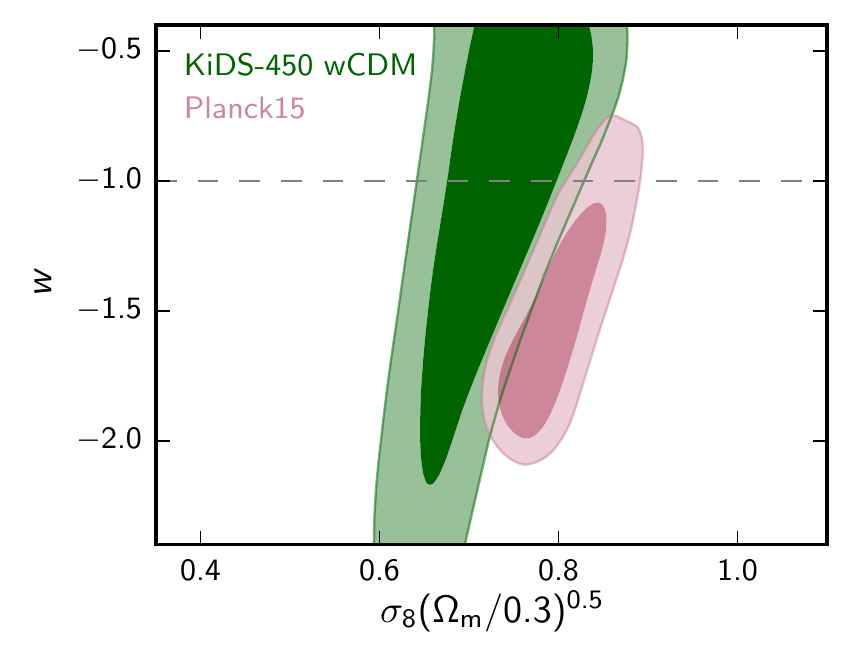}
\vspace{-2.3em}
\caption{\label{fig:Om_w} Marginalized posterior contours (inner 68\%~CL, outer 95\%~CL) in the $S_8$-$w$ plane from KiDS-450 (green) and Planck 2015 (pink).}
\end{figure}

While a comprehensive analysis of KiDS-450 constraints on extensions to the standard model of cosmology is beyond the scope of this paper, we include one test of the effect of allowing the equation of state parameter $w$ of the dark energy to vary. Unlike the other systematics tests described above, we allow all astrophysical parameters to vary for this test. These constraints and their dependence on $\Omega_{\rm m}$ are shown in Fig.~\ref{fig:Om_w} in comparison to the Planck results.

We find that the cosmic shear result of KiDS-450 by itself is not able to yield constraints on $w$ as evidenced by the extended contours in Fig.~\ref{fig:Om_w}. Within these large uncertainties on $w$ there is no discrepancy with previous measurements, and no indication for a deviation from a cosmological constant. 

\subsection{Comparison of $S_8$ values}
\label{sec:S_8}
\begin{figure*}
\vspace{-1em}
\resizebox{14.5cm}{!}{{\includegraphics{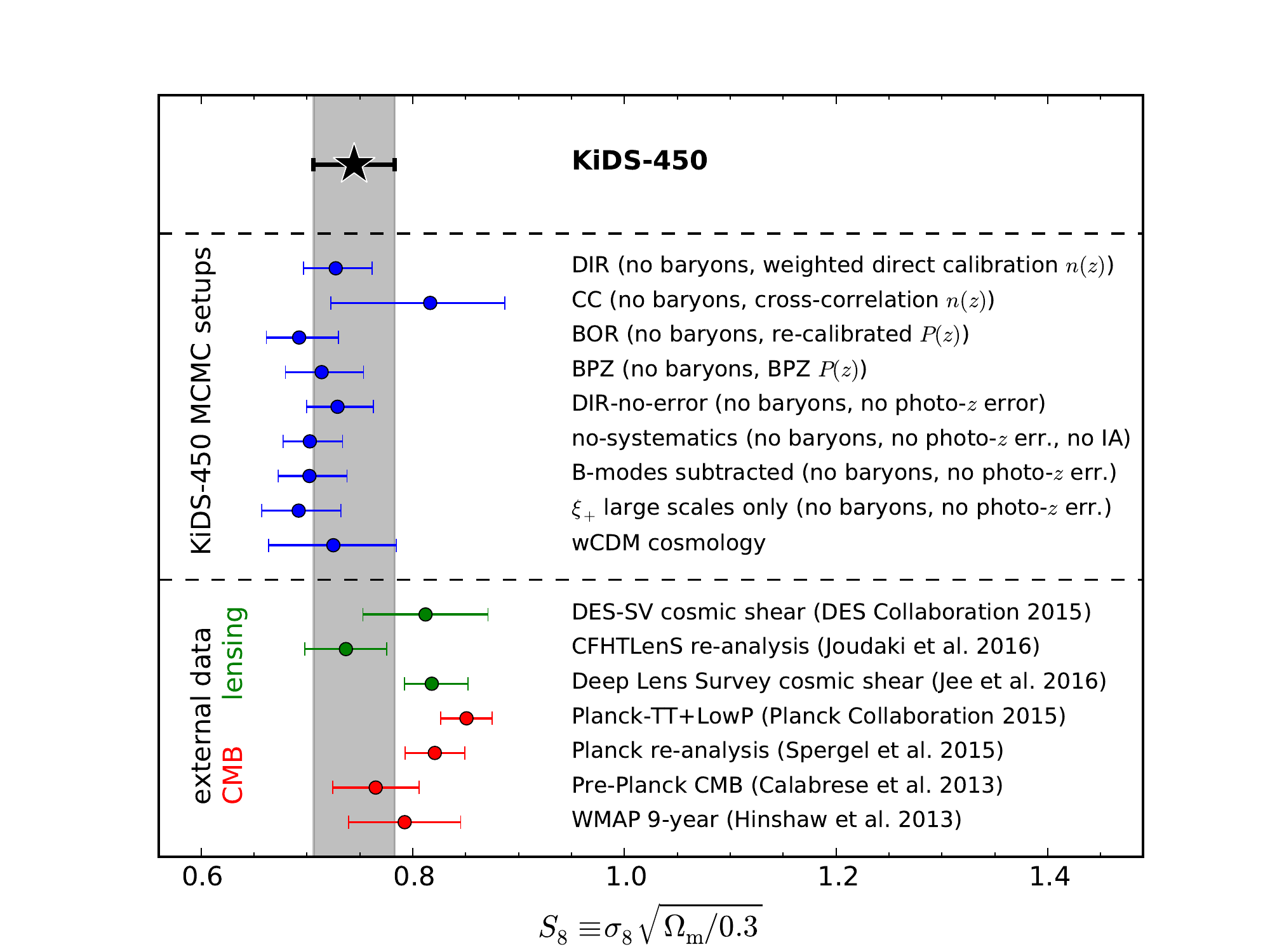}}}
\caption{\label{fig:S8}Constraints on $S_8$ for the different runs considered in the KiDS-450 analysis as well as several literature measurements. The grey band indicates the $1\sigma$ constraints from our primary analysis.  Note that most of the runs which test for systematic errors (blue data points) switch off some of the astrophysical or redshift systematics. Hence not all data points shown here are fully comparable. For numerical values of the plotted data points see Table~\ref{tab:primary_param}.}
\end{figure*}

In Fig.~\ref{fig:S8} we compare the constraints on $S_8$ for the different setups listed in Table~\ref{tab:MCMC_setups} with our primary result, and also compare to measurements from the literature. 

We find that the different setups yield results consistent with the primary analysis. Neglecting all systematic uncertainties shifts the $S_8$ value by one standard deviation and shrinks the error bars by 30\%. The impact of the joint inclusion of the systematic uncertainties on the central value of $S_8$ is small because the separate shifts partially cancel each other. The small, subdominant effect of baryon feedback can be seen by comparing the `KiDS-450' setup to the `DIR (no baryons)' setup. If additionally the photo-$z$ errors on the weighted direct calibration are ignored, the constraints labeled `DIR-no-error' are obtained. Comparing those two models hence gives an indication of the importance of the statistical error of the photo-$z$ calibration for the total error budget. Since the $S_8$ errors for those two cases are almost identical this confirms what was already found above, namely that statistical photo-$z$ errors are subdominant in the KiDS-450 analysis. 

Switching from the weighted direct calibration to the alternative $n(z)$ estimates yields consistency with the primary results, in agreement with the findings of Section~\ref{sec:MCMC_photoz}. Extending the model by allowing for a free equation-of-state parameter $w$ increases the error on $S_8$ by about a factor of two. The central value is still fully consistent with the primary setup. The two different schemes for correcting for the B modes are consistent with the `no baryons, no photo-$z$ err.' case, as already seen in Section~\ref{sec:MCMC_Bmode}.

Comparing the KiDS-450 constraints to external datasets we find consistency with the re-analysis of CFHTLenS by \citet{joudaki/etal:2016} and pre-Planck CMB constraints by \citet{calabrese/etal:2013}. The DES-SV tomographic cosmic shear constraints \citep{abbott/etal:2015}  %\citepalias{abbott/etal:2015} 
and the WMAP nine-year results \citep{hinshaw/etal:2013} have wider error bars that are also consistent with KiDS-450, but tend towards higher $S_8$ values. A mild discrepancy of $\sim1.5\sigma$ is found with the most recent cosmic shear results from the Deep Lens Survey \citep{jee/etal:2016}, which are based on deeper, and hence harder to calibrate, data. For a full overview of the constraints obtained from older cosmic shear measurements see \citet{kilbinger:2015} and references therein.

The greatest tension, at 2.3 $\sigma$, is found when comparing to the 2015 Planck results \citep{planck/cosmo:2015}, though the tension is diminished in the \citet{spergel/etal:2015} re-analysis of the Planck data. The uncertainty on the KiDS-450 result for $S_8$ is about a factor of two larger than the uncertainty from Planck and almost identical to the uncertainty from the best pre-Planck analyses and CFHTLenS. Understanding the cause of the discordance between the latest CMB and cosmic shear datasets is an important challenge for observational cosmology.

It is interesting to compare to recent results based on alternative measurements that also constrain $\sigma_8$ and $\Omega_\rmn{m}$. For instance, the number density of massive clusters of galaxies as a function of redshift is a sensitive probe of the large-scale structure growth rate. New wide-area millimeter surveys that detect large numbers of galaxy clusters with relatively well-defined selection functions through the thermal Sunyaev-Zel'dovich effect \citep[e.g.][]{Hasselfield13, Bleem15, Planck15_XXVII}, and improvements in the calibration of cluster masses \citep{Applegate14,hoekstra/etal:2015}, have resulted in constraints on cosmological parameters of comparable power to the KiDS-450 cosmic shear results. \citet{Planck15_XXIV} use a sample of 439 clusters. Although the accuracy is still affected by uncertainties in the mass calibration, they report values for $\sigma_8$ that are lower than the best fit values from the primary CMB, but agree well with our results. Similarly \citet{deHaan16} used 377 cluster candidates from the South Pole Telescope and found $\sigma_8=0.772\pm0.029$ (for $\Omega_\rmn{m}=0.3$) in excellent agreement with our results. Similar low values for $\sigma_8$ are found in recent studies that make use of a combination of galaxy-galaxy lensing and galaxy clustering \citep{cacciato/etal:2013,mandelbaum/etal:2013,more/etal:2015}. This complementary approach does not trace the matter power spectrum directly, but instead measures the mass associated with galaxies as well as their linear density bias.

Measurements of redshift space distortions, using large spectroscopic surveys, provide another interesting avenue to study the growth rate. \citet{planck/cosmo:2015} present a compilation of constraints from redshift space distortions as a function of redshift, again indicating a preference for lower growth rates compared to the predictions from the best fit $\Lambda$CDM model to the CMB. For instance, \citet{Beutler14} use the Baryon Oscillation Spectroscopic Survey (BOSS) CMASS DR11 sample and conversion of their results at $z_{\rm eff}=0.57$ implies $\sigma_8=0.73\pm0.05$, while \citet{Samushia14} use the same data to find $\sigma_8=0.77\pm0.05$. More recent analyses of the BOSS CMASS DR12 sample \citep{gilmarin/etal:2016,gilmarin/etal:prep} confirm these results with tighter error bars. Generally, most redshift space distortion results seem to be in agreement with our measurements even if the degree of tension with the Planck results varies from study to study.

\subsection{Assessing concordance with Planck}
\label{sec:discordance}
In Section~\ref{sec:S_8} we compare measurements of $S_8$, but this does not necessarily capture the overall level of dis-/concordance between Planck and KiDS-450. In assessing the concordance between CFHTLenS and Planck, \citet{joudaki/etal:2016} found that concordance tests grounded in the deviance information criterion (DIC; Section~\ref{sec:cosmo_priors}) and Bayesian evidence largely agreed, with the former enjoying the benefit of being more readily obtained from existing MCMC chains. We therefore follow this approach and assess the level of concordance between the two datasets $D_1$ and $D_2$ by computing
\begin{equation}
{\mathcal{I}}(D_1, D_2) \equiv \exp\{{-{\mathcal{G}}(D_1, D_2)/2}\}, 
\label{diceqn}
\end{equation}
where
\begin{equation}
{\mathcal{G}}(D_1, D_2) = {{{\rm{DIC}}(D_1 \cup D_2)} - {{\rm{DIC}}(D_1) - {{\rm{DIC}}(D_2)}}},
\end{equation}
and ${{\rm{DIC}}(D_1 \cup D_2)}$ is the DIC of the combined dataset. Thus, $\log \mathcal{I}$ is constructed to be positive when the datasets are concordant and negative when the datasets are discordant. The significance of the concordance test follows Jeffreys' scale \citep{jeffreys}, such that $\log \mathcal{I}$ values in excess of $\pm 1/2$ are `substantial', in excess of $\pm 1$ are `strong', and in excess of $\pm 2$ are `decisive'.

For our primary analysis we find that $\log \mathcal{I} = -0.79$, which translates into substantial discordance between KiDS-450 and Planck. This is consistent with the level of discordance inferred from the respective $S_8$ constraints. Note that we only use the Planck ``TT + lowP'' data for these comparisons. If we included Planck polarisation data as well (``TT + TE + EE + lowP'') the discordance would be even more pronounced.
%6

\section{DISCUSSION}
\label{sec:discussion}
The KiDS-450 dataset analysed here represents one of the most powerful cosmic shear surveys to date. Its combination of area, depth, and image quality is unprecedented, and this results in one of the most accurate and precise 
cosmological constraints from cosmic shear to date. In view of this precision, understanding systematic uncertainties becomes more important than in any previous such analysis. The treatment of systematic errors in the shear and photo-$z$ measurements of KiDS-450 is based on the most advanced methods described in the literature.  After accounting for residual uncertainties in these calibrations, KiDS-450 yields a constraining power on cosmological parameters similar to CFHTLenS.

The results presented in Section~\ref{sec:results} reveal a tension between Planck and KiDS-450 constraints on the matter density and the normalisation of the matter power spectrum. While the $2.3$-$\sigma$ level tension in the combined parameter $S_8$ is similar compared to previous analyses like CFHTLenS, there is now less room for explaining this tension with photometric redshift errors that were either unaccounted for or not considered as rigorously in the past. 
The reduced $\chi^2$ value of $\chi^2_{\rm eff}/{\rm dof}=1.3$ for our primary analysis indicates that our model is a reasonable fit.  Traditionally weak lensing analyses have focused on possible systematic errors in the shear measurements, and there are now a number of techniques that are able to achieve calibration uncertainties on the order of a per cent (see \citealt{mandelbaum/etal:2015} for a recent compilation). This level of accuracy is adequate for ground-based surveys like KiDS.  Attention is therefore shifting to the other main observable, the photometric redshifts.

The calibration of the source redshift distribution remains one of the main uncertainties in the analysis. In this work we compared three different calibration techniques and we found consistent cosmological results. Our primary method (DIR) is conceptually straightforward and statistically sufficiently powerful for present purposes, but relies on available deep spectroscopic surveys that span the full range of colours and magnitudes of KiDS galaxies -- something that is only beginning to be the case with current data.  For the alternative galaxy clustering based method (CC) the errors on the calibration are so large that the tension with Planck disappears. However, we believe that this calibration technique is currently the most problematic, both in terms of statistical power and in terms of systematic errors. Hence this apparent consistency should not be over-stressed. Estimating the cross-correlations from a much larger area in the future will not only yield better statistics but also alleviate some of the systematic problems discussed in Appendix~\ref{sec:sysCC}. Interestingly, the best-fit $\chi^2_{\rm eff}$ increases by $\sim10$ when switching from either the DIR or CC redshift distributions to the BOR or BPZ distributions (see Table~\ref{tab:derived_param}). This could be an indication that indeed the two re-calibrated $n(z)$ (DIR, CC) are a better representation of the data compared to the two sets of stacked $P(z)$ (BOR, BPZ). All three re-calibration approaches suffer from sample variance in the spec-$z$ calibration sample due to its finite size that we do not explicitly take into account. For KiDS-450 we estimate that this sample variance is subdominant to other sources of error on the cosmological parameters though.

We have found a small but significant B-mode signal at small angular scales. Its existence hints some aspect of the data that is not well understood, but ironically only the statistical power of KiDS makes it possible to detect such a low-level B-mode signal.  We assessed the impact of the measured B modes on our cosmological constraints by excluding the small angular scales from the analysis and by subtracting them from the E-mode signal under certain reasonable assumptions (Appendix~\ref{sec:EB}). In both cases we found no significant difference in the inferred cosmological parameters compared to our primary measurements. Given the small amplitude of the measured B modes, it seems unlikely that an improved understanding would lead to full consistency with all external datasets considered here. 

It is interesting to compare our findings to several recent re-analyses of the Planck data and to results from earlier CMB measurements. The general picture is that those re-analyses and independent measurements are more consistent with our findings than the Planck 2015 results. As for whether the main systematic problems are on the side of the weak lensing measurements or whether some aspect of the (much higher signal-to-noise) CMB measurements have to be revised remains a topic for further investigation. In order to make our results compatible with the Planck 2015 constraints one would have to assume the unlikely scenario that strong systematic errors have been overlooked. In particular one would require a multiplicative shear bias of $m\sim0.16$ or a photo-$z$ bias of $\Delta z\sim0.14$ that have been left unaccounted for in this lensing analysis.
These numbers are significantly larger than our estimated errors on the $m$ correction and the DIR photo-$z$ calibration.  Even if we had underestimated these errors, the main conclusions of this paper would not significantly change.  For example adopting a three-fold increase in the error on our shear calibration correction $\sigma_m$ would only increase our error on $S_8$ by $\sim15\%$ (see Sect.~\ref{sec:PS_marginal} and Fig.~\ref{fig:PS_marginal}), still resulting in a $\sim2$-$\sigma$ tension with respect to Planck. A similar argument holds for the residual sample variance in the photo-$z$ calibration.
%7

\section{SUMMARY AND OUTLOOK}
\label{sec:summary}
In this paper we present the first tomographic cosmic shear analysis of the Kilo Degree Survey using almost one third of the final data volume ($\sim$\SI{450}{\square\degree}). We make use of state-of-the-art data analysis tools like the \textsc{theli} pipeline for image reduction of the lensing data, the \textsc{Astro-WISE} system for multi-colour reduction and measurements, a new self-calibrating version of \emph{lens}fit for shear measurements, and three different photo-$z$ calibration methods based on deep spectroscopic surveys. For the estimation of measurement errors and sample variance we employ a redundant approach by estimating two independent covariance matrices for our data vector using an analytical and numerical approach. Our theoretical model mitigates the impact of astrophysical systematic effects related to intrinsic galaxy alignments and baryon feedback. The analysis was fully blinded with three different shear estimates in the catalogues.  Unblinding occurred right before submission of this paper.

The high-level data products used in this paper are publicly available at \url{http://kids.strw.leidenuniv.nl/}. This release includes the shear correlation functions, the covariance matrices, redshift distributions from the weighted direct calibration and their bootstrap realisations, and the full Monte Carlo Markov Chains for the primary analysis. 

Our findings are:
\begin{enumerate}
\item We find a best-fit value for $S_8\equiv\sigma_8\sqrt{\Omega_{\rm m}/0.3}=0.745\pm0.039$ assuming a flat $\Lambda$CDM model using weak external priors. The uncertainty on this parameter combination is within a factor of two of that derived from Planck alone although constraints on $\sigma_8$ and $\Omega_{\rm m}$ separately are much tighter for Planck. These findings are in tension with the Planck 2015 results at the $2.3$-$\sigma$ level but consistent with previous cosmic shear analyses and a number of other literature measurements. \item We use three different photo-$z$ calibration methods which yield slightly different redshift distributions.  When the uncertainty in each calibration is included in the cosmological analysis we find consistent cosmological constraints. 
\item For our primary results we use the photo-$z$ calibration method which makes the fewest assumptions and hence is most likely to have the best control of systematic errors. This direct calibration technique has uncertainties that are subdominant compared to the measurement errors. As an independent cross-check we also estimate cosmological constraints based on an alternative calibration technique that uses angular cross-correlations. The statistical uncertainties resulting from this technique dominate the error budget because of small areal coverage, resulting in weaker constraints that are, however, compatible with Planck, previous cosmic shear analyses, and other literature measurements. Further checks using the uncalibrated photo-$z$ probability distributions, or a re-calibrated version, give results that do not differ significantly from the primary analysis.
\item The multiplicative shear calibration estimated from a suite of dedicated end-to-end image simulations is on the order of 1 per cent with a statistical error of 0.3 per cent and a systematic uncertainty of 1 per cent. This calibration is a factor of 4 smaller than the one applied to the CFHTLenS shear measurements thanks to the new self-calibrating version of \emph{lens}fit.
\item The additive shear calibration estimated empirically from the data by averaging galaxy ellipticities in the different KiDS patches and redshift bins is, on average, a factor of 2 smaller than the one derived for the CFHTLenS analysis. We calibrate each $\sim$\SI{100}{\square\degree} patch and each tomographic bin separately for this effect.
\item We use two independent covariance matrices, one estimated analytically and the other one from $N$-body simulations, and find that they agree on small scales. On large scales the $N$-body method underestimates the covariance due to missing super-sample covariance (SSC) terms. Thus we use the analytical estimate for the final results.   This is the first time that the full SSC contribution is calculated and used in an observational cosmic shear analysis. 
\item We measure significant but low-level shear B modes on the smallest angular scales used in our analysis. This hints to some as yet uncorrected systematic error in the data. Making reasonable assumptions about the behaviour of these B modes or restricting the analysis to large scales with $\theta \geq 4.2'$, we show that the cosmological conclusions, in particular the tension found with Planck, are not affected.
\item We constrain the amplitude of the intrinsic alignments to $A_{\rm IA}=1.10\pm0.64$ assuming no luminosity or redshift dependence. The tension with the previously reported negative amplitude $A_{\rm IA}=-1.18^{+0.96}_{-1.17}$ from a joint analysis of CFHTLenS with WMAP7 \citep[see][]{heymans/etal:2013}, can be fully understood in terms of our improved knowledge of the true redshift distribution for low photometric redshift galaxies.
\item We extend our analysis to $w$CDM models, varying the equation-of-state parameter $w$.  We find full consistency with a cosmological constant.
\end{enumerate}

Data acquisition for the Kilo-Degree Survey is on-going, and we will revisit this cosmic shear measurement with future data releases. The European VIKING \citep[VISTA Kilo-Degree Infrared Galaxy;][]{edge/etal:2013} survey complements KiDS in five near-infrared bands. Inclusion of VIKING data will lead to better photo-$z$ and allow us to efficiently use redshifts $z>1$. These better photo-$z$ will also make it possible to divide the source sample into more tomographic bins giving better resolution along the line of sight. The redshift calibration will benefit as well since the mapping from 9-dimensional magnitude space ($ugriZYJHK_{\rm s}$) to redshift is better defined, with fewer colour-redshift degeneracies than in the 4-dimensional case presented here.

As the survey grows and the statistical precision increases further, it will become crucial to obtain a better understanding of the small-scale B modes and derive a correction scheme. We are currently investigating different hypotheses but as the presence of small-scale B modes does not impact the conclusions that we can draw from this KiDS-450 analysis, we leave a detailed investigation of the source of the B modes to a subsequent analysis.

Future progress in cosmic shear measurements will rely heavily on external datasets, in particular deep spectroscopic calibration fields. Ideally the weighted direct calibration used here should be carried out in a redundant way by using multiple independent spectroscopic surveys from different instruments and telescopes as well as from many different lines of sight. Filling up gaps in high-dimensional magnitude space by using a technique as described in \cite{masters/etal:2015} will greatly help to reach the full potential of cosmic shear measurements. In general, shallower and wider surveys are easier to calibrate with this technique compared to deeper, narrower surveys which is an important constraint for future observation plans.

If the tension between cosmological probes persists in the future, despite increasingly accurate corrections for systematic errors, modification of the current concordance model will become necessary \citep[see for example][]{riess/etal:2016}.   It is still too early to make the case for such extended models based on the KiDS-450 data alone, but in this cosmic shear study we see no evidence that this tension can be attributed to systematic errors in the weak lensing results.
%8

\section*{Acknowledgments}

We thank Matthias Bartelmann for being our external blinder, revealing which of the three catalogues analysed was the true unblinded catalogue at the end of this study.  We thank Martin Kilbinger for the {\sc athena} correlation function measurement software and the {\sc nicaea} theoretical modelling software and Antony Lewis for the {\sc camb} and {\sc CosmoMC} codes.  We thank members of the KiDS $i$-band eyeballing team Margot Brouwer, Marcello Cacciato, Martin Eriksen and Cristobal Sif\'on, for all their ground-work in road-testing the data verification checks that were subsequently used for the $r$-band analysis presented in this paper. We thank Maciek Bilicki and Marcello Cacciato for useful comments that improved the manuscript, Marika Asgari for useful discussions,  Aku Venhola and Reynier Peletier who kindly took the VST observations of the DEEP2 fields and Peter Capak for providing additional spectroscopy of galaxies to test our direct calibration analysis in the COSMOS field.  We thank Erminia Calabrese and Renee Hlozek for providing likelihood chains from the \citet{calabrese/etal:2013} and \citet{spergel/etal:2015} CMB analyses. We thank the zCOSMOS team for making their full, non-public redshift catalogue available for photo-$z$ tests.

This work was financially supported by the DFG (Emmy Noether grants Hi 1495/2-1 and Si 1769/1-1; TR33 `The Dark Universe'), the BMWi (via DLR, project 50QE1103), the Alexander von Humboldt Foundation, the ERC (grants 240185, 279396, 647112 and 670193), the Seventh Framework Programme of the European Commission (Marie Sklodwoska Curie Fellowship grant 656869 and People Programme grant 627288), the ERDF (project ERDF-080), the STFC (Ernest Rutherford Research Grant ST/L00285X/1, and Ernest Rutherford Fellowship ST/J004421/1), NWO (grants 614.001.103, 614.061.610, 022.003.013 and 614.001.451), NOVA, the Target program (which is supported by Samenwerkingsverband Noord Nederland, European fund for regional development, Dutch Ministry of economic affairs, Pieken in de Delta, Provinces of Groningen and Drenthe), NSERC, CIfAR and the ARC.
We acknowledge the use of ASTAC time on Swinburne's gSTAR and SwinSTAR computing clusters. We are grateful to the Lorentz Center for hosting our workshops.

Based on data products from observations made with ESO Telescopes at the La Silla Paranal Observatory under programme IDs 177.A-3016, 177.A-3017 and 177.A-3018, and on data products produced by Target/OmegaCEN, INAF-OACN, INAF-OAPD and the KiDS production team, on behalf of the KiDS consortium.

{\small \textit{Author Contributions:} All authors contributed to the development and writing of this paper. The authorship list is given in three groups: the lead authors (HHi, MV, CH, SJ, KK), followed by two alphabetical groups. The first alphabetical group includes those who are key contributors to both the scientific analysis and the data products. The second group covers those who have either made a significant contribution to the data products, or to the scientific analysis.}

\bibliographystyle{mnras}

\bibliography{CosmicShearPaper}

\clearpage
\appendix % begins the appendices
\addcontentsline{toc}{part}{Appendices} % adds a part entry in the TOC
\adjustptc % fixes the parttoc counter ptc

\section*{List of Appendices}

\mtcsettitle{parttoc}{} % blanks the parttoc title
\parttoc % adds a partial toc for the appendices

\section[Requirements on the shear and photo-$z$ calibrations]
{REQUIREMENTS ON THE SHEAR AND PHOTO-$z$ CALIBRATIONS}
\label{sec:app_A}
\newcommand{\ave}[1]{\langle #1 \rangle}
\newcommand{\Ave}[1]{\Big\langle #1 \Big\rangle}
\renewcommand{\d}[0]{{\rm d}}
\renewcommand{\vec}[1]{{\textbf{#1}}}
\newcommand{\tr}[1]{{\rm tr}{#1}}
\newcommand{\Ref}[1]{(\ref{#1})}
Given the statistical power of KiDS-450 it is important to ask the question how well we need to calibrate the shear and photo-$z$ estimates. We use a Fisher matrix formalism to get such an estimate of the required calibration.  As a fiducial model the following analysis adopts a standard $\Lambda\rm CDM$ cosmology with parameters $\Omega_{\rm m}=0.2905$, $\Omega_{\rm b}=0.0473$, $\sigma_8=0.826$, $h=0.69$, $n_{\rm s}=0.969$, and $\Omega_\Lambda=1-\Omega_{\rm m}$ from \cite{hinshaw/etal:2013}.

\subsection{Fisher matrix}
As a general approach to the problem, we imagine a vector $\vec{x}$ of random data points that is fitted by a model with parameters $\vec{p}=(p_1,p_2,\ldots,p_{N_{\rm p}})$. Here we employ two fitting parameters $\vec{p}=(\Omega_{\rm m},S_8)$. The data are subject to random noise $\vec{n}$ as defined by
\begin{equation}
  \vec{x} = \vec{m}(\vec{p})+\vec{n}(\vec{p})\;,
\end{equation} 
where $\vec{m}$ is the predicted model vector. The random noise vanishes on average, $\ave{\vec{n}(\vec{p})}=0$, and the covariance of noise is $\mat{C}(\vec{p})\equiv\ave{\vec{n}(\vec{p})\vec{n}(\vec{p})^{\rm T}}$. Thus, for a Gaussian noise model the covariance $\mat{C}$ fully defines the noise properties.  A perfect, non-degenerate model reproduces the noise-free data vector for one particular set of parameters $\vec{p}_{\rm true}$, such that $\ave{\vec{x}}=\vec{m}(\vec{p}_{\rm true})$. In the following, $\vec{p}_{\rm true}$ denotes our fiducial cosmology.

From noisy data, we infer the parameters $\vec{p}$ up to a statistical uncertainty $\Delta\vec{p}$ determined by the data likelihood ${\cal L}(\vec{x}|\vec{p})$.  The Cram\`er-Rao lower bound provides a lower limit for the parameter uncertainty,
\begin{equation}
  \Ave{(\Delta p_i)^2}\ge
  [\mat{F}^{-1}]_{ii}
\end{equation}
through the Fisher matrix $\mat{F}$ which for a Gaussian likelihood is given by
\begin{multline}
  \label{eq:fishermat}
  F_{ij}:=
  -\left.\Ave{\frac{\partial^2\ln{{\cal L}(\vec{x}|\vec{p})}}{\partial p_i\,\partial p_j}}\right|_{\vec{p}=\vec{p}_{\rm true}}
  \\
  =\frac{1}{2}\tr{\big(
    \mat{C}^{-1}\,\mat{C}_{,i}\,\mat{C}^{-1}\,\mat{C}_{,j}+
    \mat{C}^{-1}\left[ \vec{m}_{,i}^{}\,\vec{m}^{\rm
        T}_{,j}+\vec{m}_{,j}^{}\,\vec{m}^{\rm T}_{,i} \right]\big)}\;
\end{multline}
for the matrix components of $\mat{F}$, where
\begin{equation}
\label{eq:Cgrad}
  \mat{C}_{,i}:=
  \left.\frac{\partial{\mat{C}(\vec{p})}}{\partial p_i}\right|_{\vec{p}=\vec{p}_{\rm true}}~;~
  \vec{\vec{m}}_{,i}:=
  \left.\frac{\partial{\vec{m}(\vec{p})}}{\partial p_i}\right|_{\vec{p}=\vec{p}_{\rm true}}
\end{equation}
\citep{taylor/etal:2007}. 
The diagonal elements $[\mat{F}^{-1}]_{ii}$ are the square of the Fisher error of $p_i$, $\sigma(p_i)$, whereas the off-diagonals $[\mat{F}^{-1}]_{ij}$ quantify the covariance between $p_i$ and $p_j$. 

For KiDS-450, we express $\vec{m}$ by a tomography of shear power spectra, similar to \citet{2002PhRvD..66h3515H}, with 30 logarithmic bins covering angular wave numbers $\ell$ between 280 and 5000.  For a model of the noise covariance $\mat{C}$ we apply \citet{joachimi08} using the effective number density and ellipticity dispersion as listed in table~\ref{tab:tomo_bins}. With this setup we obtain Fisher errors of $\sigma(\Omega_{\rm m})=0.104$ and $\sigma(S_8)=0.033$, as well as a Pearson correlation of $r=-0.91$ between them. Note that these errors are in good agreement with those obtained from our `no systematics' MCMC analysis but, predictably, are slightly smaller than what we find once we allow for other uncertainties (see Table~\ref{tab:derived_param}).

\subsection{Bias due to calibration errors}
We imagine a model $\vec{m}(\vec{p})$ that has a set of additional nuisance or calibration parameters \mbox{$\vec{q}=(q_1,q_2,\ldots,q_{N_{\rm q}})$} that are constrained by external information rather than constrained by the data $\vec{x}$. With nuisance parameters included, both the model and the noise covariance are also functions of $\vec{q}$, henceforth denoted by $\vec{m}(\vec{p}|\vec{q})$ and $\mat{C}(\vec{p}|\vec{q})$. By $\vec{q}_{\rm true}$ we denote the values of nuisance parameters in the fiducial model. 

If $\vec{q}_{\rm true}$ is known, nothing changes in comparison to the foregoing Fisher formalism; we just set $\vec{m}(\vec{p})=\vec{m}(\vec{p}|\vec{q}_{\rm true})$ and $\mat{C}(\vec{p})=\mat{C}(\vec{p}|\vec{q}_{\rm true})$. If, on the other hand, we adopt biased calibration parameters $\vec{q}=\vec{q}_{\rm true}+\delta\vec{q}$, then the (average) likelihood function will, to linear order, be shifted by
\begin{equation}
  \label{eq:responserel}
  \delta\vec{p}=
  -\mat{F}^{-1}(\vec{p}_{\rm true})\,\mat{G}\,\delta\vec{q}
\end{equation}
\citep[cf. the appendix of][]{taylor/etal:2007}. Here we have introduced the pseudo-Fisher matrix $\mat{G}$, whose elements $G_{ij}$ are defined as in Eq.~\ref{eq:fishermat} but where the partial derivatives are with respect to the nuisance parameters $q_i$,
\begin{equation}
  \mat{C}_{;i}:=
  \left.\frac{\partial{\mat{C}(\vec{p}_{\rm true}|\vec{q})}}{\partial q_i}\right|_{\vec{q}=\vec{q}_{\rm true}};\!\!~
  \vec{\vec{m}}_{;i}:=
  \left.\frac{\partial{\vec{m}(\vec{p}_{\rm true}|\vec{q})}}{\partial q_i}\right|_{\vec{q}=\vec{q}_{\rm true}}\;.
\end{equation}

\begin{figure}
  \includegraphics[width=\hsize]{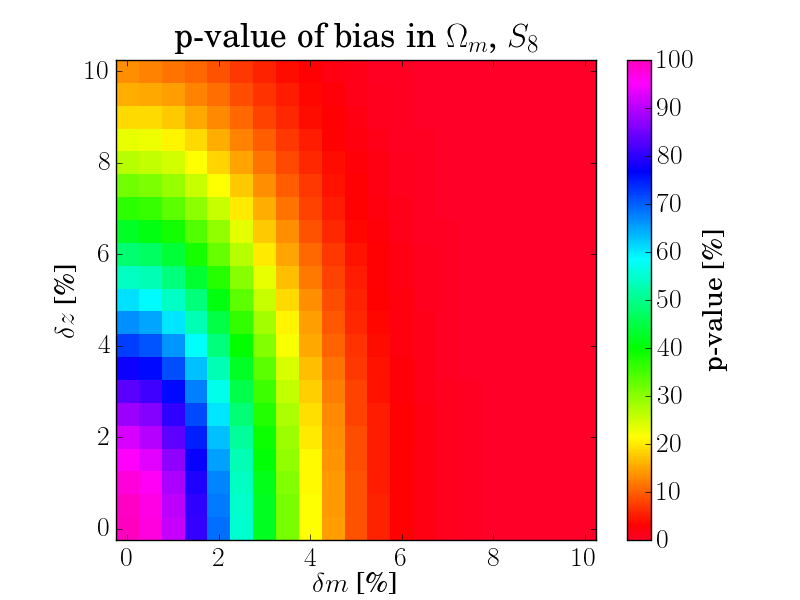}\\
  \includegraphics[width=\hsize]{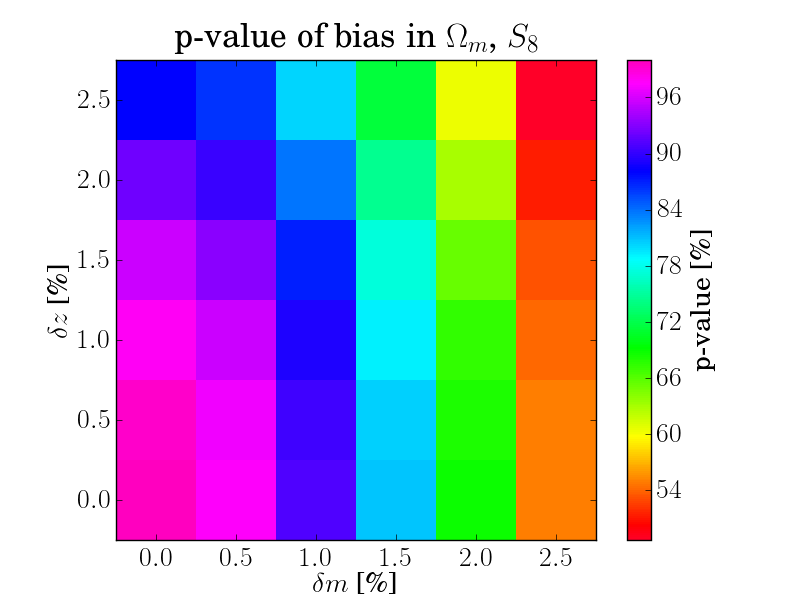}\\
  \caption{\label{fig:PS_tolerance}The value of $p_{68}$, the bias in the cosmological parameters $\Omega_{\rm m}$ and $S_8$ if the error in the multiplicative shear calibration $\delta m$ and the error in the mean redshift of the tomographic bins $\delta z$ is not corrected. The lower panel zooms in to the relevant region of parameter space.}
\end{figure}

\subsection{Tolerance limits}
\label{sec:tolerance_limits}
For an accurate model fit, we require the typical bias \mbox{$\delta\vec{p}=-\mat{F}^{-1}\,\mat{G}\,\delta\vec{q}$} to be small in comparison to the expected statistical error $\Delta\vec{p}$. For the following assessment, we assume for $\Delta\vec{p}$ a multivariate Gaussian probability density with covariance $\mat{F}^{-1}$ and zero mean. The probability density function (PDF) of $\delta\vec{q}$ shall also obey Gaussian statistics, now with a variance of $\sigma_i^2$ for each component $\delta q_i$. As test statistics for the significance of a $\delta\vec{p}$ relative to the typical distribution of statistical errors $\Delta\vec{p}$, we use
\begin{eqnarray}
  \label{eq:chi2}
  \Delta\chi^2(\delta\vec{q})&:=&
  \delta\vec{p}^{\rm T}\,\mat{F}(\vec{p}_{\rm true})\,\delta\vec{p}\\
  &=&
  \delta\vec{q}^{\rm T}
  \left(\mat{G}^{\rm T}\,\mat{F}^{-1}(\vec{p}_{\rm true})\,\mat{G}\right)
  \delta\vec{q}\;.
\end{eqnarray}
Given a PDF model for $\delta\vec{q}$, the statistics $\Delta\chi^2(\delta\vec{q})$ follows a distribution for which the $68$th percentile $\Delta\chi^2_{68}$, given by
 \begin{equation}
   P(\Delta\chi^2(\delta\vec{q})\le\Delta\chi^2_{68})=0.68\;,
 \end{equation}
quantifies the spread of values. We compute the values of $\Delta\chi^2_{68}$ for a range of models for $\delta\vec{q}$ which differ in $\sigma_i^2$. For each model, we then consider the impact of a bias $\delta\vec{q}$ negligible relative to statistical errors $\Delta\vec{p}$ if the corresponding value of $\Delta\chi^2_{68}$ is small compared to the distribution \mbox{$\Delta\chi^2_0:=\Delta\vec{p}^{\rm T}\vec{F}(\vec{p}_{\rm true})\Delta\vec{p}$}, or if the probability of \mbox{$\Delta\chi^2_0\ge\Delta\chi^2_{68}$} is large. We assess this by computing numerically the $p$-value
 \begin{equation}
  p_{68}=P\left(\Delta\chi^2_0\ge\Delta\chi^2_{68}\right)
\end{equation}
by doing Monte-Carlo realisations of $\Delta\chi^2_0$ based on random values of $\Delta\vec{p}$. A large value of $p_{68}$ thus indicates that statistical errors $\Delta\vec{p}$ are large in comparison to systematic errors $\delta\vec{p}$: the bias is negligible.

For KiDS-450, we consider two types of calibration errors $\delta\vec{q}$ in each of the four tomographic bins: a systematic shift of the redshift distribution, $z=z_{\rm true}(1+\delta z)$, and a systematic error of shear values, $\gamma=\gamma_{\rm true}(1+m)$. The photo-$z$ errors are slightly correlated between the tomographic bins $i$ and $j$ according to the correlation matrix
\begin{equation}
  [\mat{r}]_{ij}=
  \left(
    \begin{array}{cccc}
      1 & +0.03 & -0.02 & +0.04 \\
      +0.03 & 1 & -0.03 & +0.01 \\
      -0.02 & -0.03 & 1 & +0.08 \\
      +0.04 & +0.01 & +0.08 & 1
    \end{array}
  \right)\;,
\end{equation}
which we determined by bootstrapping the data. The systematic errors of shear, however, are strongly correlated. We adopt a correlation coefficient of $r=0.99$ between all bins throughout. This represents a worst case scenario in terms of deriving requirements for the shear bias calibration uncertainty. We indeed expect a strong correlation between the multiplicative biases in the four tomographic bins, as they share very similar distributions in terms of signal-to-noise and galaxy size, which are the main parameters used to characterise the shear bias (see Appendix \ref{sec:app_shear_tests}). The resulting $p$-values as a function of uncertainty of calibration errors are shown in the top panel of Fig.~\ref{fig:PS_tolerance}. Here we assume the photo-$z$ error to be the same in each bin as given by the value on the y-axis. The $p$-value drops below 95 per cent for photo-$z$ and shear calibration errors with RMS uncertainties $\sigma_{\delta z},\sigma_m\sim1.5$ per cent for all four bins.

In addition, we consider uncertainties that mimic our calibration precision in KiDS-450 where the uncertainties differ across the tomographic bins. The correlation of errors are as before. The systematic redshift error (taking the relative error of the mean from column 8 in Table~\ref{tab:tomo_bins}) is $\{\sigma_{\delta z,i}\}=\{4.8,2.8,1.4,0.7\}$ per cent, and the shear bias is calibrated to a precision of either $\sigma_m=0.5$ per cent (optimistic) or $\sigma_m=1$ per cent (pessimistic) in all bins. We find $p_{68}=0.66(0.71)$ in the pessimistic (optimistic) scenario. This implies that the low-level systematics that we have identified and calibrated could bias our results such that we need to also marginalise over our uncertainty in the measured calibration.

\begin{figure}
  \includegraphics[width=\hsize]{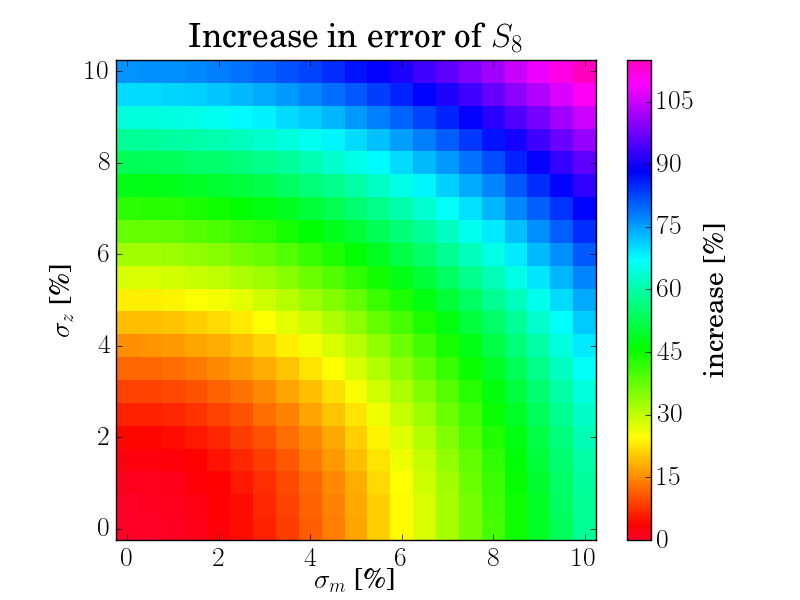}\\
  \includegraphics[width=\hsize]{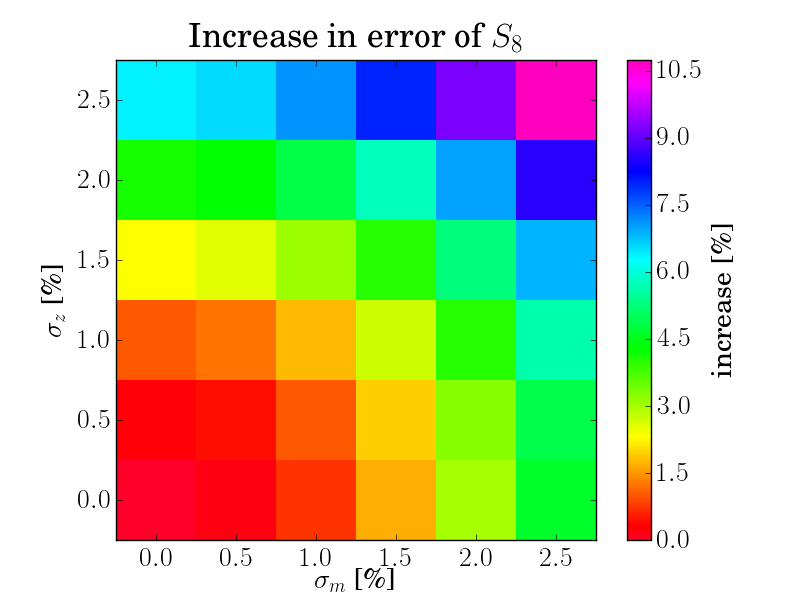}\\
  \caption{\label{fig:PS_marginal}The percentage increase in the errors of the composite parameter $S_8$ if the uncertainties in the multiplicative shear calibration, $\sigma_m$, and the mean redshifts, $\sigma_z$, are marginalised over.  The lower panel zooms in to the relevant region of parameter space.}
\end{figure}

\subsection{Marginalising calibration errors}
\label{sec:PS_marginal}
We consider the possibility that the uncertainty of $\delta\vec{q}$ is directly accounted for in the statistical errors of model parameters $\vec{p}$. Thus we do not set the calibration parameters to a specific value but instead marginalize over the uncertainty in $\vec{q}$. For this discussion, we assume that the PDF of $\vec{p}$ is well approximated by a Gaussian density: ${\cal N}(\vec{p}_{\rm true},\mat{F}^{-1})$; it has the mean $\vec{p}_{\rm true}$ and the covariance $\mat{F}^{-1}(\vec{p}_{\rm true})$. In addition, we define a Gaussian prior PDF of the calibration error $\delta\vec{q}$, namely ${\cal N}(0,\mat{C}_{\rm q})$, with zero mean and covariance $\mat{C}_{\rm q}=\ave{\delta\vec{q}\,\delta\vec{q}^{\rm T}}$.  The calibration uncertainty corresponds to the systematic error in $\vec{p}$ space that has the covariance
\begin{equation}
  \mat{C}_{\rm p}=
  \ave{\delta\vec{p}\,\delta\vec{p}^{\rm T}}=
  \mat{F}^{-1}\,\mat{G}\,\mat{C}_{\rm q}\,\mat{G}^{\rm T}\,\mat{F}^{-1}\,,
\end{equation}
because $\delta\vec{p}=-\mat{F}^{-1}\,\mat{G}\,\delta\vec{q}$. Marginalising with respect to $\delta\vec{p}$ hence adds extra uncertainty to \mbox{${\cal N}(\vec{p}_{\rm true},\mat{F}^{-1})$} which we obtain by convolving this PDF with the kernel ${\cal N}(0,\mat{C}_{\rm q})$.  This results in the Gaussian ${\cal N}(\vec{p}_{\rm true},\mat{F}_{\rm m}^{-1})$ with broadened covariance
\begin{equation}
  \mat{F}_{\rm m}^{-1}=
  \mat{C}_{\rm p}+\mat{F}^{-1}=
  \mat{F}^{-1}\,
  \left(
    \mat{G}\,\mat{C}_{\rm q}\,\mat{G}^{\rm
      T}+\mat{F}
    \right)\,
    \mat{F}^{-1}\;.
\end{equation}

We utilise the marginal Fisher matrix $\mat{F}_{\rm m}$ to assess the relative growth of the Fisher error of the composite parameter $S_8$ due to marginalisation.  For this, we plot the fractional increase $\sigma(S_8)/\sigma_0-1$ of $\sigma(S_8)=\sqrt{[\mat{F}_{\rm m}^{-1}]_{22}}$ relative to $\sigma_0=\sqrt{[\mat{F}^{-1}]_{22}}$. 

The situation for KiDS-450 is shown in Fig.~\ref{fig:PS_marginal}. It can be seen that the errors on the relevant cosmological parameters increase by $\sim2$ per cent for shear and photo-$z$ calibrations that are known to $\sim1$ per cent if those uncertainties are marginalised over. Again it is assumed that the photo-$z$ errors are the same in all bins.

For the calibration precision of KiDS-450, i.e. photo-$z$ errors of $\{\sigma_{\delta z,i}\}=\{4.8,2.8,1.4,0.7\}$ per cent in the four bins and a shear calibration error of $\sigma_m=0.5$ per cent (optimistic) or $\sigma_m=1$ per cent (pessimistic), we find the corresponding value of $\sim8.1$ and $\sim8.7$ per cent in the optimistic and pessimistic scenario, respectively. This error is dominated by the uncertainty of the photo-$z$ calibration in the first tomographic bin. The actual increase in the error on $S_8$ that we find when switching from the ``DIR no error'' case to the ``DIR'' case (see Table~\ref{tab:MCMC_setups}) is of the order of $\sim3$ per cent. This  suggests that the linear model we are using here in this Fisher analysis by shifting the redshift distribution coherently around is too pessimistic compared to the complex changes in the shape of the redshift distributions, especially for the first tomographic bin. The results in this section can hence be understood as upper limits on how much systematic errors in the shear and photo-$z$ calibration compromise the statistical power of KiDS-450.
%A

\section[Photometric calibration with stellar locus regression]
{PHOTOMETRIC CALIBRATION WITH STELLAR LOCUS REGRESSION}
\label{sec:app_SLR}

Photometric homogeneity is an important requirement for a large imaging survey such as KiDS. It is difficult to attain because the observations consist of many separate tiles observed over a time span of years, and in conditions that are not always fully photometric. For this reason KiDS tiles overlap slightly with their neighbours, so that sources common to adjacent tiles can be used to cross-calibrate the individual tiles' photometric zero points. However, as Fig.~\ref{fig:footprint} shows, the KiDS-450 data are still quite fragmented, especially outside the main contiguous areas in G9, G12, G15, G23 and GS and therefore tile overlaps are inadequate to obtain homogeneous photometry across the full dataset. The results of the tile-by-tile KiDS photometric calibration described in Section~\ref{sec:AW} and \cite{dejong/etal:2015} are reported in Table~\ref{tab:SLR1} finding a scatter in the $(u-r,g-r,r-i)$ colours of $(0.04,0.04,0.06)$ with respect to the SDSS DR9 photometry, as well as an average offset of $(0.005,0.005,0.015)$ mag.  In addition we find that some outlier tiles can display magnitude residuals up to 0.1 in $g$, $r$ and $i$ and up to 0.2 in $u$ \citep{dejong/etal:2015}. 

\begin{table*}
\caption{\label{tab:SLR1}Main results from the Stellar Locus Regression (SLR) applied to the KiDS \textsc{GAaP} photometry. Comparisons to SDSS use its DR9 PSF photometry \citep{ahn/etal:2012}. Column 2: indicator of achieved colour homogeneity prior to applying Stellar Locus Regression. Listed are mean and standard deviation of the colour residual per tile when subtracting the SDSS and KiDS stellar photometry. Column 3: Derived SLR corrections. Column 4: same as Col. 2 but now after SLR has been applied.  For comparison and to judge the intrinsic scatter of the method we also apply SLR to the SDSS photometry and then compare to the original SDSS photometry in Column 5.}
\begin{tabular}{lrrrr}
\hline
colour & KiDS$-$SDSS before SLR & SLR offset & KiDS$-$SDSS after SLR & (SDSS+SLR)$-$SDSS\\
           & [mmag] & [mmag] & [mmag] & [mmag] \\
\hline
${\rm d}(u-r)$ &   $4\pm42$ & $71\pm87$ & $80\pm64$ & $26\pm38$\\
${\rm d}(g-r)$ &   $6\pm38$ & $14\pm75$ & $9\pm12$ & $11\pm10$\\
${\rm d}r$       & $11\pm28$ & $-$ & $11\pm28$ & N.A.\\
${\rm d}(r-i)$  & $-16\pm56$ & $-21\pm60$ & $-6\pm11$ & $4\pm6$\\
\hline
\end{tabular}
\end{table*}

In order to improve the photometric calibration, particularly in KiDS-S where there is no overlap with SDSS photometry, we make use of the fact that the majority of stars display a well-defined photometric `stellar locus': a tight relation in colour-colour space that varies little across the sky outside the Galactic plane \citep{ivezic/etal:2004, high/etal:2009}. Matching the observed loci to the fiducial intrinsic locus therefore offers the possibility to achieve colour homogeneity for the KiDS-450 tiles without using the photometry of objects in the overlap regions of different exposures. We follow the usual nomenclature and refer to this approach as Stellar Locus Regression (SLR).

\subsection{Implementation and results}
We apply the SLR to the KiDS \textsc{GAaP} \citep[Gaussian Aperture and Photometry;][]{kuijken:2008} photometry. The first step is to determine `principal colours': linear combinations of $u-g$, $g-r$ and $r-i$ that align with the characteristic straight regions of each stellar locus in colour-colour space. Following the approach of \citet{ivezic/etal:2004} we define four principal colours \citep[see Table~\ref{tab:SLR2} for the fitting coefficients which are taken from][]{ivezic/etal:2004}:
\begin{description}
\item $s$ (straight region in  $u-g$ , $g-r$);  
\item $x$ (straight red region in $g-r$ , $r-i$);
\item $w$ (straight blue region in $g-r$, $r-i$);
\item $k$ (straight region in $u-r$ vs $r-i$).
\end{description}
For each principal colour $c$, $P1c$ and $P2c$ denote the colour projected along/perpendicular to the stellar locus, respectively. Any deviation from the fiducial stellar locus reveals itself as a non-zero $P2$ colour. Since there are only three independent colours in the dataset, we choose to line up the stellar loci by perturbing only the $u-r$, $g-r$ and $r-i$ colours in each tile, and to leave the $r$-band zero points unchanged. Indeed, analysis of the per-tile calibration residuals shows the $r$-band to be the most homogeneous \citep{dejong/etal:2015}.

\begin{table}
\caption{\label{tab:SLR2}Coefficients from \citet{ivezic/etal:2004} that define the principal colours $s$, $x$, $w$ and $k$. Listed are the coefficients defining the principal colour perpendicular to the straight locus. For example $P2s = -0.249\,u + 0.794\,g - 0.555\,r + 0.234$.}
\begin{tabular}{lrrrrr}
\hline
Principal & $u$ & $g$ & $r$ & $i$ & constant\\
colour & & & & & \\
\hline
$P2s$ & $-$0.249 & 0.794 & $-$0.555 & & 0.234\\
$P2x$ & & 0.707 & $-$0.707 & & $-$0.988\\
$P2w$ & & $-$0.227 & 0.792 & $-$0.567 & 0.050\\
$P2k$ & 0.114 & & $-$1.107 & 0.994 & $-$0.0420\\
\hline
\end{tabular}
\end{table}

Given the small differences between the KiDS and SDSS photometric systems \citep{dejong/etal:2015} we use the same intrinsic locations for $s$, $x$ and $w$ as \cite{ivezic/etal:2004}. The fourth, redundant principal colour $k$ is used as an additional guard for fitting robustness. 

\begin{figure}
\includegraphics[width=\hsize]{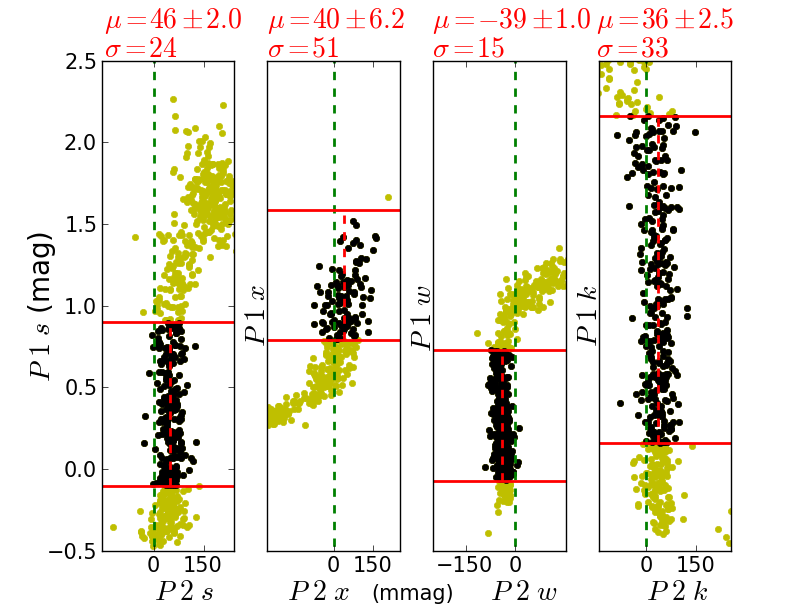}
\caption{\label{fig:SLR1}Principal colours $s$, $w$, $x$ and $k$ for the pointing KiDS\_3.6\_-34.1. The initial selection of stars (all points) is identical for $s$, $w$, $x$ and $k$. From this parent sample the stars on the locus are selected (black). The inferred median $P2$ offsets, its formal error and the standard deviation of the locus point sources are shown above each panel.}
\end{figure}

\begin{figure}
\includegraphics[width=\hsize]{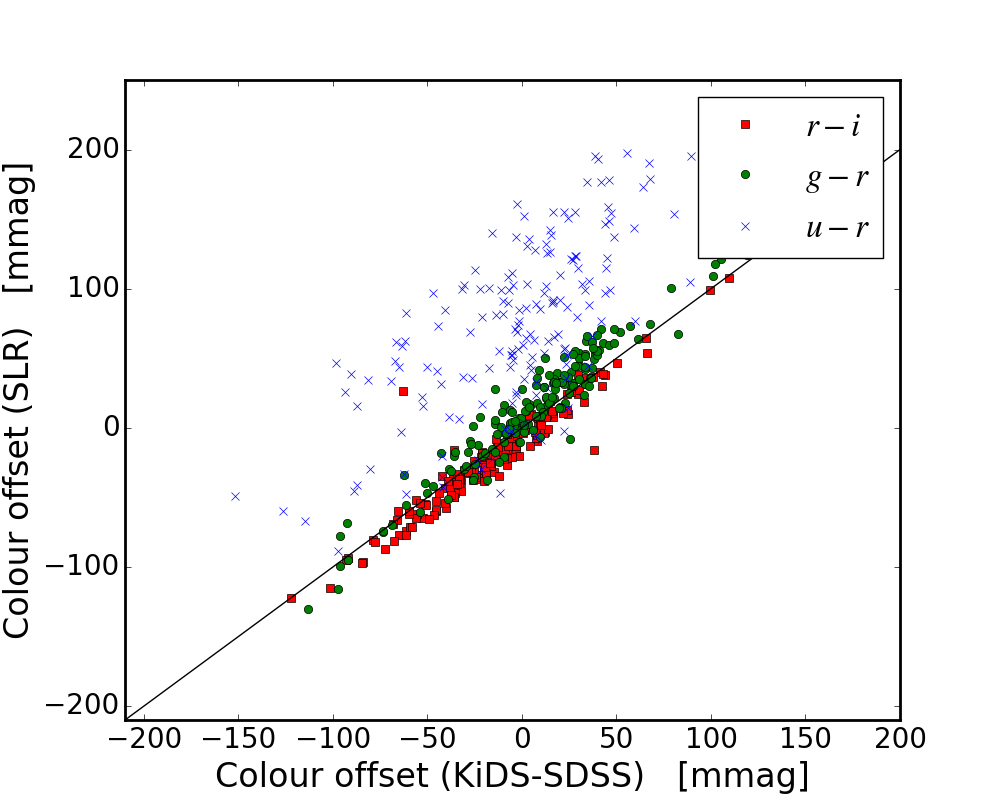}
\caption{\label{fig:SLR2}The offsets in the $u-r$, $g-r$ and $r-i$ colours predicted by SLR vs.~the predictions from a direct comparison with SDSS. The mean and standard deviation of the distributions are listed in Table~\ref{tab:SLR1}.}
\end{figure}

In each tile we select bright point sources with $r<19.0$  on the Gaussianized coadds using the following criteria:
\begin{enumerate}
\item 2pix $<$ {\verb FWHM  }$<$ 4pix
\item {\verb Flag_r } $=$ 0 
\item {\verb IMAFLAGS_ISO_r } $=$ 0 
\end{enumerate}
The first criterion selects point-like objects and the second and third remove sources with compromised photometry (see \citealt{dejong/etal:2015} for precise definitions). For each point source the \textsc{GAaP} magnitude is corrected for Galactic extinction using the $E(B-V)$ colour excess from \cite{schlegel/etal:1998} in combination with a standard $R_V=3.1$ Galactic extinction curve. This assumes that most of the stars used for calibration are outside the dust disk of the Milky Way. For each principal colour we then compute the $P1$ and $P2$ components of these sources, as shown in Fig.~\ref{fig:SLR1}. We identify sources near the straight region of the stellar locus by setting a fixed range around the median $P1$ and requiring $|P2-{\rm median}(P2)|<\sim200{\rm mmag}$. Per tile the median $P2$ values of these sources are finally converted into three colour offsets ${\rm d}(u-r)$, ${\rm d}(g-r)$, ${\rm d}(r-i)$ and applied to the KiDS \textsc{GAaP} magnitudes before they are fed to the photo-$z$ code.  Column 3 of Table~\ref{tab:SLR1} lists the distribution of the resulting offsets which are also shown in Fig.~\ref{fig:SLR2}. For comparison and to judge the intrinsic scatter of the method we also apply SLR to the SDSS photometry and then compare to the original photometry. Those findings are reported in the last column of Table~\ref{tab:SLR1}.

We find that stellar locus regression on KiDS \textsc{GAaP} photometry significantly improves the photometric stability over the survey area in the $g-r$ and $r-i$ colours. The fluctuations with respect to SDSS decrease from $\sim0.04$ and $\sim0.06$ mag in $g-r$ and $r-i$, respectively, to roughly 0.01 mag in both colours. The calibration of the $u-r$ colour, however, does not improve: to the contrary, the scatter after SLR ($\sim0.06$ mag) is slightly larger than before SLR ($\sim0.04$ mag). Also a significant offset of $\sim0.08$ mag is introduced. We attribute these problems to metallicity variations in the stellar sample which results in a variable stellar locus, making this technique fundamentally problematic for near-UV data \citep[similar findings are reported by][]{high/etal:2009}.   In the final column of Table~\ref{tab:SLR1}, we see that applying SLR to the actual SDSS data also degrades the calibration of the $u-r$ colour.

We argue that the photo-$z$ calibration (Section~\ref{sec:photoz_calibration}) is in no way compromised by the \emph{offset} in the $u-r$ calibration since the spectroscopic calibration fields are also calibrated with SLR and hence share this offset. The $\sim0.06$ mag \emph{fluctuation} of the $u-r$ photometry leads to a tile-to-tile difference in the photo-$z$ bias, and hence in the redshift distribution.  The redshift distributions estimated from the four calibration deep spec-$z$ fields are still applicable on average if the fluctuations between those four fields are comparable to the full survey.  Also given the relatively large errors of the individual $u$-band measurements, we do not expect this $0.06$\,mag fluctuation in $u-r$ to have any major consequences for the applicability of the redshift calibration.
%B

\section[Photo-$z$ calibration analysis]
{PHOTO-$z$ CALIBRATION ANALYSIS}
\label{sec:app_z_tests}
\subsection{VST imaging of deep spec-\textit{z} fields}
\label{app:deepzdata}

We calibrate the KiDS-450 tomographic redshift distributions through the analysis of deep spectroscopic datasets from the literature.  In order to extend our spectroscopic overlap we incorporate VST observations of a number of deep spectroscopic fields that fall outside the main KiDS survey footprint (Table~\ref{tab:specz_obs_cond}).  The DIR calibration procedure described in Section~\ref{sec:DIR} re-weights the spec-$z$ catalogue such that it represents the colour and magnitude properties of the photometric catalogue. Magnitude \emph{errors} will inevitably be affected by noise variations across the survey area caused by variations in seeing, exposure time, atmospheric extinction, moon phase and distance, etc.   It is therefore important that the observing setup and conditions of the imaging observations in the deep spectroscopic fields are representative of the main KiDS survey observations such that the re-weighting scheme, determined from the spec-$z$ fields, is valid in its application to the full KiDS-450 area.

Table~\ref{tab:specz_obs_cond} summarises the observing conditions in the four fields, in comparison to the mean observing conditions in KiDS-450.  It demonstrates that these data are indeed typical in terms of exposure time and seeing.  The PSF ellipticity and size variations between the observations are taken care of by our galaxy photometry pipeline.

\begin{table}
\caption{\label{tab:specz_obs_cond}Observing conditions in the spec-$z$ fields compared to the main KiDS-450 survey. Central coordinates of the \SI{1}{\square\degree} VST observations for each field are listed under the names of the fields.}
\begin{tabular}{lcrr}
\hline
field & band & seeing FWHM & exposure time\\
 & & [arcsec] & [s]\\
\hline
KiDS-450 mean    & $u$ & 1.01 & 1000 \\
                 & $g$ & 0.87 &  900 \\
                 & $r$ & 0.70 & 1800 \\
                 & $i$ & 0.83 & 1200 \\
\hline
COSMOS$^a$           & $u$ & 0.74 & 1000 \\
 $(150.08,2.20)$                & $g$ & 1.05 &  900 \\
                 & $r$ & 0.57 & 1800 \\
                 & $i$ & 1.04 & 1200 \\
\hline
CDFS$^b$            & $u$ & 1.09 & 1200 \\
 $(53.40,-27.56)$                & $g$ & 0.56 & 1080 \\
                 & $r$ & 0.55 & 1800 \\
                 & $i$ & 0.93 & 1200 \\
\hline
DEEP2-2h$^c$         & $u$ & 0.92 & 1000 \\
  $(37.16,0.50)$               & $g$ & 0.91 &  900 \\
                 & $r$ & 0.72 & 1800 \\
                 & $i$ & 0.70 & 1200 \\
\hline
DEEP2-23h$^c$        & $u$ & 1.08 & 1000 \\
  $(352.00,0.00)$               & $g$ & 0.96 &  900 \\
                 & $r$ & 0.64 & 1800 \\
                 & $i$ & 0.73 & 1200 \\
\hline
\end{tabular}
$^a$The COSMOS field is observed as part of the KiDS survey. \\
$^b$Data from the VOICE project \citep{vaccari/etal:2012}.\\ 
$^c$Data taken during OmegaCAM guaranteed time (Nov.~2015).
\end{table}

\subsection{Statistical errors in DIR and CC calibrations}
\label{sec:DIR_stat_err}

\begin{figure}
\includegraphics[width=\hsize]{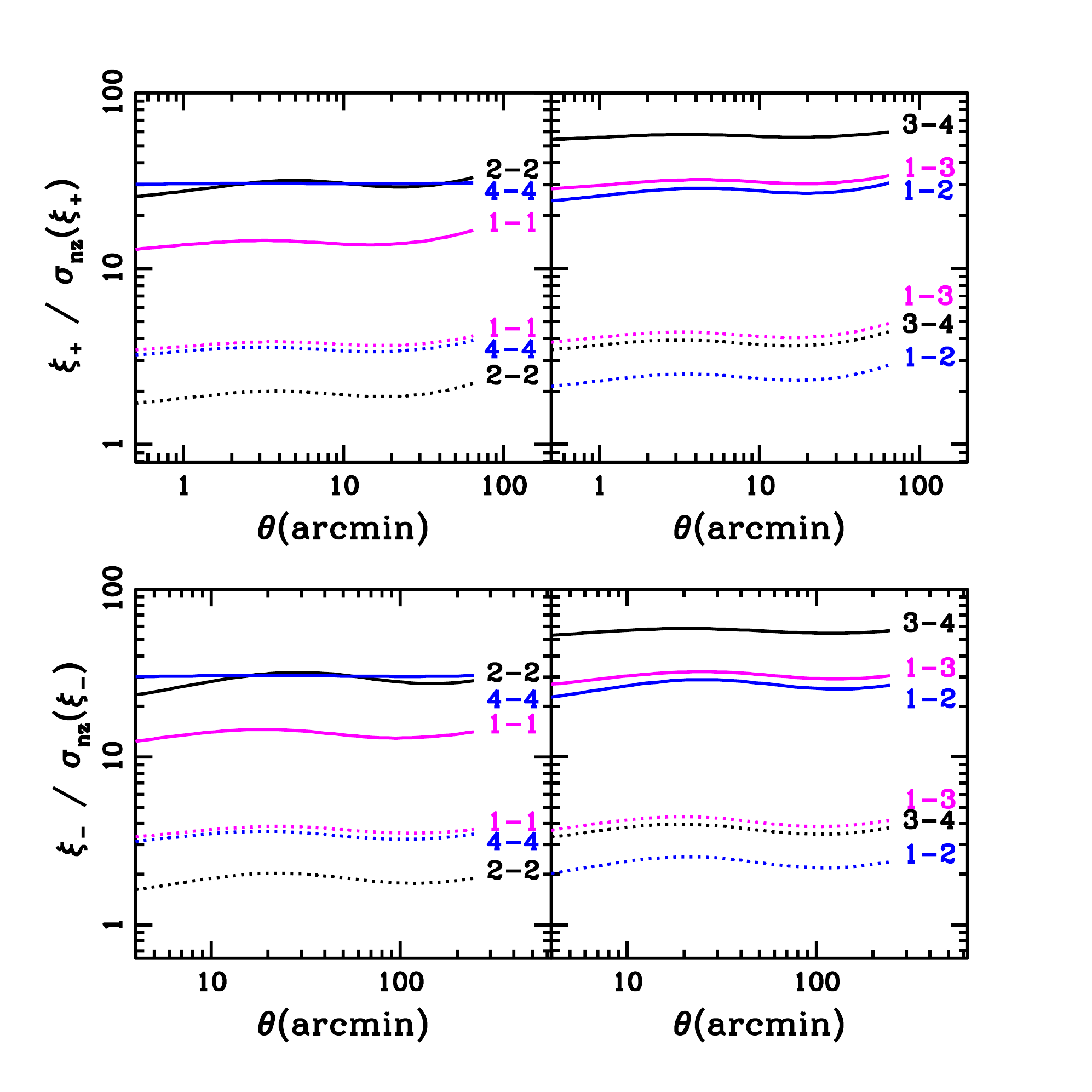}
\caption{\label{fig:SN_DIR_CC} Redshift distribution uncertainty:  signal-to-noise estimates of $\xi_+$ (upper) and  $\xi_-$ considering only the statistical noise $\sigma_{\rm nz}$ that arises from the uncertainty in the $n(z)$ as measured by the weighted direct calibration (DIR, solid) and by the cross-correlation calibration (CC, dotted).   This can be compared to the actual signal-to-noise on measurements of $\xi_\pm$ in the KiDS-450 data that are presented in Fig.~\ref{fig:covSN} for the same sample of tomographic bins, labelled `$i-j$' where $i,j=1\dots4$.}
\end{figure}

Given the importance of the photometric redshifts for the interpretation of the tomographic cosmic shear measurements, we present an assessment of how the uncertainty in the $n(z)$ that we estimate from the KiDS-450 data propagates into errors on our theoretical model of the shear correlation function $\xi_\pm$.   Statistical errors on both the weighted direct photo-$z$ calibration and the cross-correlation photo-$z$ calibration are estimated using 1000 bootstrap samples created from the full spec-$z$ catalogue of 23\,088 objects.   For each bootstrap realisation of the tomographic redshift distributions, we calculate a theoretical model for $\xi_\pm$ for a fixed fiducial cosmology.  The variance between the resulting models provides an estimate of the uncertainty on $\xi_\pm$, denoted $\sigma_{\rm nz}$,  that arises purely from our uncertainty in the $n(z)$.    Fig.~\ref{fig:SN_DIR_CC} shows the signal-to-noise ratio of $\xi_\pm$, with the noise given by $\sigma_{\rm nz}$, for a selection of 6 out of the 10 tomographic bin combinations used in our analysis.    The statistical noise from the weighted direct calibration estimate (DIR, shown solid) is significantly lower than the statistical noise from the cross-correlation calibration (CC, shown dotted), reflecting the lower precision of the latter technique given the small-area spectroscopic surveys that we can cross-correlate with.

Fig.~\ref{fig:SN_DIR_CC} can be compared to the actual signal-to-noise of measurements of $\xi_\pm$, presented in Fig.~\ref{fig:covSN} for the same sample of tomographic bins.    For the DIR calibration we see that the $\sigma_{\rm nz}$ statistical errors are subdominant to the noise in the shear correlation function measurements on all scales.   For the CC calibration, however, the $\sigma_{\rm nz}$ statistical errors are greater than the shot noise and sample variance in the data.    The uncertainty on the CC calibrated $n(z)$ therefore significantly limits the cosmological information that can be extracted from the cosmic shear analysis as shown in Fig.~\ref{fig:Om_s8_pzerr}. We note that the spec-$z$ catalogues used here are amongst the deepest and most complete surveys that are currently available. In the absence of new deeper spectroscopic surveys, Fig.~\ref{fig:SN_DIR_CC} represents a limit on the precision of all lensing surveys, not just KiDS-450.

\subsection{Systematic error analysis}
\subsubsection{Weighted direct calibration (DIR)}
\label{sec:sysDIR}
In principle, the weighted direct calibration method should be relatively free from systematic errors, provided the magnitude measurements and spectroscopic redshifts are accurate. The only requirement is that the spec-$z$ sample spans the full extent of the magnitude space that is covered by the photometric sample, and that the mapping from magnitude space to redshift is unique. In the following we describe the tests that we have undertaken to verify that we have met these requirements.

\begin{figure*}
\includegraphics[clip=true, trim=0.5cm 0cm 2cm 1cm, width=0.32\textwidth]{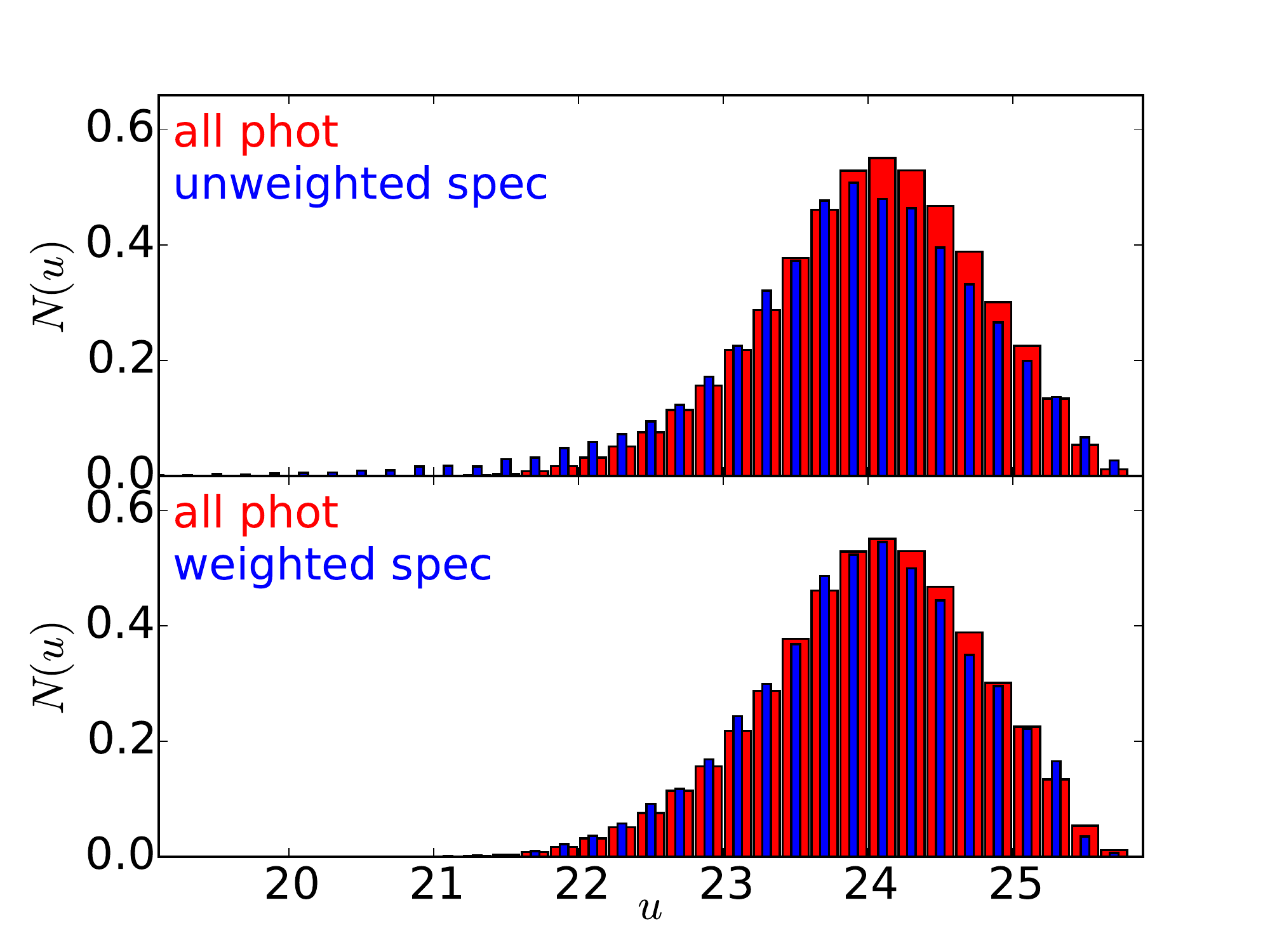}
\includegraphics[clip=true, trim=0.5cm 0cm 2cm 1cm, width=0.32\textwidth]{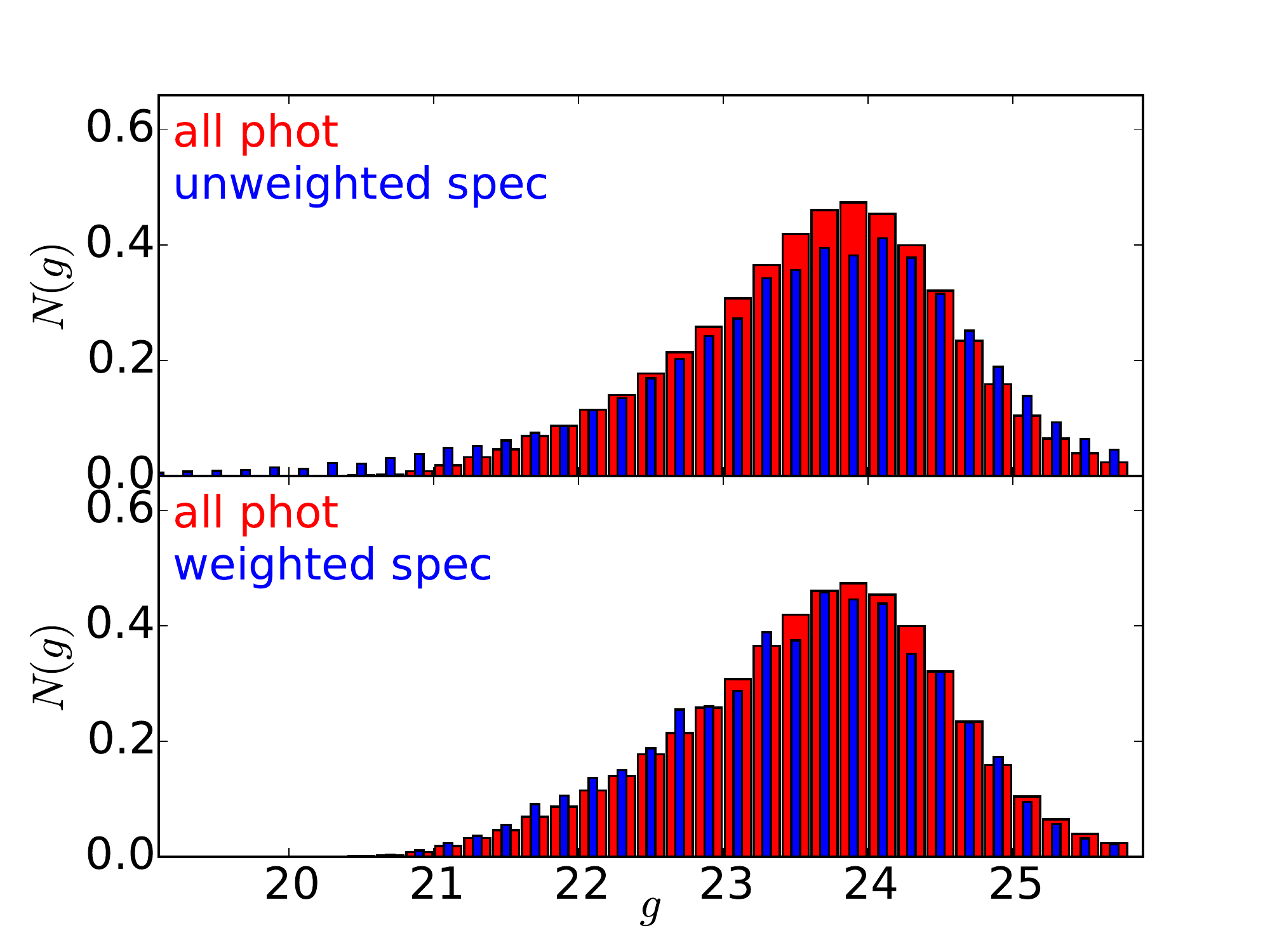}
\includegraphics[clip=true, trim=0.5cm 0cm 2cm 1cm, width=0.32\textwidth]{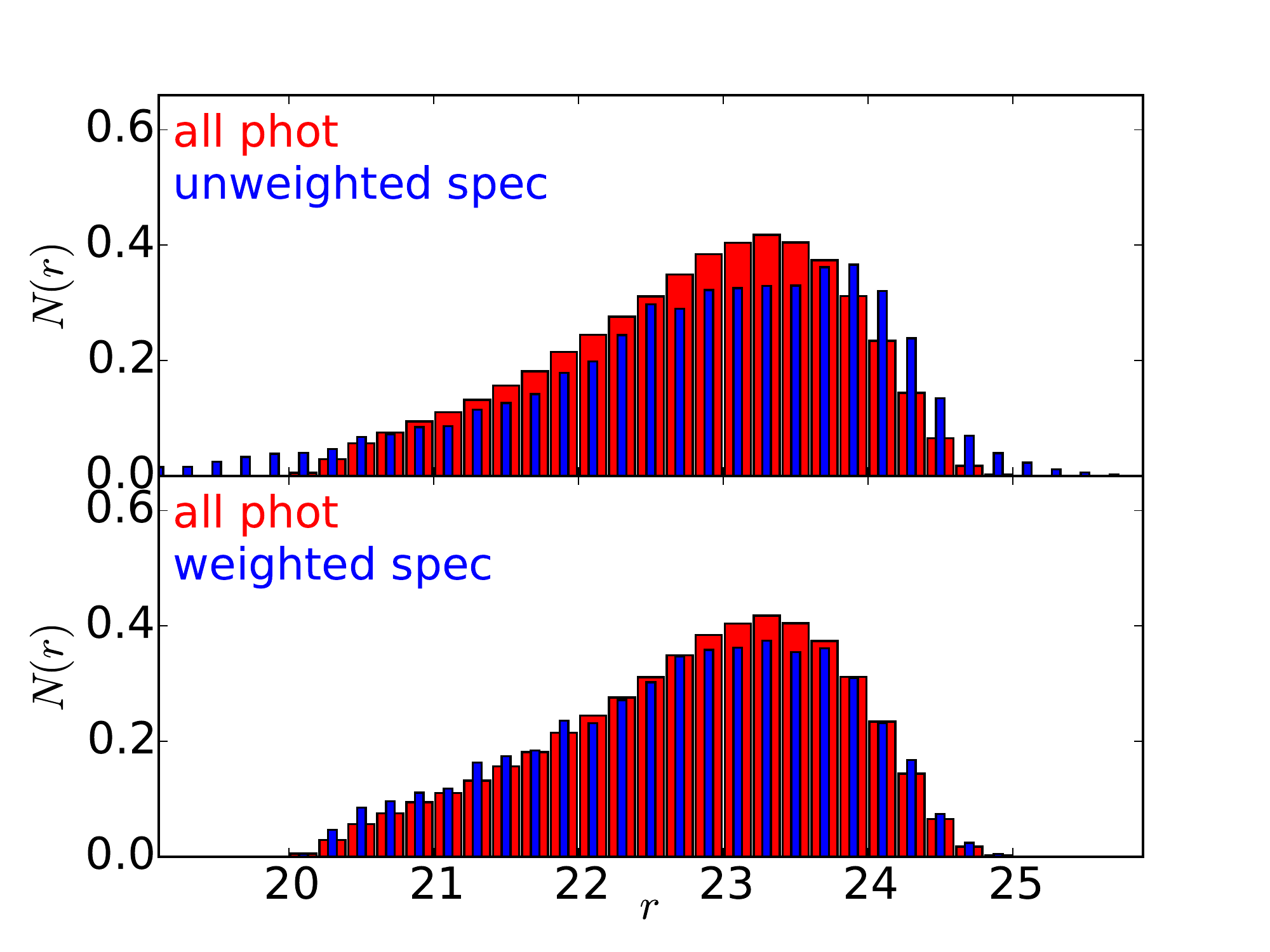}
\includegraphics[clip=true, trim=0.5cm 0cm 2cm 1cm, width=0.32\textwidth]{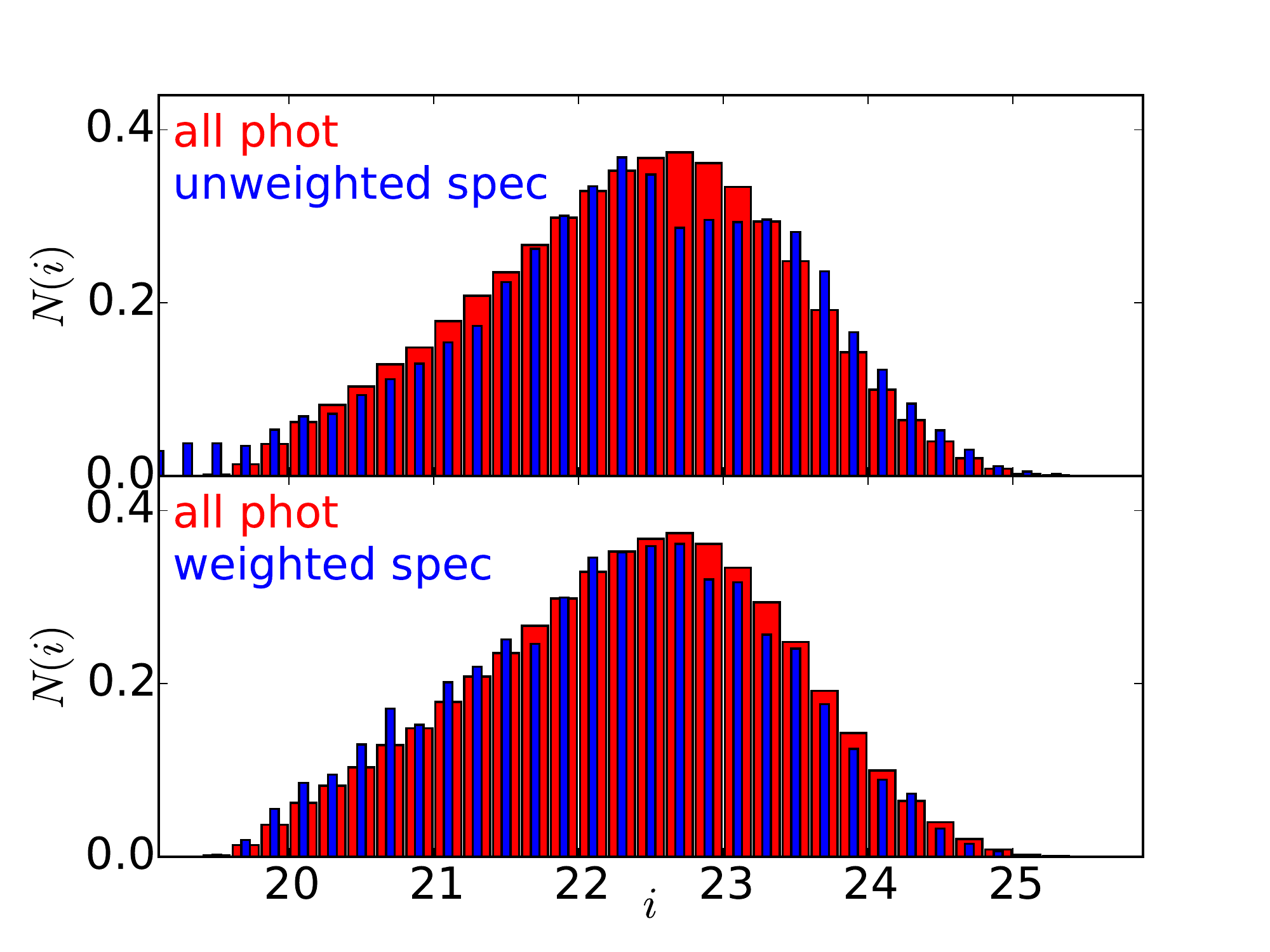}
\includegraphics[clip=true, trim=0.5cm 0cm 2cm 1cm, width=0.32\textwidth]{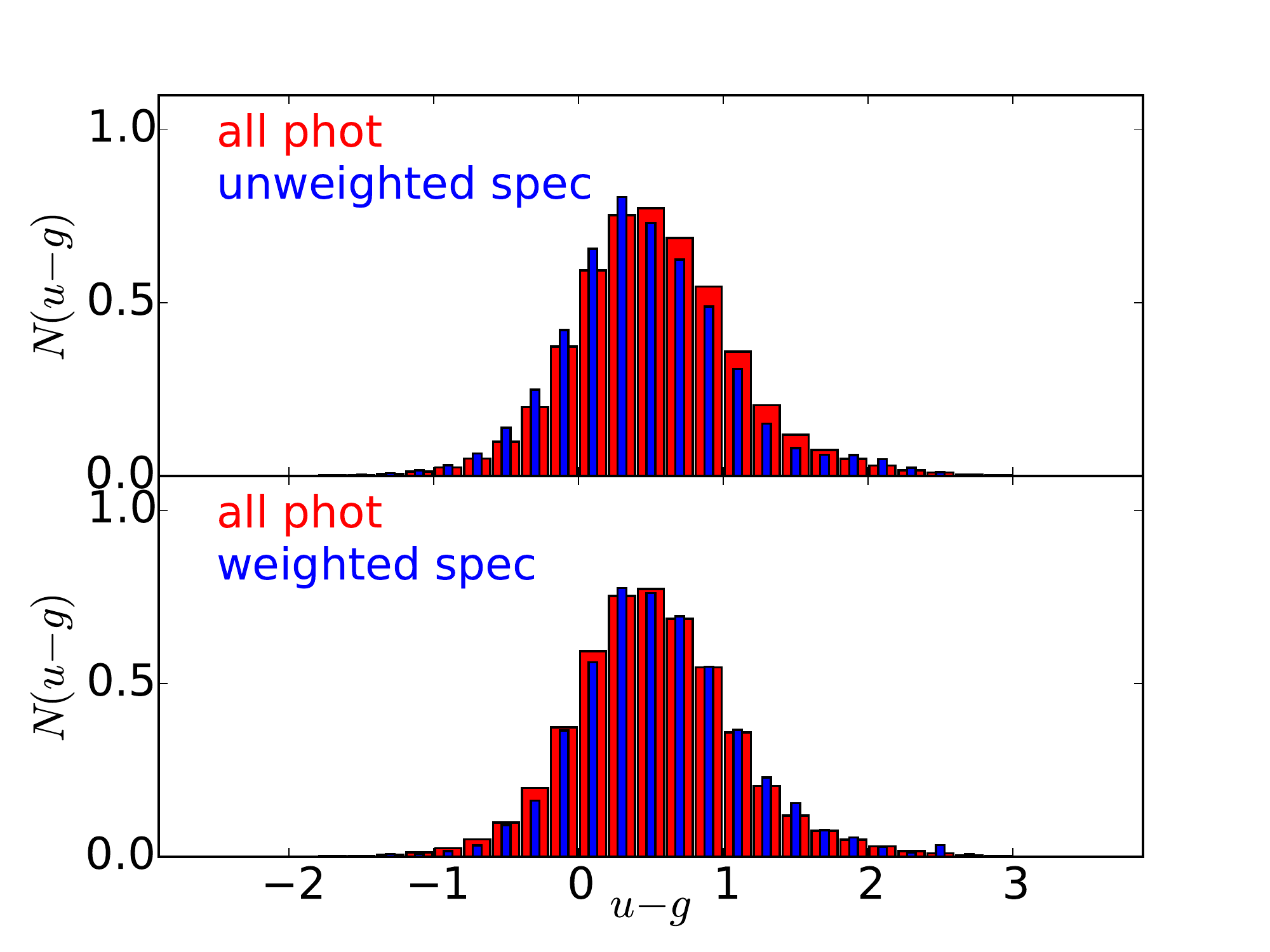}
\includegraphics[clip=true, trim=0.5cm 0cm 2cm 1cm, width=0.32\textwidth]{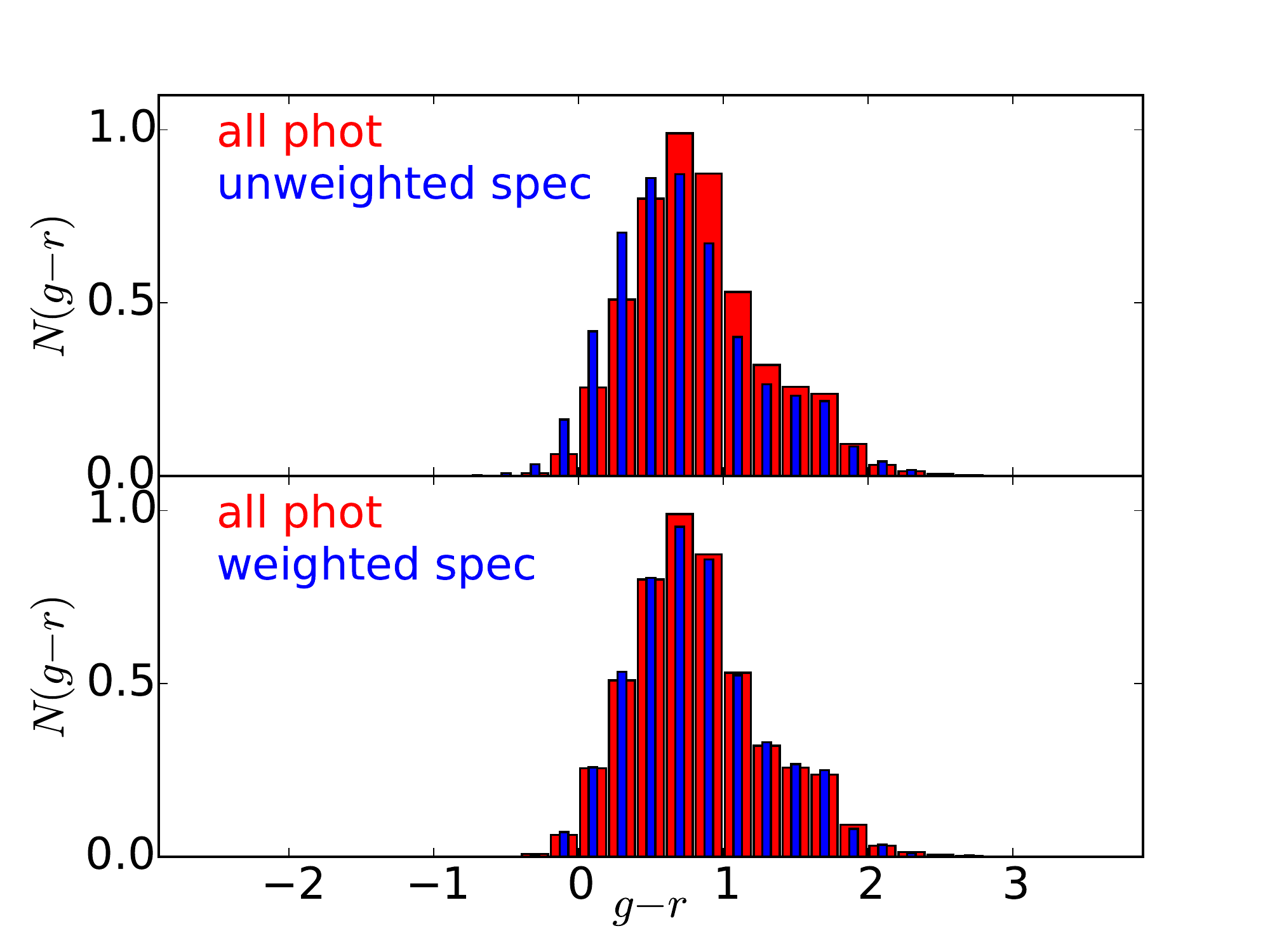}
\includegraphics[clip=true, trim=0.5cm 0cm 2cm 1cm, width=0.32\textwidth]{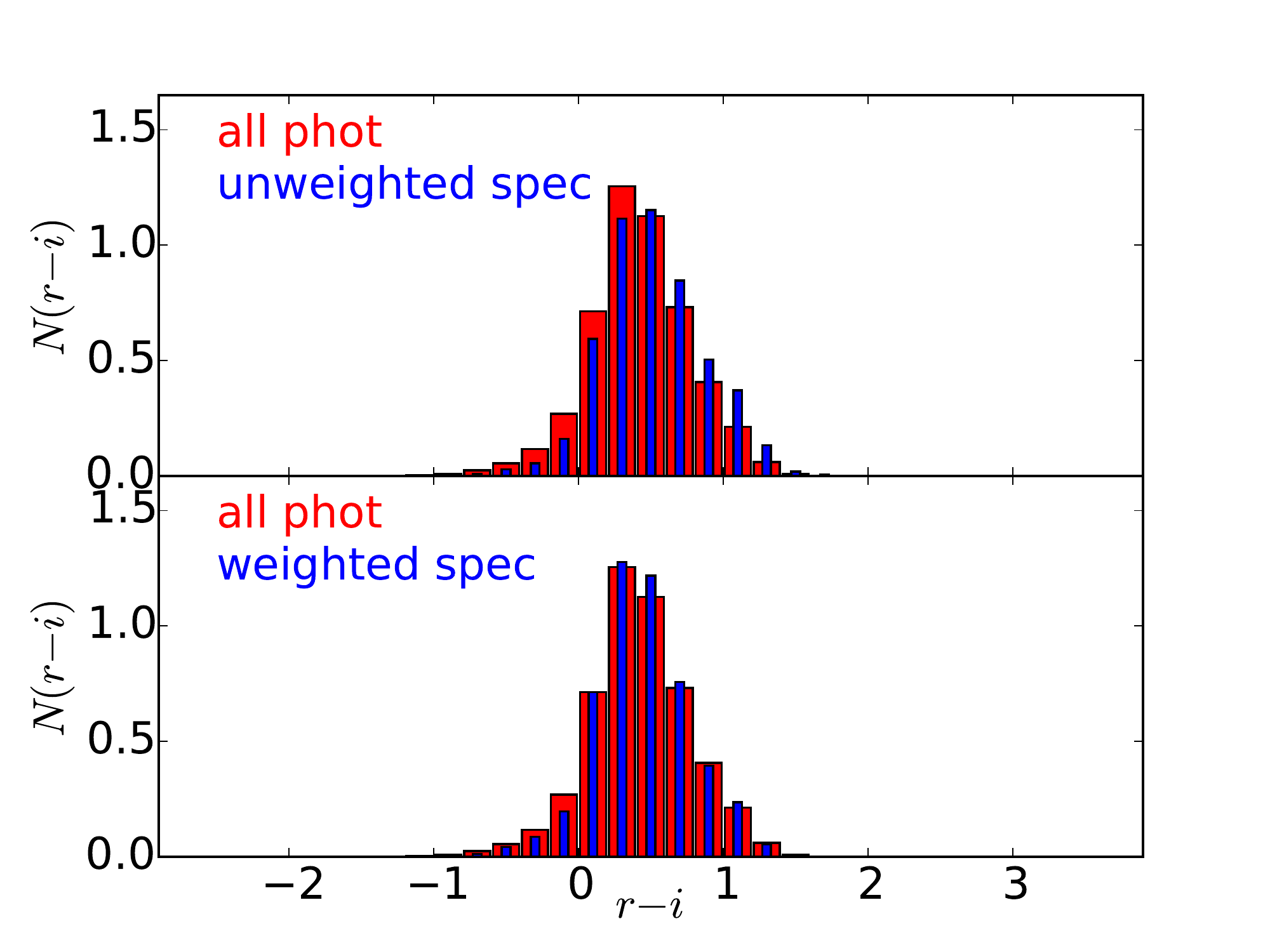}
\includegraphics[clip=true, trim=0.5cm 0cm 2cm 1cm, width=0.32\textwidth]{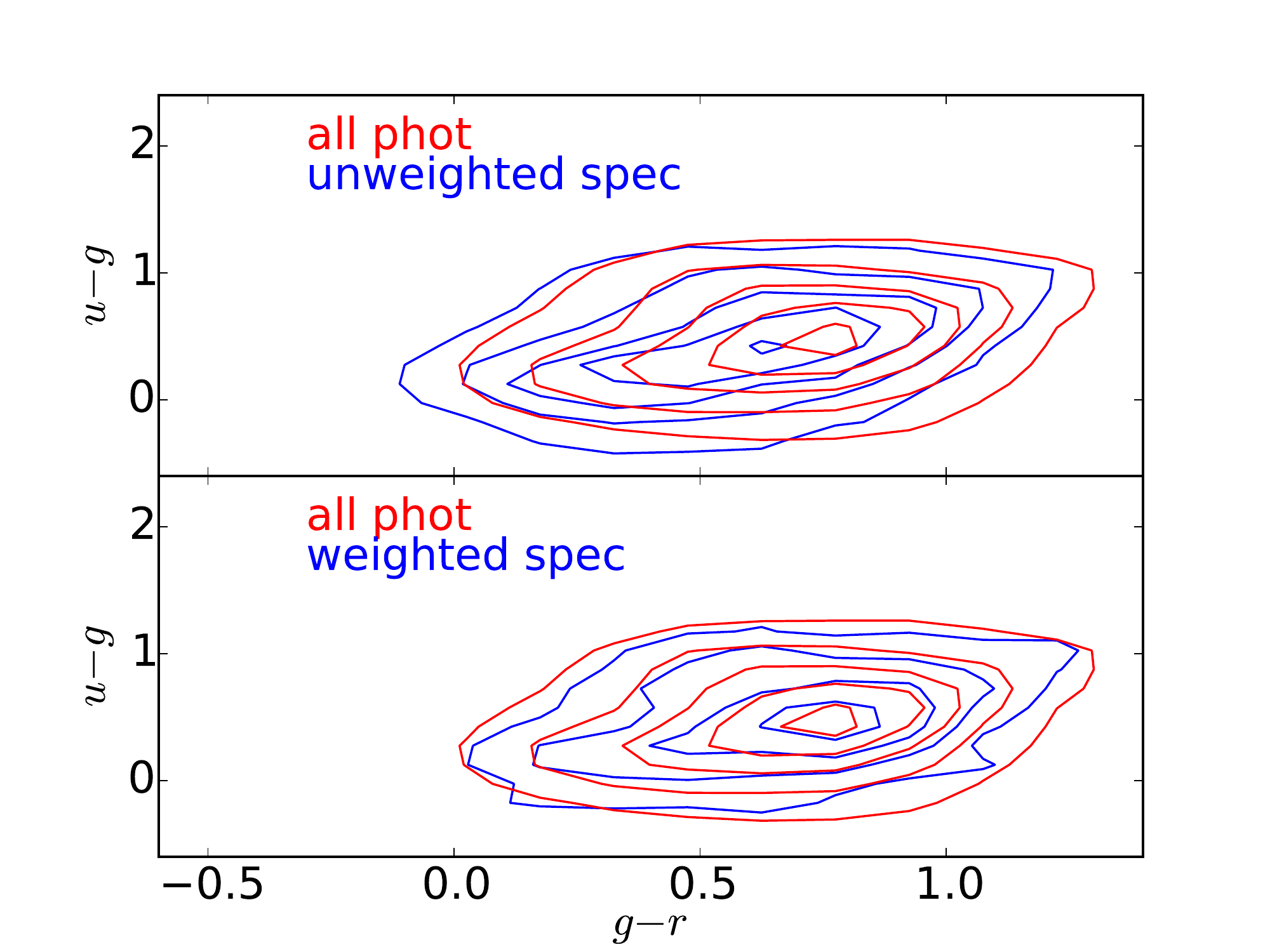}
\includegraphics[clip=true, trim=0.5cm 0cm 2cm 1cm, width=0.32\textwidth]{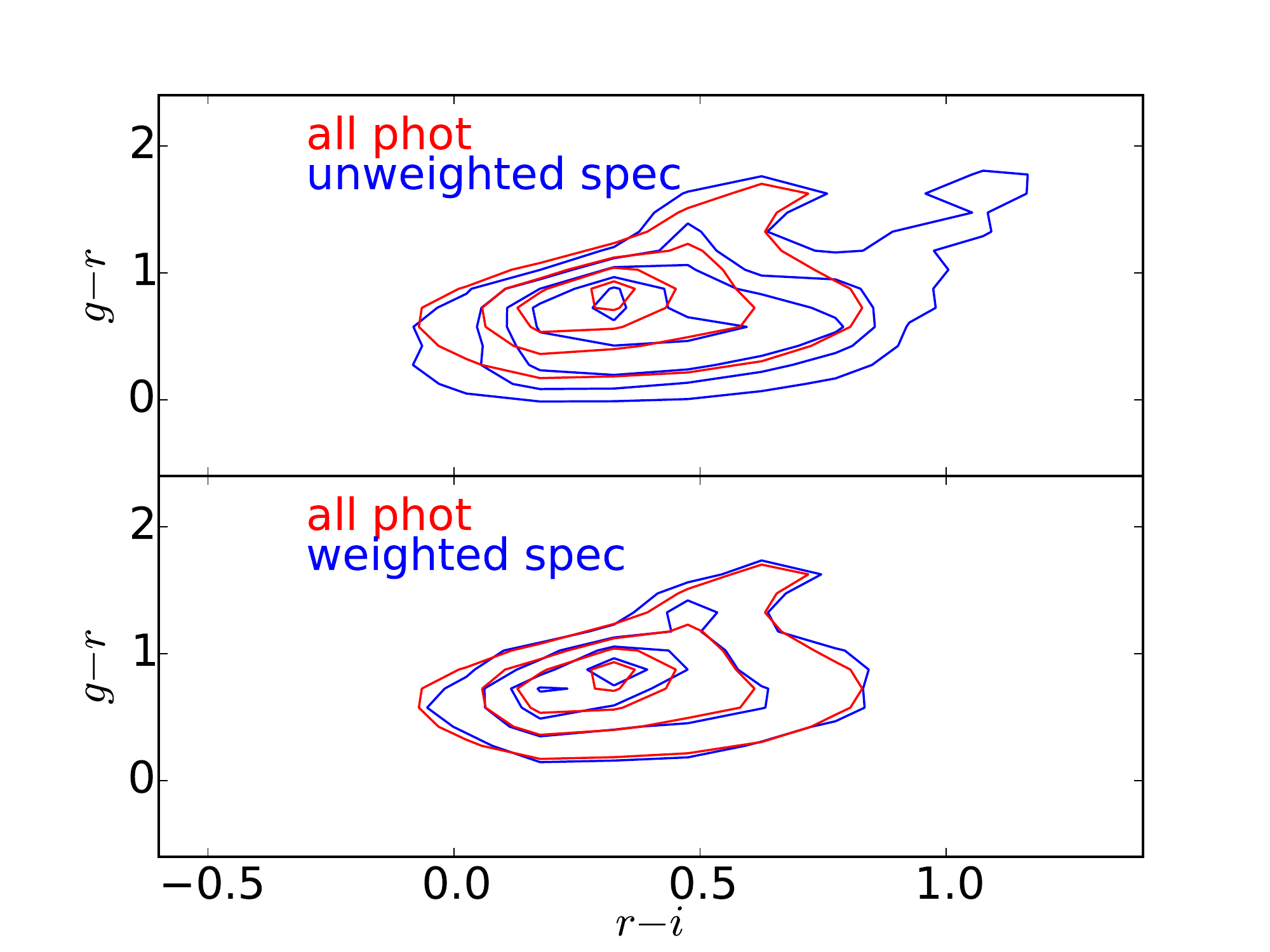}
\caption{\label{fig:col_space_projections}Magnitude and colour distributions of the photometric sample (red) and the spectroscopic sample (blue) before (upper panels) and after re-weighting (lower panels). The first four panels show the magnitude distributions in the $ugri$-bands, the next three panels show the one-dimensional colour distributions in the three independent colours $u-g$, $g-r$, and $r-i$, and the last two panels show two-dimensional colour distributions in $u-g$ vs. $g-r$ and $g-r$ vs. $r-i$. While the match between the photometric and spectroscopic distributions is fairly good even without re-weighting the spec-$z$ sample the match improves significantly after re-weighting.}
\end{figure*}

In KiDS-450 we work in four-dimensional $(u,g,r,i)$ magnitude space. Fig.~\ref{fig:col_space_projections} shows the distribution of photometric and spectroscopic objects in different projections of this colour space. The spec-$z$ sample is shown before and after re-weighting. Any significant mismatch between the re-weighted distribution of spec-$z$ objects and the photometric objects would indicate a violation of the first requirement that the spec-$z$ sample must span the extent of the phot-$z$ sample. No obvious deviations are found if the full spec-$z$ sample is used. Interestingly, if we run the re-weighting algorithm with the COSMOS spec-$z$ catalogue only, there is a very significant mismatch at faint magnitudes. This suggests that there are not enough faint high-$z$ galaxies in the z-COSMOS catalogue that could be up-weighted to match the distribution of the photometric catalogue. Including the DEEP2 and CDFS catalogues cures this problem and leads to the distributions shown in Fig.~\ref{fig:col_space_projections}. We re-run the same test for the four tomographic bins individually, finding a good match for all bins in all bands.

The requirement of a unique mapping from magnitude space to redshift cannot be tested easily. Given that we are working with four bands only, there is certainly some concern that this requirement is not completely fulfilled. If this was the case there would be regions in magnitude space that correspond to several very different redshift ranges. These phenomena are also called colour-redshift degeneracies \citep[see e.g.][]{benitez:2000}. This is one of the reasons why we limit the cosmic shear analysis to photometric redshifts of $0.1<z_{\rm B}\le 0.9$. As indicated in fig.~12 of \cite{kuijken/etal:2015} the outlier rate of our photo-$z$ is very low in this photometric redshift range. While this could be also caused by spectroscopic incompleteness, this result is confirmed by analysing simulated mock KiDS photometry catalogues. Given these results we are confident that the combination of a highly complete spec-$z$ sample (as indicated by Fig.~\ref{fig:col_space_projections}) and a conservative photo-$z$ range means that we meet the requirements for the weighted direct calibration.

Another possible source of systematic error in the DIR calibration is sample variance due to the finite size of the spectroscopic training sample. The $n(z)$ are clearly affected by this sample variance as can be seen from their non-smooth shape in Fig.~\ref{fig:z_dist_comp}. However, the relevant question is whether this sample variance in the photo-$z$ calibration contributes significantly to the total error budget of the cosmological parameters of interest. \citet{cunha/etal:2012} estimate the effect of sample variance in the redshift calibration for DES from simulations. Their results are not directly applicable to KiDS-450 as their simulated survey covers an area of 5000\,deg$^2$, goes deeper ($n_{\rm eff}=12$arcmin$^{-2}$), reaches out to $z=1.35$, and employs 20 tomographic bins. They also concentrate on the equation-of-state parameter $w$ instead of $S_8$. They find that the uncertainty in $w$ due to photo-$z$ calibration is larger than their statistical error ($\sigma_w=0.035$) by a factor of $\sim4$ if they use the same magnitude weighting technique as our DIR method (called $p(z)_w$ in their paper) and train this technique with a spec-$z$ survey covering a single square degree and $\sim10^4$ training galaxies (see their table 2). The $S_8$ parameter is somewhat more sensitive (within a factor of $\sim2$) to redshift errors than $w$ \citep[see][]{huterer/etal:2006}. However, given the more modest statistical power of KiDS-450 (compare their $\sigma_w=0.035$ to our $\sigma_w\sim1$) and our larger spec-$z$ calibration sample that originates from four widely separated fields we estimate that any leakage from the spec-$z$ sample variance into our photo-$z$ calibration is subdominant to our statistical uncertainties. 

Similar conclusions can be reached by looking at the results of \citet{vanwaerbeke/etal:2006}. They look at the more pessimistic case of direct photo-$z$ calibration with spec-$z$ but without magnitude weighting. For a cosmic shear survey of 200\,deg$^2$ area, $n_{\rm eff}=20$, $z_{\rm max}\approx2$ and a spec-$z$ calibration sample from 4\,deg$^2$ they find that for angular scales $\theta>10'$ the errors on the shear measurement (from shape noise and survey sample variance) dominate over the errors from the redshift calibration.

In order to further reduce sample variance in the redshift calibration we plan to observe additional calibration fields that are covered by deep, public spectroscopic surveys. This will be necessary to keep pace with the growing KiDS survey and the shrinking statistical uncertainties.

\subsubsection{Calibration from cross-correlations (CC)}
\label{sec:sysCC}
Calibrating the redshift distributions in the different tomographic bins with the help of angular cross-correlations has the great benefit that it does not require a representative sample of objects with spectroscopic redshifts. However, there are several systematic errors that can affect such a clustering redshift recovery. The cross-correlations are relatively robust against an angular selection function in one of the two samples (i.e. the photometric and spectroscopic sample) as long as the angular selection functions of both samples are not correlated themselves, e.g. because both mask out true structures like stars. The auto-correlation functions, which are needed to calibrate the typically unknown galaxy bias, are however heavily affected. One can therefore only use samples where this angular selection function (or weighted footprint) is precisely known. Depending on colour/photo-$z$ selections and noise properties this can become highly non-trivial (see e.g., \citealt{morrison/etal:2015} for an analysis of angular selection effects in CFHTLenS). In our particular case this means that we can only use DEEP2 and -- with some caveats -- zCOSMOS in this analysis. Here all correlation functions are estimated with the publicly available STOMP library \citep{scranton/etal:2002}\footnote{\url{https://github.com/ryanscranton/astro-stomp}}.

The other major systematic uncertainty in this method is related to the correction for the galaxy bias. For the spec-$z$ sample the galaxy bias can be estimated robustly as a function of redshift if the angular selection function is known. However, the fine bins that are typically used for the cross-correlations contain too few galaxies to yield a robust estimate of the auto-correlation function and hence the galaxy bias. We are therefore forced to use wider redshift bins for the auto-correlation, and interpolate the results. For the spectroscopic sample we estimate the projected auto-correlation function, $w_{\rm p}(r_{\rm p})$ as defined in Eq.~9 of \cite{matthews/etal:2010}. Following their method we fit a power-law to this projected auto-correlation:
\begin{equation}
w_{\rm p}(r_{\rm p}) \propto \left(\frac{r_{\rm p}}{r_0}\right)^{-\gamma}\,,
\end{equation}
and interpolate the fitted parameters $r_0$  and $\gamma$ to the redshift of each cross-correlation bin. For these projected auto-correlation function measurements we choose redshift bins of constant comoving width so that we do not have to measure the clustering scale length $r_0$ in absolute terms but just relatively between the different bins.

For the photometric sample we estimate the angular auto-correlation function with the estimator of \cite{landy/szalay:1993} and also fit a power law:
\begin{equation}
w(\theta)=A \, \theta^{1-\gamma}-C\,,
\end{equation}
with $A$ being the amplitude and $C$ being the integral constraint. The relative clustering strengths are then adjusted with the method laid out in \citet[][see their Eq.~13]{matthews/etal:2010}.

\begin{figure}
\includegraphics[width=\hsize]{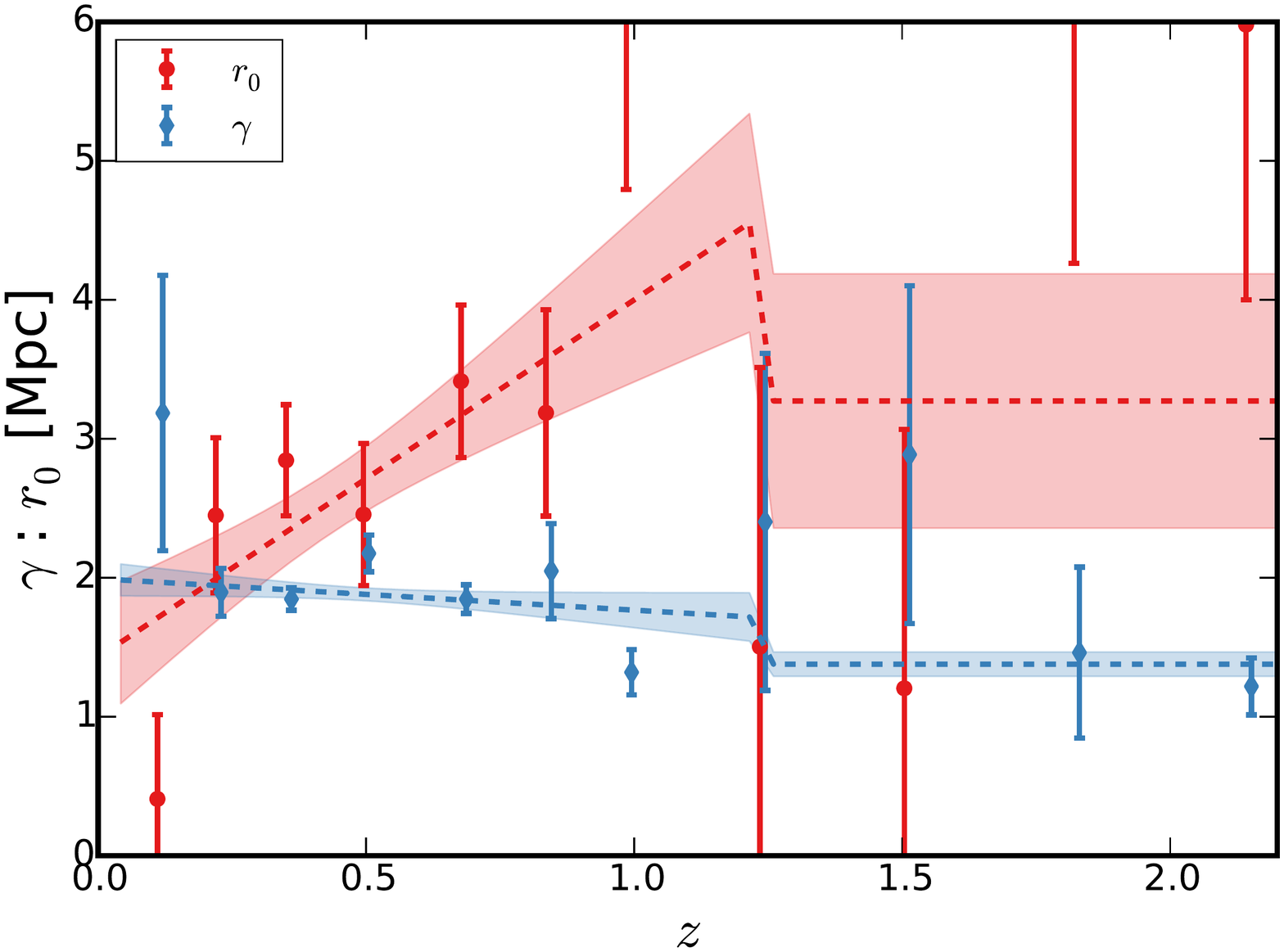}
\caption{\label{fig:power_law_z}Relative clustering scale length $r_0$ (red) and slope $\gamma$ (blue) of power-law fits to the auto-correlation functions as a function of redshift in the COSMOS spec-$z$ catalogue. The dashed lines show parametric fits to the data, consisting of a linear function for $0<z<1.2$ and a constant at $z>1.2$.}
\end{figure}

Due to observational effects, a spec-$z$ sample might show sudden breaks in its properties at certain redshifts. For example, the zCOSMOS sample changes abruptly at a redshift of $z\sim1.2$ where the main features that are used for redshift measurement leave the optical window and only certain types of galaxies can still be detected. This makes the interpolation harder and requires a judicious choice of model to describe the redshift dependence of the galaxy bias, which we capture through the free variables $r_0$ and $\gamma$. After some experimentation we adopt a two-part model for the redshift dependence of these parameters (Fig.~\ref{fig:power_law_z}) in the COSMOS field: a linear dependence from $0<z<1.2$ and a constant at $z>1.2$. For DEEP2 we use a linear relation for both parameters in the range $0.5<z<1.5$.

The estimation of the auto-correlation function for the photometric sample is even more problematic. If the photometric selection yields a single-peaked true redshift distribution the galaxy bias can be corrected for following the same methodology as the spec-$z$ analysis \citep{schmidt/etal:2013}. However, photo-$z$ selections tend to yield highly non-Gaussian true redshift distributions with multiple peaks. Measuring the angular auto-correlation function of such a sample yields a projected mix of multiple auto-correlation functions of galaxies at different redshifts. Unfortunately, the galaxies in the multiple peaks of the true redshift distribution are typically of different types, with different bias. This mix of galaxy populations in photo-$z$ tomographic bins is an inherent problem of the clustering redshift recovery method.

\begin{figure}
\includegraphics[clip=true,trim=3cm 0cm 5cm 1.5cm,width=0.49\textwidth]{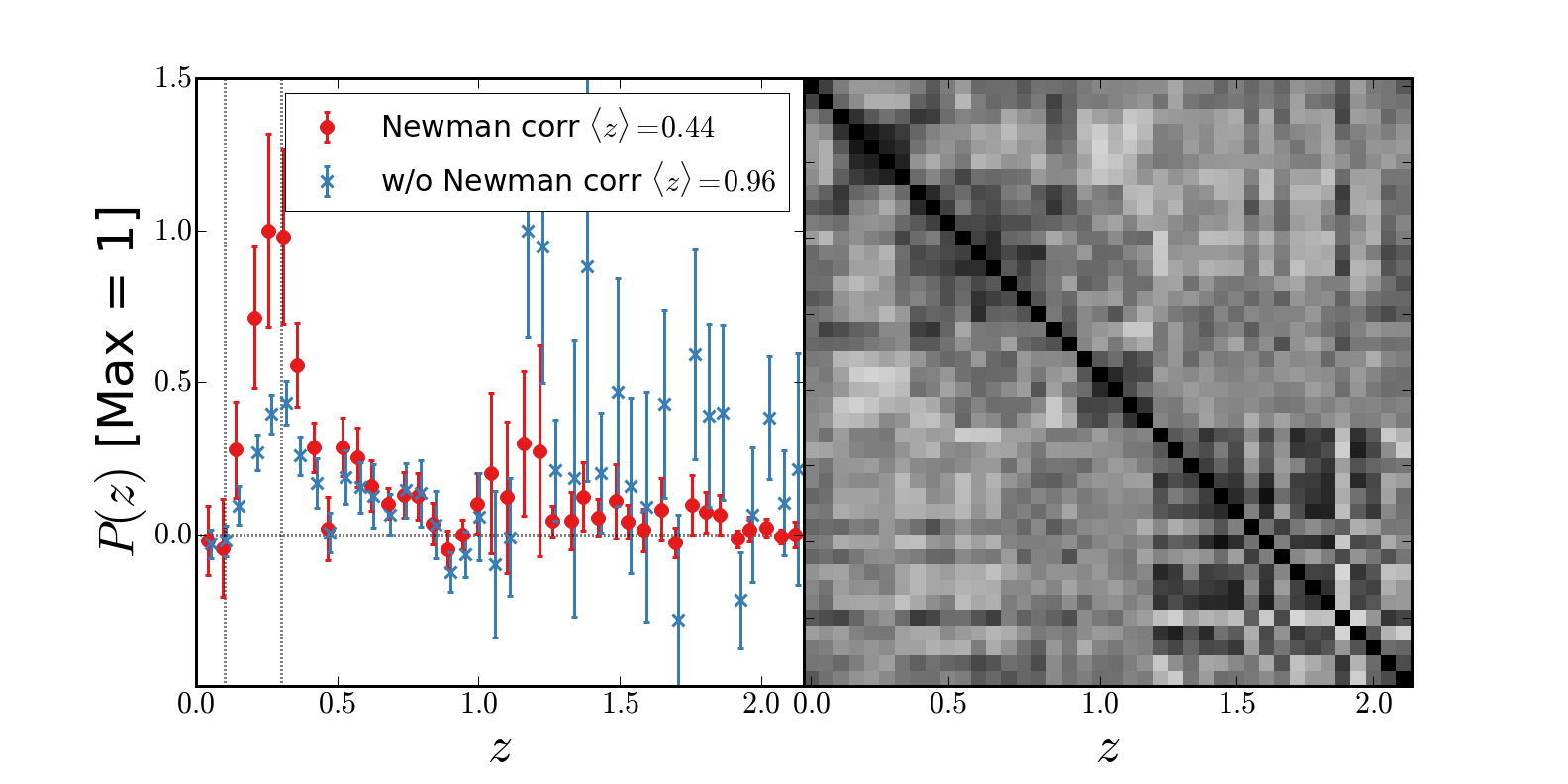}
\includegraphics[clip=true,trim=3cm 0cm 5cm 1.5cm,width=0.49\textwidth]{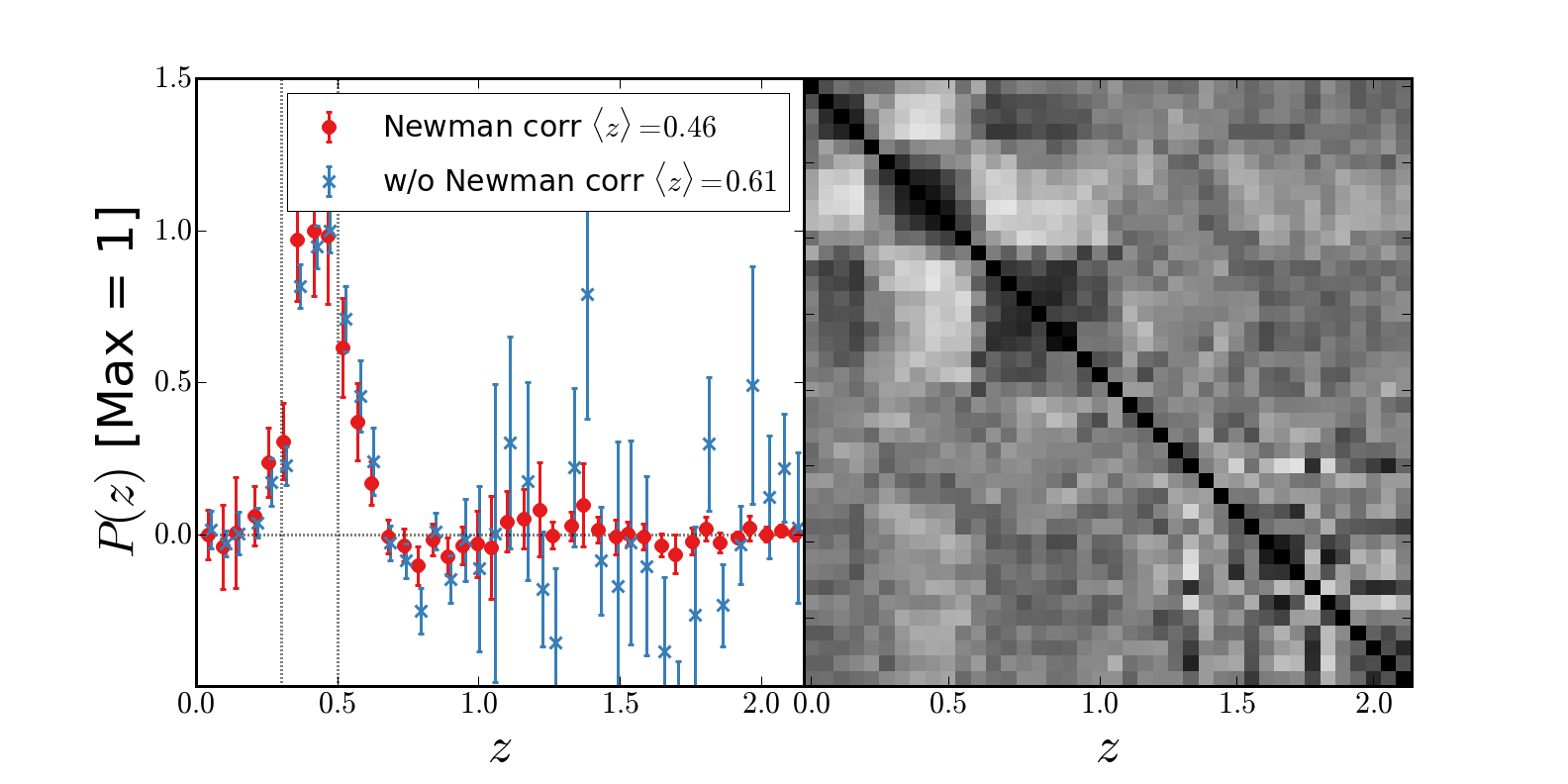}
\includegraphics[clip=true,trim=3cm 0cm 5cm 1.5cm,width=0.49\textwidth]{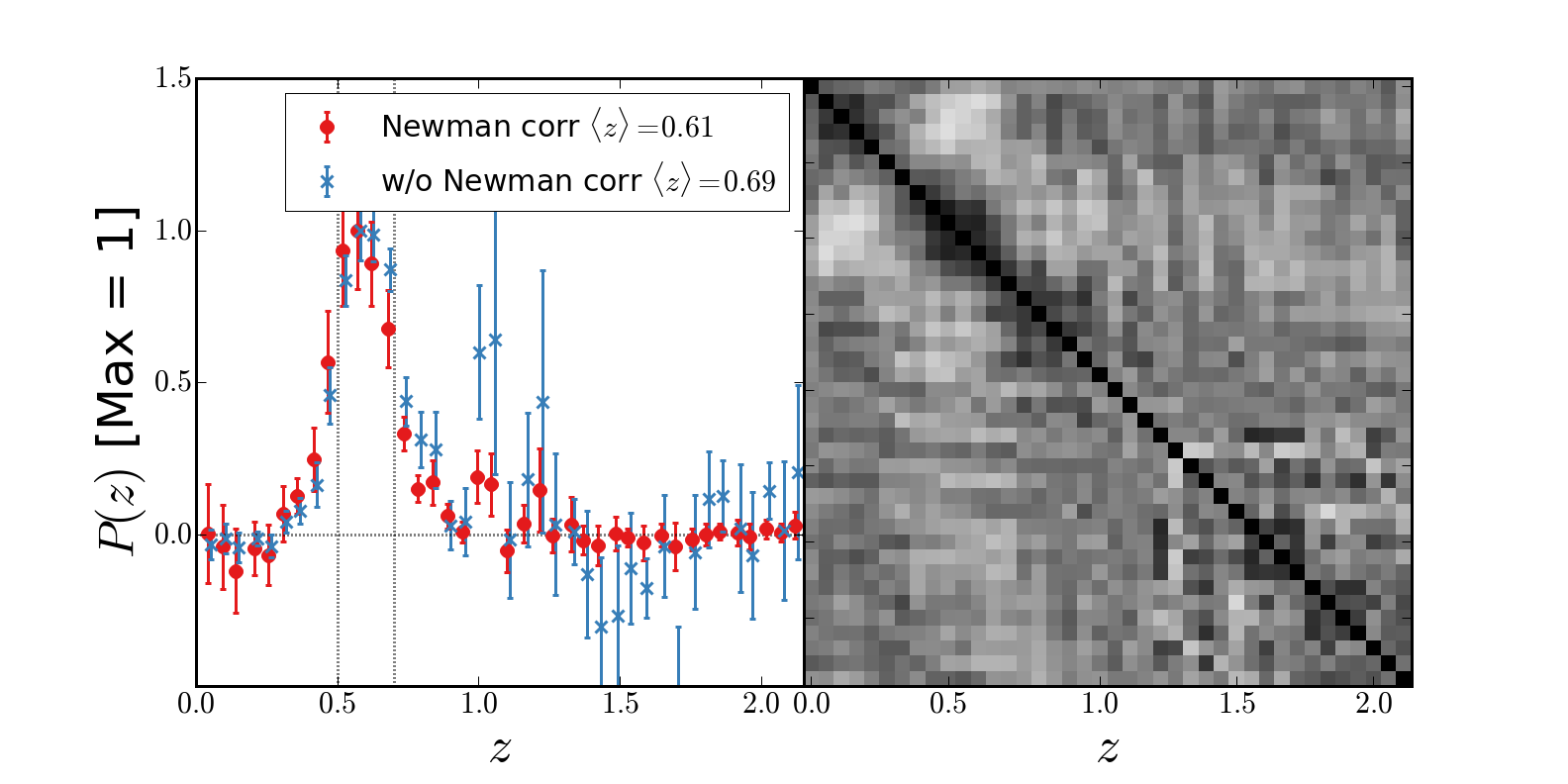}
\includegraphics[clip=true,trim=3cm 0cm 5cm 1.5cm,width=0.49\textwidth]{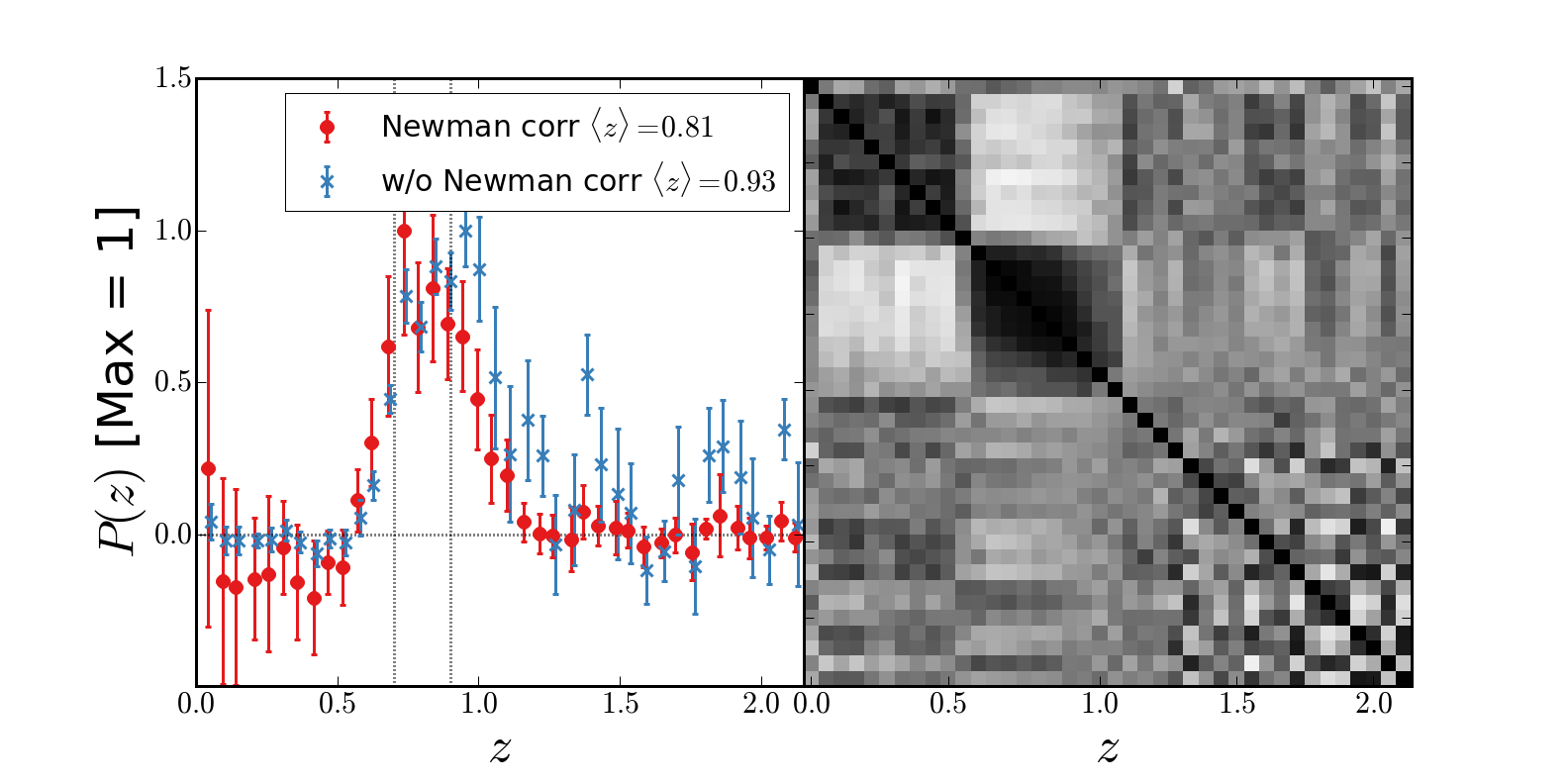}
\caption{\label{fig:z_dist_cross}
Redshift distribution recovery as determined by the cross-correlation technique in the COSMOS field.
Left-hand panels: Recovered redshift distributions of the four tomographic bins before (blue crosses) and after (red circles) correction for the galaxy bias. The $z_\rmn{B}$ limits of each tomographic bin are indicated by vertical lines. Right-hand panels: correlation matrices of the red data points with white corresponding to a value of corr$_{ij}=-1$ and black to corr$_{ij}=1$. Significant correlations are introduced by the bias correction. Note that these are not the final redshift distributions from this method: to generate those we combine similar estimates from DEEP2 with COSMOS in an optimally weighted way.}
\end{figure}

The importance of correcting for the galaxy bias is shown in Fig.~\ref{fig:z_dist_cross} where the raw redshift recovery (i.e. just the amplitude of the cross-correlation function) is compared to the final bias-corrected redshift distribution for COSMOS. The correction for the galaxy bias, proposed by \cite{newman:2008}, essentially tilts the redshift distribution around some pivot redshift. We find that the amplitude of the high-$z$ outlier populations -- and hence also the amplitude of the resulting shear correlation function -- is sensitive to this correction.

We do not further quantify the systematic error in the CC distributions from the effects discussed above since the weighted direct calibration (DIR) has a significantly higher precision (see Fig.~\ref{fig:SN_DIR_CC}). In the future when the overlapping area with deep spectroscopic surveys has increased,  cross-correlation techniques will become competitive with other approaches of photo-$z$ calibration and it will become mandatory to model, estimate, and correct for these uncertainties. The unknown bias of an outlier population might be the ultimate limit to the accuracy of this method. The required measurement of precise selection functions for the spectroscopic and photometric samples will also pose a formidable challenge.

\subsubsection{Re-calibration of the photometric \textit{P}(\textit{z}) (BOR)}
\label{sec:sysBOR}
\begin{figure}
\includegraphics[width=\hsize]{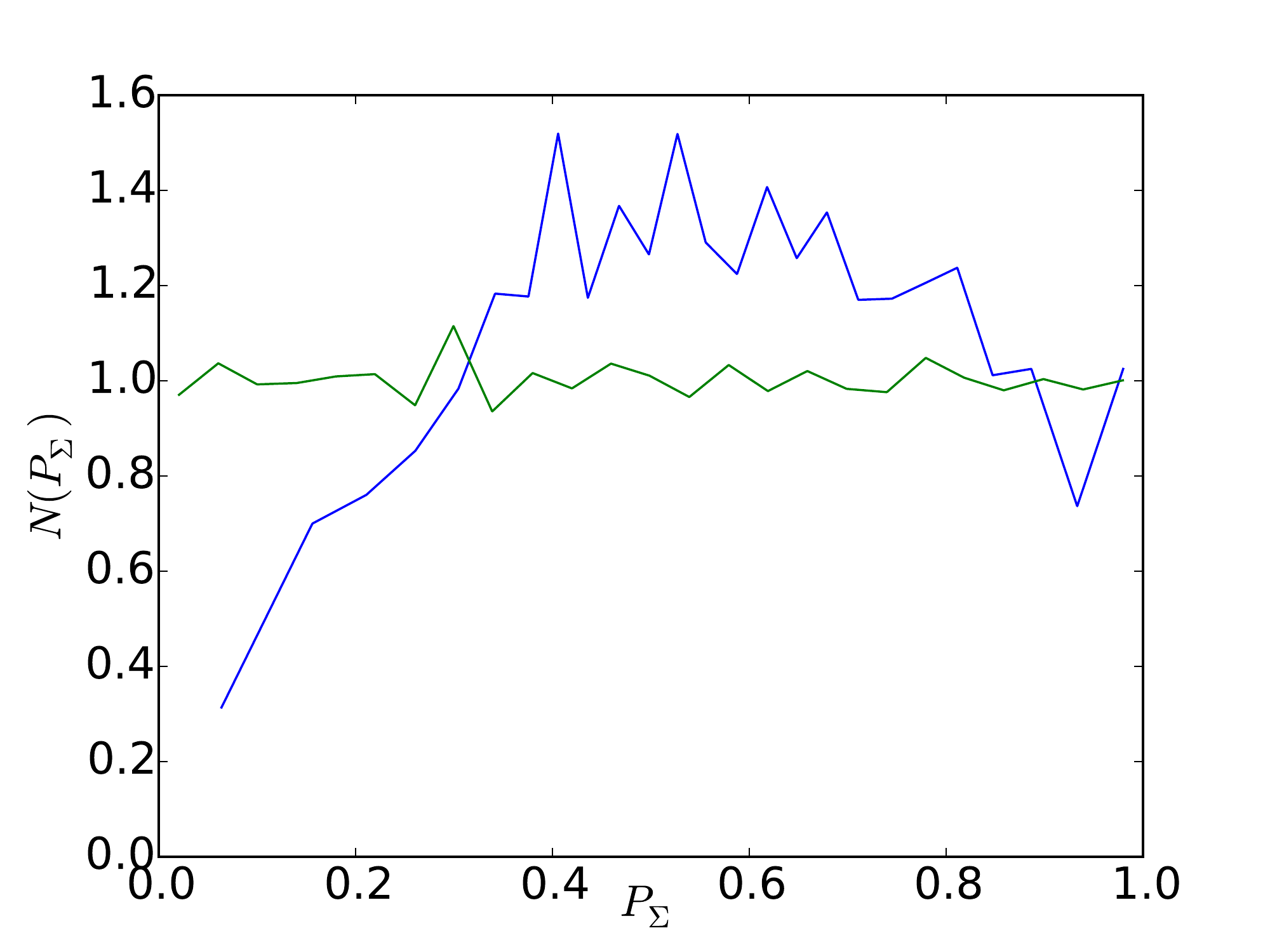}
\caption{\label{fig:N_of_P}Distribution of $P_{\Sigma}$ before (blue) and after (green) re-calibration with the method by Bordoloi et al. (2010).}
\end{figure}

The main criterion for quality control in the photo-$z$ re-calibration method by \cite{bordoloi/etal:2010} is the shape of the $N(P_{\Sigma})$ (Eq.~\ref{eqn:PBOR}) which should be flat after a successful re-calibration. In Fig.~\ref{fig:N_of_P} this distribution is shown before and after application of the method. As can be seen the distribution of $P_{\Sigma}$ is indeed flat after re-calibration of the $P(z)$. A more in-depth study of systematic effects of this technique can be found in \cite{bordoloi/etal:2012}. We do not explore this method further in the present analysis.

\subsection{Galaxy-galaxy-lensing shear-ratio test}
Assuming a fixed cosmology, there is a clear prediction for the lensing signal around galaxies (galaxy-galaxy-lensing; GGL), where the amplitude depends on mass and cosmology, and the signal should scale for sources in the different tomographic bins, given their $n(z)$. Here we test this redshift scaling with lenses with spectroscopic redshifts from BOSS \citep{alam/etal:2015} and from GAMA \citep{driver/etal:2011}. This `shear-ratio' test \citep{jain/etal:2003,kitching/etal:2015,schneider:2016} is similar to the tests described in \cite{heymans/etal:2012} (see section~6 and fig.~12 of that paper) and \cite{kuijken/etal:2015} (see section~5.2 and fig.~18 of that paper). As the GGL signal is rather insensitive to the choice of cosmology, this analysis can be used to verify the redshift distributions and the redshift scaling of the shear calibration correction $m$ (Eq.~\ref{eqn:mandc}).

\begin{figure}
\includegraphics[width=\hsize]{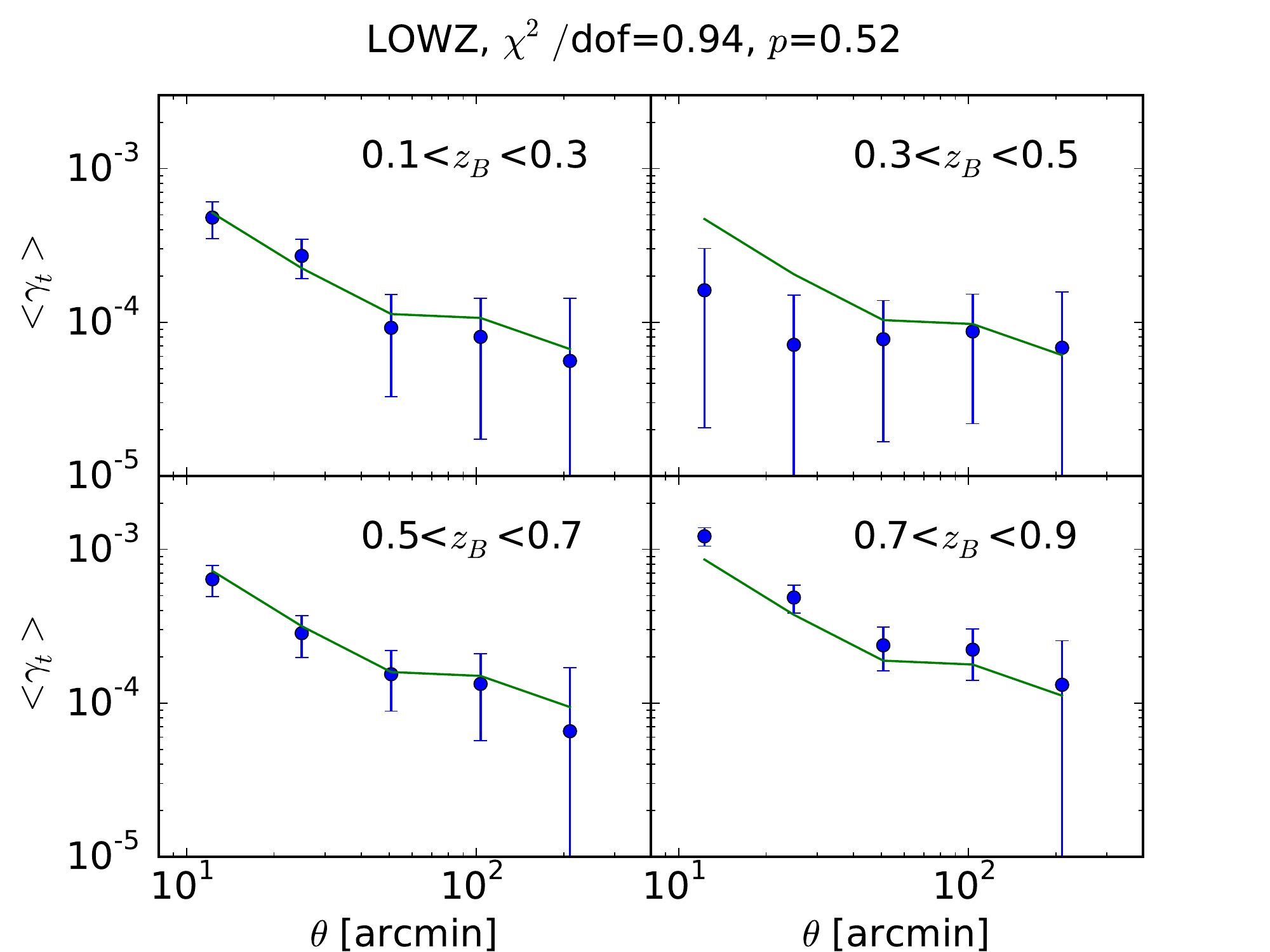}\\
\includegraphics[width=\hsize]{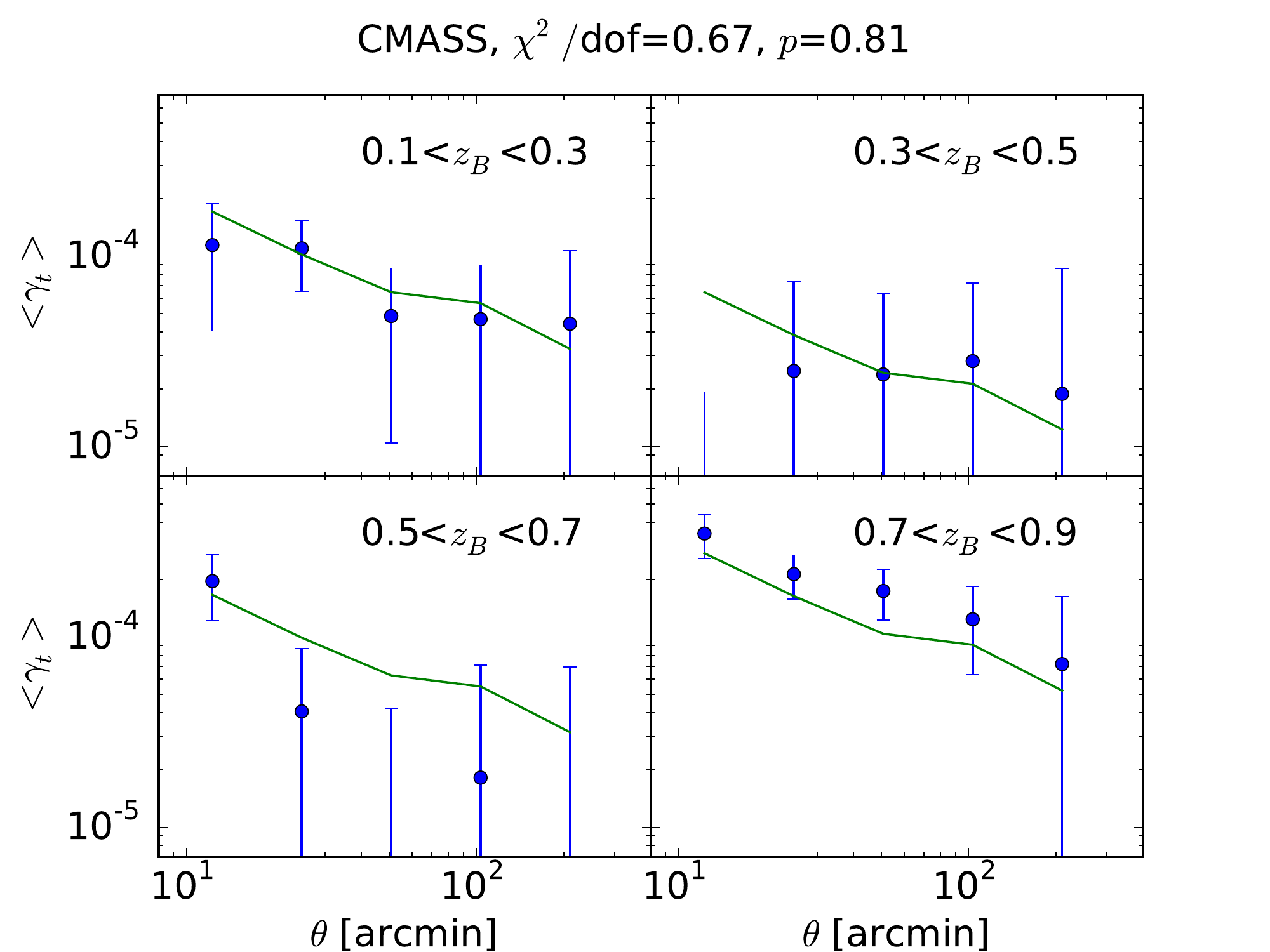}\\
\includegraphics[width=\hsize]{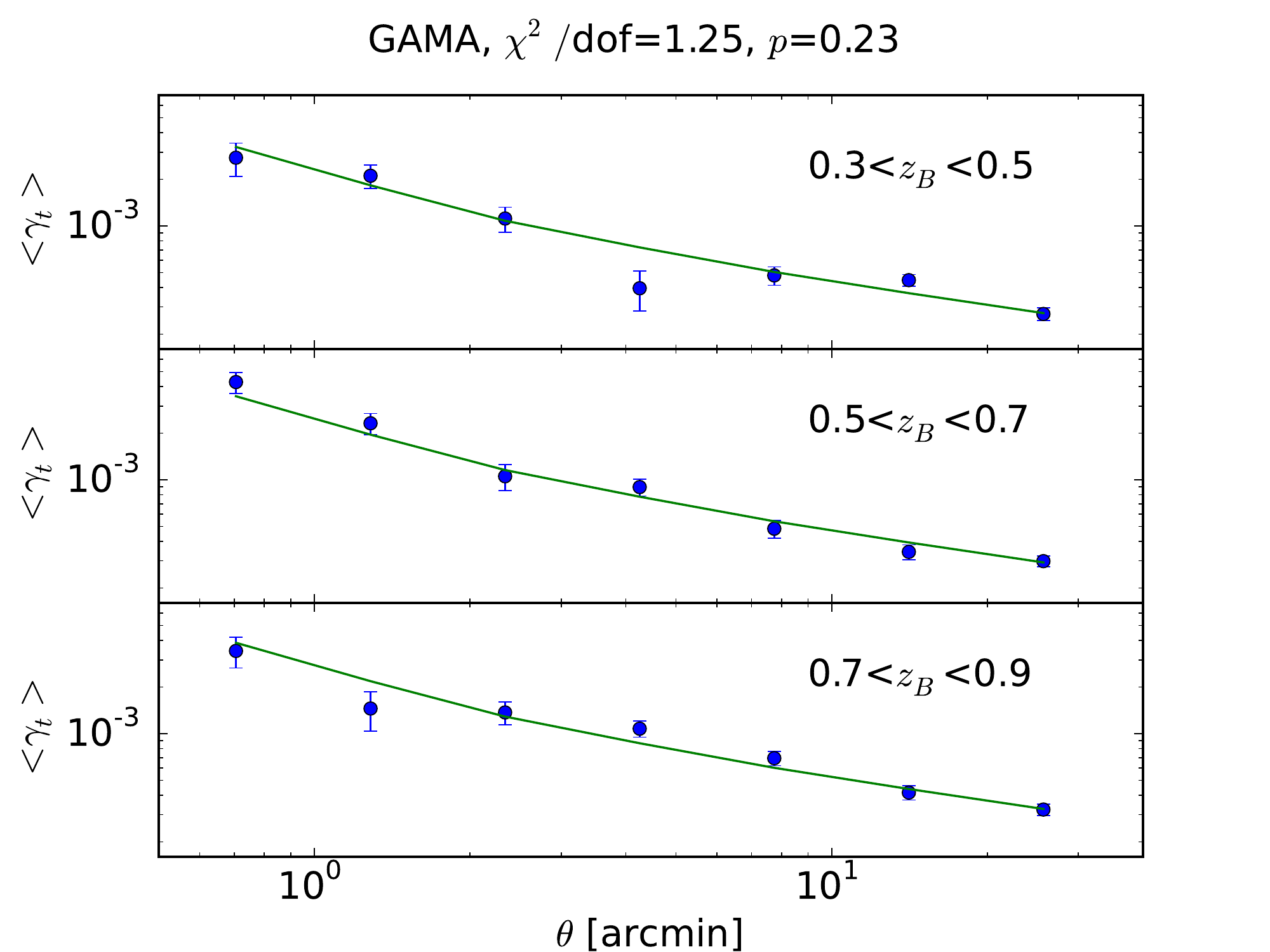}
\caption{\label{fig:GGL}Galaxy-galaxy-lensing signals using three lens samples: BOSS LOWZ (upper) and BOSS CMASS (middle) and GAMA galaxies with $z_{\rm l}<0.2$ (lower).  The GGL signal is measured using sources in the different tomographic bins (as labelled).   Shown is the measured tangential shear and a best fit lens model that leaves the amplitude at each angular scale free but constrained by the lensing efficiency $\beta$ (see Eq.~\ref{eq:beta}) for the different tomographic bins given their DIR redshift distributions. For the BOSS samples which overlap with most tomographic bins in redshift, we only show the large scales where lens-source clustering is unimportant.  For the GAMA sample with $z_{\rm l}<0.2$ we show the small-scale signal for the second, third, and fourth tomographic bin which do not overlap significantly with the lenses in redshift.  The goodness of fit is listed in each panel label.}
\end{figure}

For three lens samples; BOSS LOWZ with $0.15<z_\rmn{l}<0.43$, BOSS CMASS with $0.43<z_\rmn{l}<0.7$ and GAMA selected with $z_\rmn{l}<0.2$, we measure the azimuthally averaged tangential shear $\gamma_{\rm t}$ around the lenses, in bins of angular separation.  We make four measurements for each lens sample, using the KiDS-450 source galaxies from each of the four tomographic bins used in our cosmic shear analysis (see Table~\ref{tab:tomo_bins}).  The measured GGL signal is presented in Fig.~\ref{fig:GGL}, where, as expected, we see $\gamma_{\rm t}$ increasing with the redshift of the source galaxies. Note that the SNR when using sources from the first tomographic bin with the CMASS lenses is higher than for sources in the second and third bin because of the high-$z$ tail of the DIR redshift distribution for the first bin.

We fit a maximally-flexible lens model to all data points of one lens sample simultaneously, in which we leave the amplitudes at each angular scale free (five angular scales in the BOSS analyses and seven in the GAMA analysis). We compare the model to the data by multiplying by the appropriate predictions of the lensing efficiency $\beta$ in each tomographic bin: 
\begin{equation}
\label{eq:beta}
\beta=\int_0^\infty {\rm d}z_\rmn{l} \, n_\rmn{l}(z_\rmn{l}) \int_{z_\rmn{l}}^\infty {\rm d}z_\rmn{s} \, n_\rmn{s}(z_\rmn{s}) \frac{D(z_\rmn{l},z_\rmn{s})}{D(0,z_\rmn{s})}\,, 
\end{equation}
where $D(z,z')$ is the angular diameter distance between redshifts $z$ and $z'$, $n_\rmn{l}(z)$ is the redshift distribution of the lenses, and $n_\rmn{s}(z)$ the redshift distribution of one of the source samples. This test is independent of any properties of the lens sample and hence represents a clean shear-ratio test.    For BOSS we use a covariance matrix  for the correlated $\gamma_{\rm t}$ measurements estimated from the simulated mock catalogues described in Section~\ref{sec:mockcov}.  For GAMA, where we focus on small-scale correlations, we use an analytical covariance matrix, as described in \citet{viola/etal:2015}.   The resulting best-fit models give $p$-values of ca.~50 and 80 per cent for the LOWZ and CMASS samples, respectively and ca.~20 per cent for GAMA, indicating that the scaling of the observed GGL signal is fully consistent with the expectations given the redshift distributions estimated with the DIR technique.

Note that we choose different angular scales for the BOSS and GAMA measurements, and that for the GAMA analysis we do not measure the GGL signal for the lowest tomographic bin.  This is because for the BOSS lenses there is significant redshift overlap between the lens samples and most of the tomographic bins.  As these BOSS galaxies are strongly biased we expect to find a significant dilution of the GGL signal at small scales; the galaxies at these close separations are more likely to be clustered with the lens such that the $n(z)$ for our source sample becomes $\theta$ dependent. This effect is similar to the dilution seen in cluster lensing studies, where it is usually corrected for via a multiplicative boost factor \citep[see e.g.][]{hoekstra/etal:2015}.   While in principle one could estimate a similar correction for the shear-ratio test shown here, such an analysis lies beyond the scope of this paper and would introduce new unknowns.   Instead we perform the GGL analysis for our BOSS sample only at large $\theta$ where the dilution effect is minimal because the multiplicative boost factor goes to unity \citep{hoekstra/etal:2015}.

In order to take advantage of the high signal-to-noise ratio at small angular scales to provide a more stringent test, and to avoid additive systematics that could in principle bias the large-scale results\footnote{We apply an empirical $c$-correction in our BOSS GGL analysis by subtracting the tangential shear signal around random points from the actual tangential shear around the lens galaxies \citep[see e.g.][]{vanuitert/etal:2016}. This correction is not necessary for our GAMA measurements as we use small scales only.} we analyse the lowest GAMA redshift sample with $z_{\rm l}<0.2$ in order to reduce redshift overlap with the 2nd, 3rd, and 4th tomographic bins (see Fig.~\ref{fig:z_dist_comp}).  With this GAMA selection, lens-source clustering should be negligible given the DIR $n(z)$ allowing us to use the high signal-to-noise small angular scales for a clean shear-ratio test. 

In addition to the DIR analysis presented here, we also analysed the alternative redshift distributions described in Section~\ref{sec:photoz_calibration} (BPZ, BOR, CC) which we find to pass this test with similar $p$-values. Evidently this test is not sensitive enough to discriminate between these different options. 

In order to check the constraining power of our shear-ratio test we deliberately swap the redshift distributions of the lenses and sources in the modeling. While somewhat arbitrary this should yield a model that is incompatible with the measurements. Indeed we find extremely low $p$-values ($p\la10^{-4}$) allowing us to conclude that this shear-ratio test is meaningful.
%C

\section[Galaxy selection, shear calibration and tests for systematics]
{GALAXY SELECTION, SHEAR CALIBRATION AND TESTS FOR SYSTEMATICS}
\label{sec:app_shear_tests}
In this Appendix we document the more technical aspects of the KiDS-450 shear measurement, including object selection, multiplicative and additive shear calibration corrections and a range of systematic error analyses.

We create an object detection catalogue using \textsc{SExtractor} with a low detection threshold \citep{bertin/arnouts:1996}.  This catalogue contains $\sim 30$ objects arcmin$^{-2}$ and is used as the input catalogue for \emph{lens}fit.   In Sections~\ref{sec:galaxy_selection} and~\ref{sec:artefacts} we discuss the object selection that reduces this object catalogue down to 11.5 galaxies arcmin$^{-2}$ with accurate shear measurements, and an effective number density of $n_{\rm eff} = 8.53$ galaxies arcmin$^{-2}$.  This level of reduction is similar to CFHTLenS \citep[see for example figure 3 of][]{duncan/etal:2014}.   It reflects our choice to use a very low source detection threshold in the original construction of the object catalogue, using a set of criteria measured by \emph{lens}fit to decide which objects can then be used for accurate shear measurement.  The alternative of imposing a higher signal-to-noise cut at the object detection stage could result in galaxy selection bias \citep[see for example][]{fenechconti/etal:2016}.

\subsection{\emph{Lens}fit selection} 
\label{sec:galaxy_selection}

Apart from the ellipticity and associated weight, \emph{lens}fit also returns a number of extra parameters  \citep[see table~C1 of][]{kuijken/etal:2015}, including an estimate of the galaxy \texttt{scalelength}, and a \texttt{fitclass} parameter than encodes the quality of the fit. Using this information we filter the \emph{lens}fit output to remove sources with unreliable ellipticities, as follows.

Our initial selection requires a non-zero \emph{lens}fit weight, which automatically removes the following:
\begin{enumerate}
\item objects identified as point-like stars (\texttt{fitclass}  = 1);
\item objects that are unmeasurable, usually because they are too faint (\texttt{fitclass}  = $-3$);
\item objects whose marginalised centroid from the model fit is further from the \textsc{SExtractor} input centroid than the positional error tolerance of 4 pixels (\texttt{fitclass}  = $-7$);
\item objects for which insufficient data are found, for example if they fall near the edge of an image, or a defect (\texttt{fitclass}  = $-1$).   
\end{enumerate}
We further cut the following sources out of the catalogue:
\begin{enumerate}
\setcounter{enumi}{4} % continue count where it left off
\item objects for which the best-fitting galaxy model fit has a reduced $\chi^2>1.4$, indicating that they are poorly fit by a bulge plus disk galaxy model (\texttt{fitclass}  = $-4$). This cut removes 0.1 per cent of the objects outside the masked regions;
\item objects that are brighter than the brightest magnitude in our image simulations (\texttt{catmag} $\le20$);
\item probable asteroids or binary stars (see Section~\ref{sec:artefacts} below);
\item potentially blended sources, defined to have a neighbouring object within a contamination radius of 4.25 pixels from the galaxy's \textsc{SExtractor} centroid\footnote{Note that this selection is slightly stricter than the fiducial \texttt{fitclass}  = $-6$ which keeps all objects with a contamination radius $>4$ pixels.  The method used to determine a contamination radius for each object is discussed in section 3.7 of \citet{miller/etal:2013}.};
\item objects classified as duplicates (\texttt{fitclass}  = $-10$).  These objects are identified during the de-blending analysis when {\em lens}fit builds a dilated segmentation map that is used to mask out a target galaxy's neighbours. A very small fraction of targets has another input catalogue galaxy within its pixel region, owing to differing de-blending criteria being applied in the \textsc{SExtractor} catalogue generation stage and the \emph{lens}fit image analysis. In these cases \emph{lens}fit uses the same set of pixels to measure both input galaxies, which leads to the inclusion of two correlated, high-ellipticity values in the output catalogue. We therefore flag such cases and exclude them from subsequent analysis.    
\end{enumerate}
In contrast to earlier analyses, we retain objects which are large galaxies that overfill the postage stamp size of 48 pixels (\texttt{fitclass}  = $-9$) in order to avoid ellipticity selection bias in the brightest galaxy sample: for a fixed major axis scale length, such a cut would have preferentially removed galaxies oriented in the $\pm\epsilon_1$ direction.    \citet{fenechconti/etal:2016} recommend also removing objects that have a measured \emph{lens}fit \texttt{scalelength} smaller than 0.5 pixels.  This selection represents only 0.01\% of the weighted galaxy population and we did not apply this cut for our main analysis.  We have, however, verified that making this additional selection did not change our results in terms of the measured $\xi_\pm$ in Section~\ref{sec:results},  the measured additive calibration terms in Section~\ref{sec:c_term} or the measured small-scale B modes in Section~\ref{sec:EB}.

\begin{figure}
\includegraphics[width=\hsize]{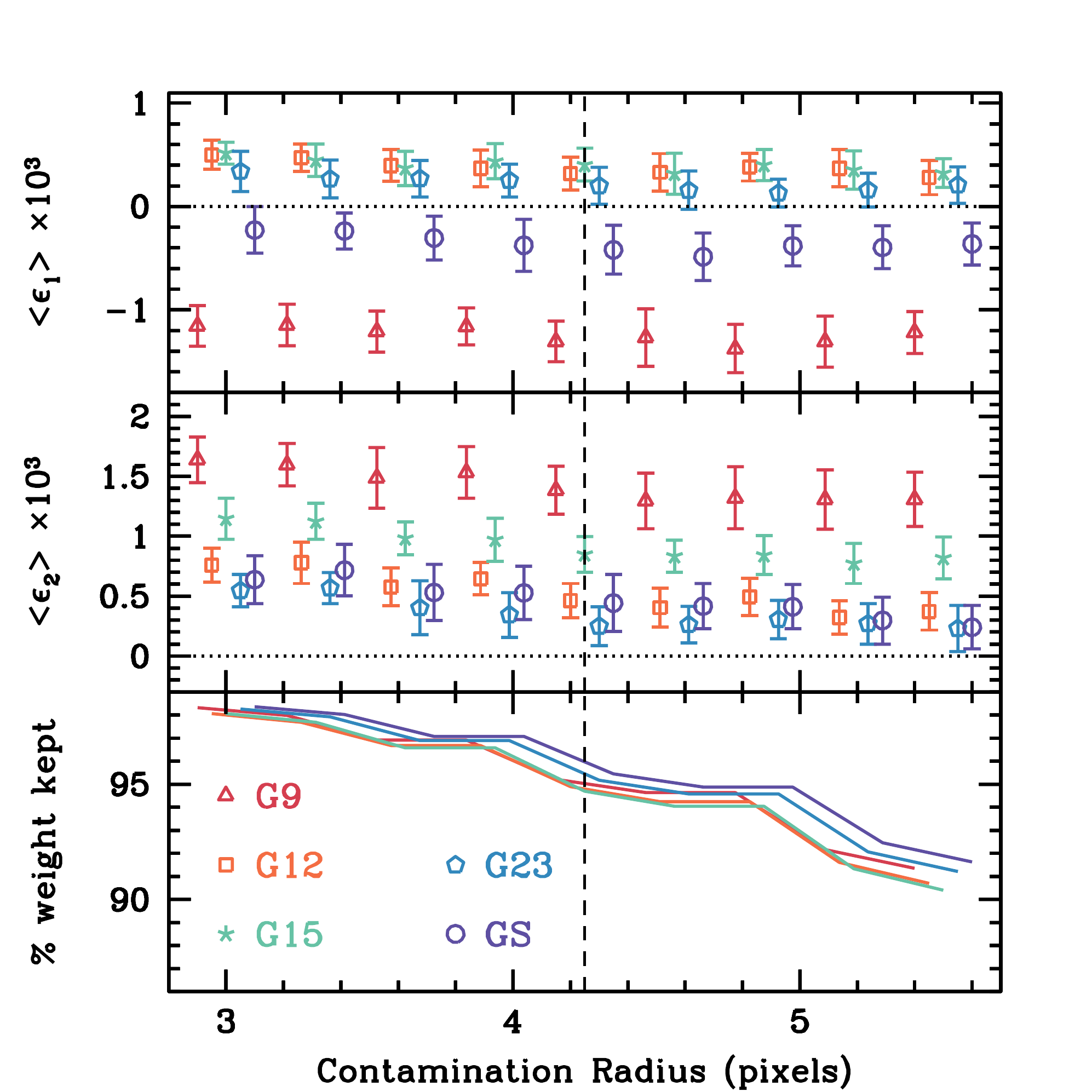}
\caption{\label{fig:set_contam_radius}Criterion for masking of neighbours: the additive calibration bias as measured from the weighted average ellipticity $\epsilon_1$ (upper) and $\epsilon_2$ (middle panel), as a function of the minimum allowed contamination radius.  The amplitude of the bias is shown for each of the five KiDS-450 patches (G9, G12, G15, G23 and GS), with points slightly offset along the horizontal axis for clarity.  The lower panel shows the decreasing effective number density of objects as the contamination radius cut increases.  Our chosen contamination radius cut of $>4.25$ pixels is shown dashed.}
\end{figure}

The contamination radius de-blending threshold was optimised to maximise the number density of galaxies whilst minimising systematic errors.  Fig.~\ref{fig:set_contam_radius} shows the additive calibration bias, as measured from the weighted average ellipticity.  The additive bias is found to slightly increase in amplitude as the minimum allowed contamination radius decreases below a contamination radius cut of $4.25$ pixels.  The lower panel shows the effect of varying the de-blending selection criterion on the weighted number density.  For galaxy-galaxy lensing studies, which are less sensitive to additive biases, a less conservative contamination radius cut could be used to increase the number density of source galaxies by $\sim 5$ per cent.

\subsection{Removing artefacts} 
\label{sec:artefacts}
\subsubsection{Asteroids}

\begin{figure}
\includegraphics[width=\hsize,trim=1cm 4cm 0 3.5cm,clip]{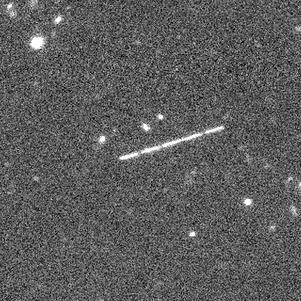}\\[0.01\hsize]
\includegraphics[width=0.24\hsize,trim=3cm 3cm 3cm 3cm, clip]{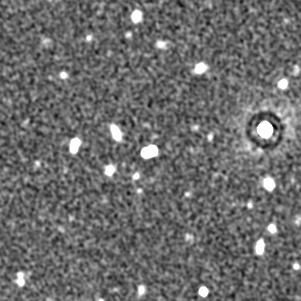}
\includegraphics[width=0.24\hsize,trim=3cm 3cm 3cm 3cm, clip]{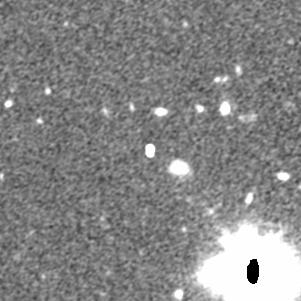}
\includegraphics[width=0.24\hsize,trim=3cm 3cm 3cm 3cm, clip]{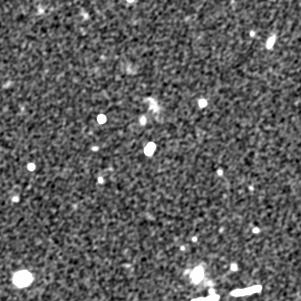}
\includegraphics[width=0.24\hsize,trim=3cm 3cm 3cm 3cm, clip]{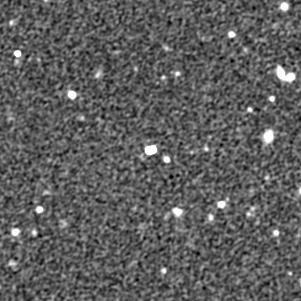}
\caption{\label{fig:asteroidpic}Examples of contaminating high-ellipticity objects that we filter out of the catalogues. Upper: an asteroid in one of the co-added images in patch G12.   Lower: four examples of bright, blended binary stars. The lower images are PSF-Gaussianized, to highlight the compact nature of these point sources.}
\end{figure}

Because the source detection catalogue is produced from stacked images, some moving objects enter the catalogue. Some of our fields (particularly G12, but also G9 and G15) lie close to the ecliptic, and individual tiles can contain as many as 100 asteroids or more, which show up as a trail of dashes, one for each sub-exposure (Fig.~\ref{fig:asteroidpic}). With typical proper motions of \SI{30}{\arcsec} per hour, asteroids appear very elongated on our six-minute $r$-band sub-exposures, and so enter the catalogue of resolved sources with high-weight \emph{lens}fit shapes. Moreover, asteroids move predominantly along the ecliptic, making them a population of coherent, bright, and very elliptical sources which biases our ellipticity correlation function. When the segments of the trail due to the different sub-exposures are de-blended, the same object may enter the catalogue up to five times, further increasing the impact of this contamination.

Eliminating the asteroids from the catalogue would ideally be done by comparing individual sub-exposures. For the present analysis, however,  we instead use the fact that they show up in the multi-colour catalogue with very characteristic colours: they are bright in the $r$-band detection image but essentially undetected in the other bands (because in those images they are no longer inside the photometric aperture at their detected position). The \textsc{bpz} photometric redshift estimates of asteroids with such erroneous `$r$-only' photometry are found to lie almost exclusively within the second tomographic bin with $0.3< z_\rmn{B}<0.5$.  

\begin{figure}
\includegraphics[width=\hsize]{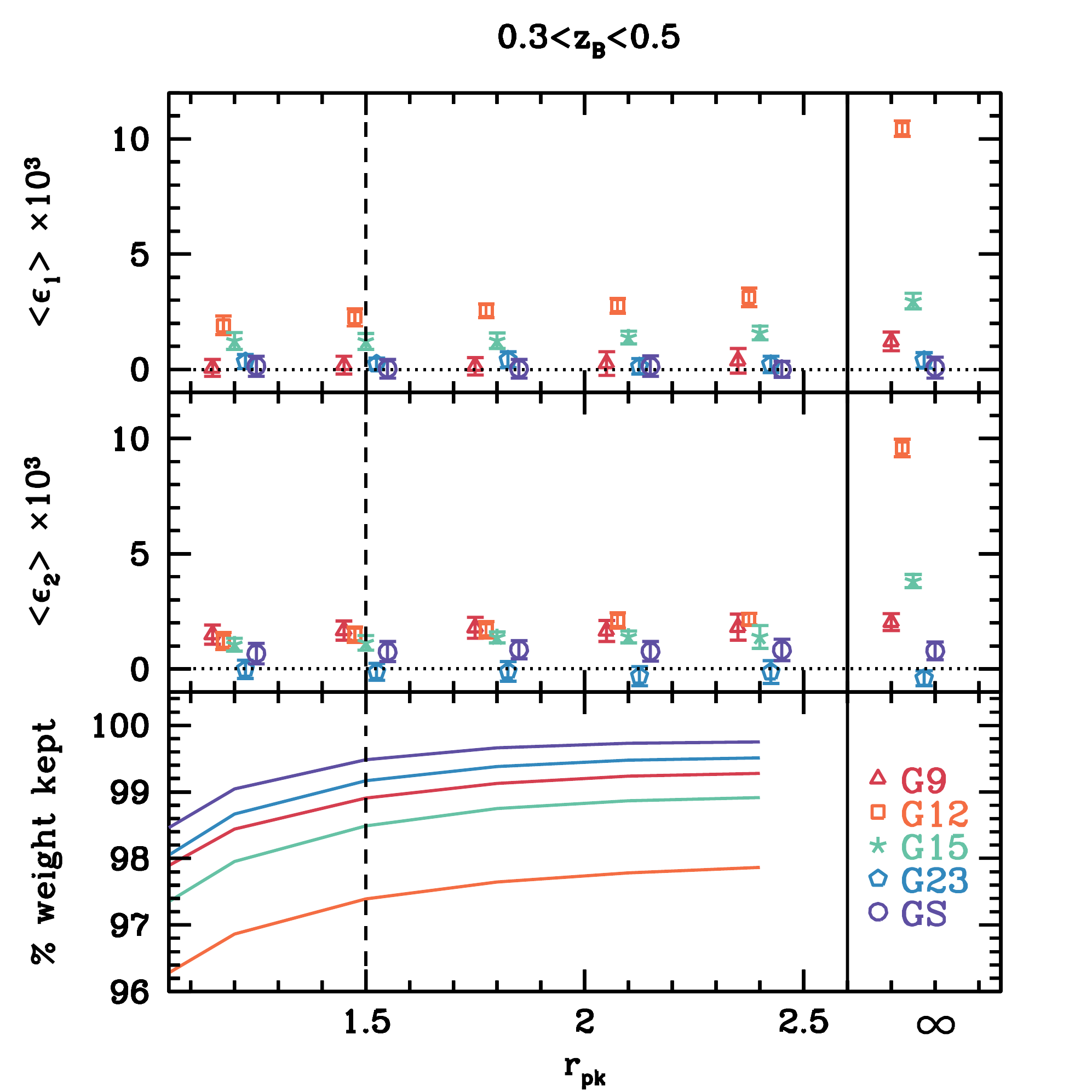}
\caption{\label{fig:asteroids}Asteroid selection criterion: the additive calibration bias for objects with $0.3< z_\rmn{B}<0.5$, measured from the weighted average ellipticity $\epsilon_1$ (upper) and $\epsilon_2$ (middle panel), as a function of the maximum allowed colour $r_{\rm pk}$ (Eq.~\ref{eqn:rpm}).  The amplitude of the bias is shown for each of the five KiDS-450 patches (G9, G12, G15, G23 and GS), with points slightly offset along the horizontal axis for clarity.  The right-most point in each panel shows the weighted average ellipticity measured for the full galaxy sample with no colour selection applied.  The lower panel shows how the effective number density of objects increases with the colour threshold.  Our chosen asteroid selection criterion removes galaxies with $r_{\rm pk}>1.5$,  shown dashed.}
\end{figure}

To identify potential asteroids we define an `$r$-peakiness' colour,
\be
r_{\rm pk} = \min(g-r,i-r) \, ,
\label{eqn:rpm}
\ee
and determine an optimal cutoff value that maximises the number density of galaxies whilst minimising systematic errors.   Fig.~\ref{fig:asteroids} shows the additive calibration bias, as measured from the weighted average galaxy ellipticity with a photometric redshift $0.3< z_\rmn{B}<0.5$.  The additive bias is found to increase in amplitude as the maximum allowed colour $r_{\rm pk}$  increases above $1.5$.     Including all objects without any asteroid de-selection can result in additive biases in the second tomographic bin as large as $\sim 0.01$ for the G12 patch which lies closest to the ecliptic.  This bias is reduced to $\sim 0.001$ after the objects with $r_{\rm pk}>1.5$ are removed from the sample.  We apply this selection to all tomographic bins.  

\begin{figure}
\includegraphics[width=\hsize,trim=1cm 0 3cm 0]{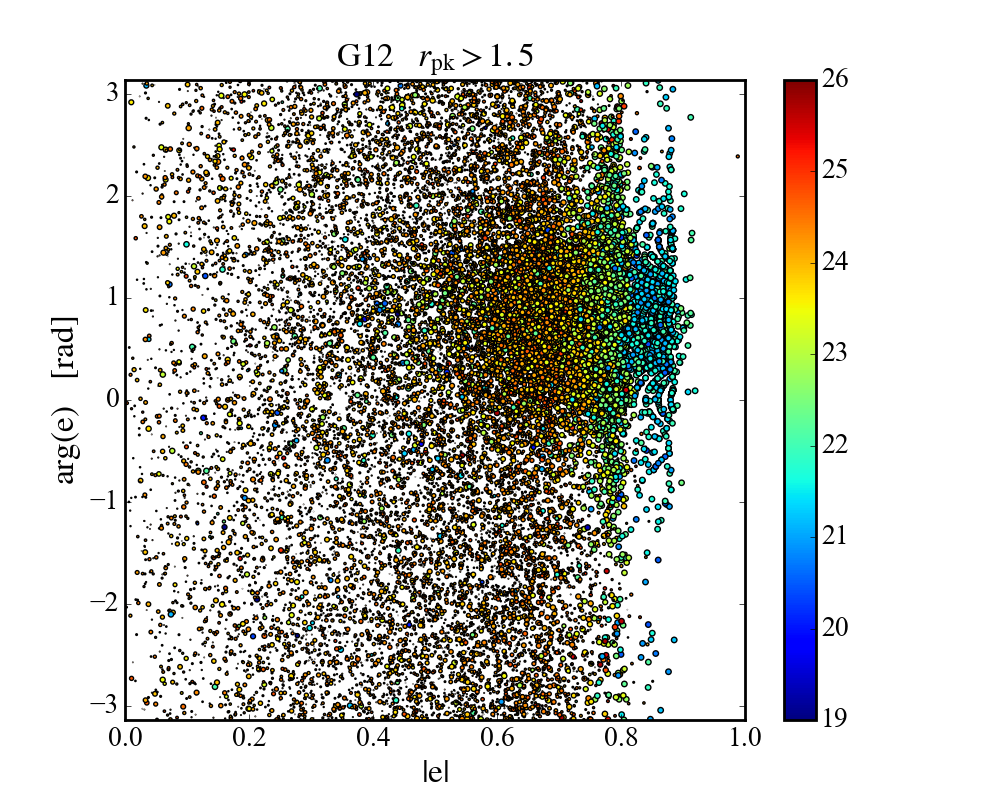}
\caption{\label{fig:asteroidcut}Sources from patch G12 that fail the asteroid cut, plotted in polar ellipticity coordinates $(|e|,\hbox{arg}(e))$. These sources are at least 1.5 magnitudes brighter in $r$ than in $g$ and in $i$.  Each symbol is colour-coded by the catalogue  $r$ magnitude, and the symbol size is proportional to \emph{lens}fit weight. }
\end{figure}

Fig.~\ref{fig:asteroidcut} shows the distribution of G12 objects with $r_{\rm pk}>1.5$ in polar ellipticity coordinates $(|e|,\hbox{arg}(e))$.  This distribution supports the theory that these artefacts are asteroids as the majority of objects excluded with this cut are highly elliptical, with a preferred orientation.

\subsubsection{Binary stars}
\begin{figure*}
\includegraphics[height=0.33\hsize, trim=0.5in 0 2.35in 0, clip]{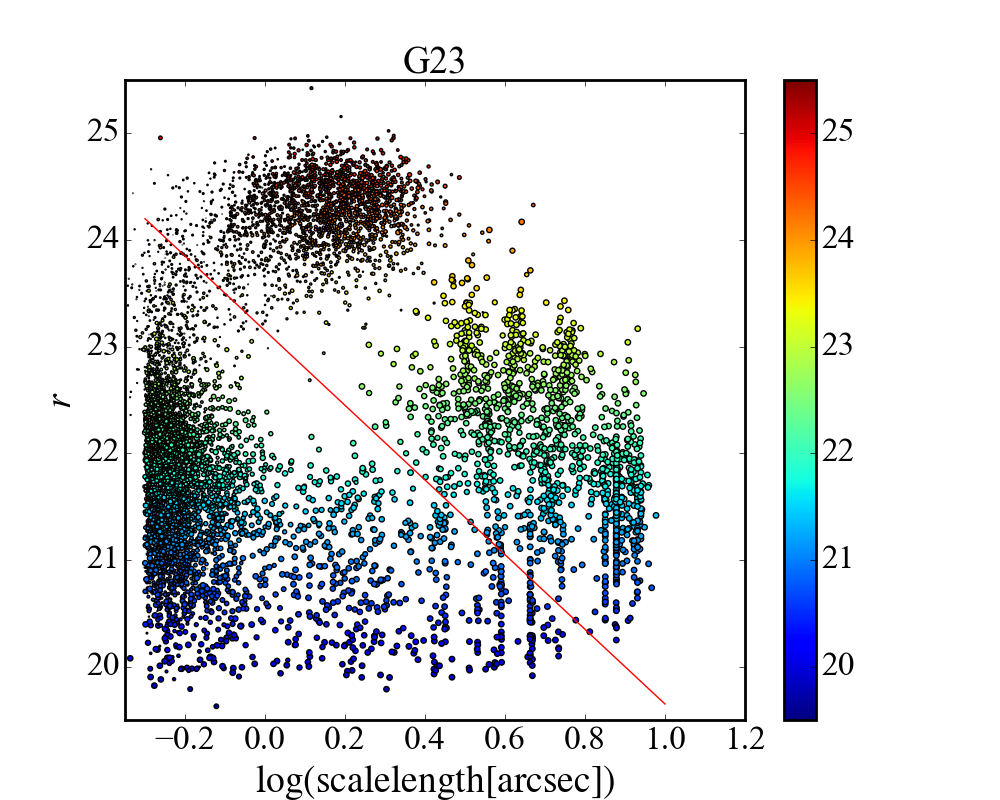}
\includegraphics[height=0.33\hsize, trim=0in 0 2.35in 0, clip]{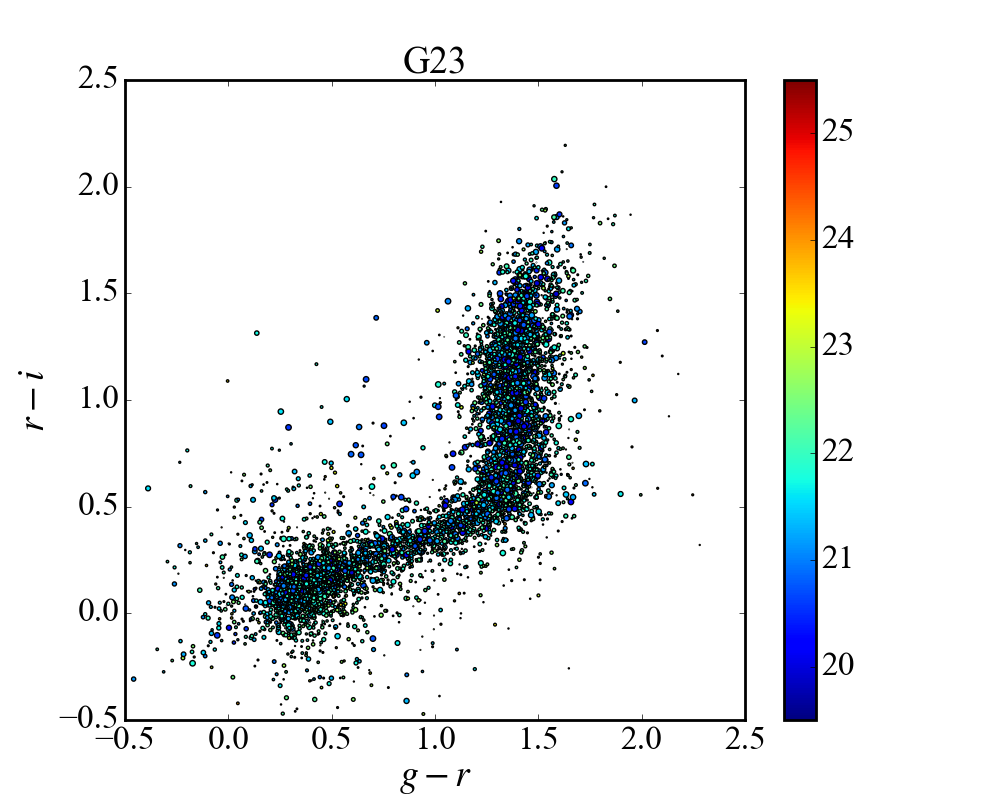}
\includegraphics[height=0.33\hsize, trim=0in 0 1.48in 0, clip]{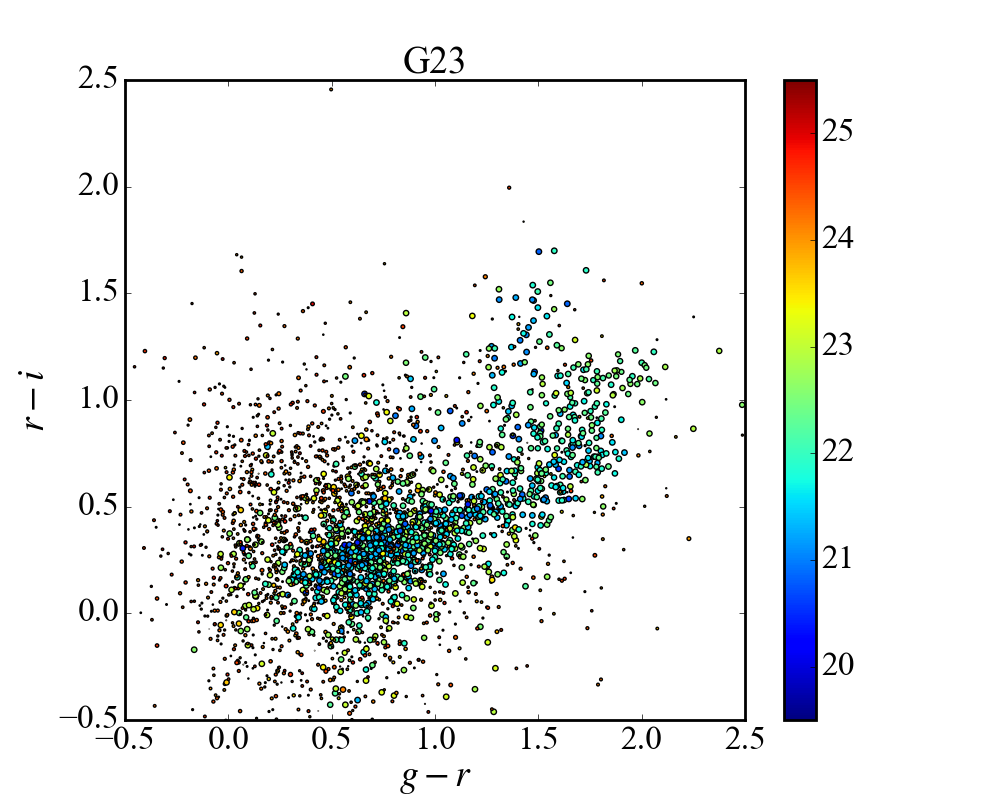}
\caption{\label{fig:binarycuts}Illustration of the cuts on high-ellipticity ($e>0.8$) objects to separate binary stars from galaxies, for one of the five patches (G23). Left: scale length vs.\ $r$-magnitude diagram with the cut indicated. Middle: $gri$ colour-colour diagram for objects below the cut. Right: same, for objects above the cut. Each symbol is colour-coded by the catalogue  $r$ magnitude and the symbol size is proportional to the \emph{lens}fit weight.}
\end{figure*}

Unresolved binary stars are another source of highly elliptical objects in our catalogues. To identify parts of parameter space where they dominate, we use the $(g-r,r-i)$ colour-colour diagrams. Any selection that predominantly picks out stars will show the characteristic stellar locus, whereas galaxy samples will not. For very high ellipticity $(e>0.8)$ it turns out that objects with measured scale length 
\begin{equation}
r_d<0.5 \times 10^{(24.2-r)/3.5} \,\hbox{pixels}
\end{equation}
have stellar colours, whereas larger objects do not (Fig.~\ref{fig:binarycuts}). We therefore apply this cut to our source catalogues as well. The fraction of sources (\emph{lens}fit weight) removed this way lies between 0.07 (0.05) and 0.33 (0.18) per cent, depending on the galactic latitude of the patches. Unlike asteroids, these binary stars do not have a preferred orientation, so their inclusion in the galaxy catalogue would generate a multiplicative and not an additive bias. A few examples of bright objects eliminated by this cut are shown in Fig.~\ref{fig:asteroidpic}.

 %D1-2
\subsection{Multiplicative shear calibration from image simulations}
\label{sec:imsim}
We use the `self-calibrating' version of \emph{lens}fit to measure galaxy shapes, quantifying any residual bias after self-calibration through the analysis of a suite of simulated images resembling the KiDS $r$-band images. We present a summary here and refer the reader to \citet{fenechconti/etal:2016} for a detailed description of the simulations and the analysis pipeline.

We use the publicly available \textsc{GalSim} software \citep{rowe/etal:2015} to simulate \SI{416}{\square\degree} of data, containing approximately 54 million galaxies, of which 16 million have a non-vanishing \emph{lens}fit weight. This number of galaxies, in combination with a shape noise cancellation scheme, allows us to achieve a statistical precision on the measurement of the shear multiplicative bias of $\sim 0.003$ which is more than a factor of 3 smaller than the requirements described in Appendix~\ref{sec:app_A}.   We simulate individual \SI{1}{\square\degree} tiles with the same resolution, focal plane footprint and five-point dither strategy as the KiDS observations (see \citealt{dejong/etal:2015}). The individual sub-exposures are coadded using \textsc{swarp} \citep{bertin:2010}.  The $r$-band magnitude distribution of the galaxies is chosen such that it matches the number counts measured in KiDS-DR1/2 and we use deep HST observations from GEMS \citep{rix/etal:2004} and UVUDF \citep{rafelski/etal:2015} to extend the distribution to 28th magnitude in order to account for the effect of undetected sources in the estimation of the shear bias \citep{hoekstra/etal:2015,bruderer/etal:2016}.  The size distribution as a function of magnitude is described in \cite{kuijken/etal:2015}. Galaxies are modelled as a linear combination of a de Vaucouleurs profile for the bulge and an exponential profile for the disk. The bulge/total flux ratio, the fraction of bulge-dominated galaxies and the galaxy ellipticity distribution are matched to the \emph{lens}fit prior \citep{miller/etal:2013}. 

The position of the galaxies is random, hence allowing us to assess the impact of blending on the shear measurements but not the impact of clustering. In order to reduce the impact of shape noise in the shear measurement, we created copies of each simulated tile in which all galaxies are rotated by 45, 90 and \SI{135}{\degree}. 

Each tile (and its three rotated copies) is simulated 8 times with a different constant shear value applied to the galaxies. The shear values are chosen such that they have the same magnitude, $|g|=0.04$, but they are rotated at 8 evenly spaced position angles. 

The galaxies are convolved with a PSF, described by a Moffat profile. The PSF is spatially constant in each simulated tile but different in each of the 5 sub-exposures. The parameters of the Moffat profile are chosen such that the PSF size and ellipticity distributions in the simulations are representative of the variations measured on KiDS-DR1/2.  In total we simulate 13 PSF sets. We include stars in the simulations, that are a perfect representation of the PSF. Their magnitude distribution is derived from the Besan\c{c}on model \citep{robin/etal:2003,czekaj/etal:2014} for stars between $20<r<25$. Bright stars are not included in the simulations as they are masked in the real data.   
 
We create an object detection catalogue using \textsc{SExtractor} \citep{bertin/arnouts:1996} to process the co-added simulated dithered exposures with the same configuration used in the analysis of the real KiDS data. The resulting detection catalogue is used as the input to \emph{lens}fit.   For each PSF set and for each of the 8 input shear values we measure the shear components $\gamma^{\rm obs}_j$ by  averaging the ellipticities of all simulated galaxies in the four rotated versions of each tile, using the recalibrated weights calculated by \emph{lens}fit.
  
Following \cite{heymans/etal:2006} we parameterise the shear bias in terms of a multiplicative and an additive term:
\begin{equation}
\gamma^{\rm obs}_j=(1+m_j)\gamma^{\rm true}_j + c_j .
\label{eqn:mandc}
\end{equation}
 
We characterise the multiplicative and additive bias as a function of the galaxy signal-to-noise ratio (SNR) and `resolution' ($\cal{R}$), defined as:
\begin{equation}
\mathcal{R} := \frac{r^2_{\textrm{PSF}}}{\left(r_{ab}^{2}+r^2_{\textrm{PSF}}\right)}.
 \label{CAL::EQU:resolution}
\end{equation}
where $r_{ab}=\sqrt{ab}$ is the circularised size of an object derived from the \emph{lens}fit measured semi-major axis $a$ and semi-minor axis $b$ of each galaxy and $r^2_{\textrm{PSF}}$ is the corresponding size of the PSF.  

In order to derive the multiplicative bias in the four KiDS-450 tomographic bins we resample the image simulations such that the distributions of simulated galaxy properties match those of the real galaxy sample in each tomographic bin. We do this by means of a two-dimensional $k$-nearest neighbour search of the simulated (SNR,$\cal{R}$) surface for all KiDS galaxies in each of the 4 tomographic bins used in this paper. As a result of the resampling each simulated galaxy is assigned an additional weight, which is the number of times a real KiDS galaxy has been matched to it.
This procedure is very similar to the DIR method used to calibrate the galaxy redshift distribution (see Section \ref{sec:photoz_calibration}). We measure the shear from the resampled simulations as a weighted average of the measured galaxy ellipticities, where the weight is the product of the `resampling' weight and the \emph{lens}fit weight. This is done for all four tomographic bins. We finally compute the multiplicative and additive shear bias by means of a linear regression (see Equation \ref{eqn:mandc}) between the measured shear and the true input shear used in the simulation.

The average multiplicative and additive bias measured in each of the four tomographic bins is very small and similar in amplitude for the two shear components. In particular we find $m=[-0.0131, -0.0107, -0.0087, -0.0217] \pm 0.01$ and $c=[3.2, 1.2, 3.7, 7.7]\pm 0.6 \times10^{-4}$.
The error on the multiplicative bias accounts for systematic and statistical uncertainties ($\sigma_{\mathrm{stat}}=0.003$). 
This systematic uncertainty comes from differences between the data and the simulations and choices in the bias estimation. It allows also for additional sources of bias which might arise from the mismatch between the galaxy model assumed by \emph{lens}fit and the true galaxy morphology. Known as `model bias' \citep{voigt/bridle:2010} this effect is not reflected in our image simulation analysis which models and analyses galaxies with a two-component S{\'e}rsic profile.  Model bias is expected to be negligible in comparison to the statistical noise of KiDS-450 \citep{kacprzak/etal:2014}. Furthermore, in the GREAT3 challenge \citep{mandelbaum/etal:2015} the self-calibrating \emph{lens}fit analysis of the image simulations created using true HST imaging suggests that the contribution from model bias is less than 0.01.  
The additive shear bias estimated from the simulations is not used to correct the shear measured from the data. An additive term is instead measured directly from the data as described in Section \ref{sec:c_term}. Hence the error on the additive bias quoted above is only the statistical uncertainty set by the volume of the simulated data and doesn't include any the systematic uncertainty.

As a way to validate our primary calibration strategy we also adopt an alternative non-parametric approach in order to derive a multiplicative bias calibration that achieves the required level of accuracy given in Appendix~\ref{sec:app_A}. This is similar to the method employed in CFHTLenS \citep{miller/etal:2013}. We compute the multiplicative and additive bias on a 20 by 20 grid in SNR and $\cal{R}$, with the bin limits chosen such that each grid cell has equal \emph{lens}fit weight, and we assign the average bias calculated in each cell to all the real galaxies falling inside that cell. This is done for each tomographic bin. We refer the reader to Section 4.5 in \cite{fenechconti/etal:2016} for further details. As the additive bias we measure in the simulations is very small compared to the bias measured in the data (see Section~\ref{sec:c_term} for a discussion), we do not apply any additive bias correction measured from the simulations to the data. 

We find consistent results between both approaches, with differences in $m$ of at most 0.003 (1 sigma) in some tomographic bins. In both cases the residual biases as function of PSF properties (pseudo-Strehl ratio, size and ellipticity) are consistent with zero. 

Following \cite{jarvis/etal:2016} we also use the image simulations to explicitly look at the print-through of the PSF shape in the galaxy shape by modelling the bias as:
\begin{equation}
\gamma^{\rm obs}_j=(1+m_j)\gamma^{\rm true}_j + \alpha_j e_j^{\star} + c_j
\end{equation} 
We find the average $|\alpha_j|$ in the four tomographic bins to be less than 0.02 for both components.   We explore the sensitivity of the recovered calibration to changes in the intrinsic ellipticity distribution used in the simulations. We find that for reasonable variations the changes in the calibration are smaller than the errors.

 %D3
\subsection{Empirical additive shear calibration}
\label{sec:c_term}

We parameterise calibration corrections to our \emph{lens}fit shear measurements in terms of an additive $c$ and multiplicative term $m$ (following Eq.~\ref{eqn:mandc}).   For a sufficiently large source sample, where the average shear $\gamma$ and intrinsic ellipticity $\epsilon^{\rm int}$  average to zero: $\langle \gamma_i + \epsilon_i^{\rm int} \rangle = 0$, the additive term can be measured directly with $c_i = \langle \epsilon_i \rangle$. 

\begin{figure}
  \includegraphics[width=\hsize]{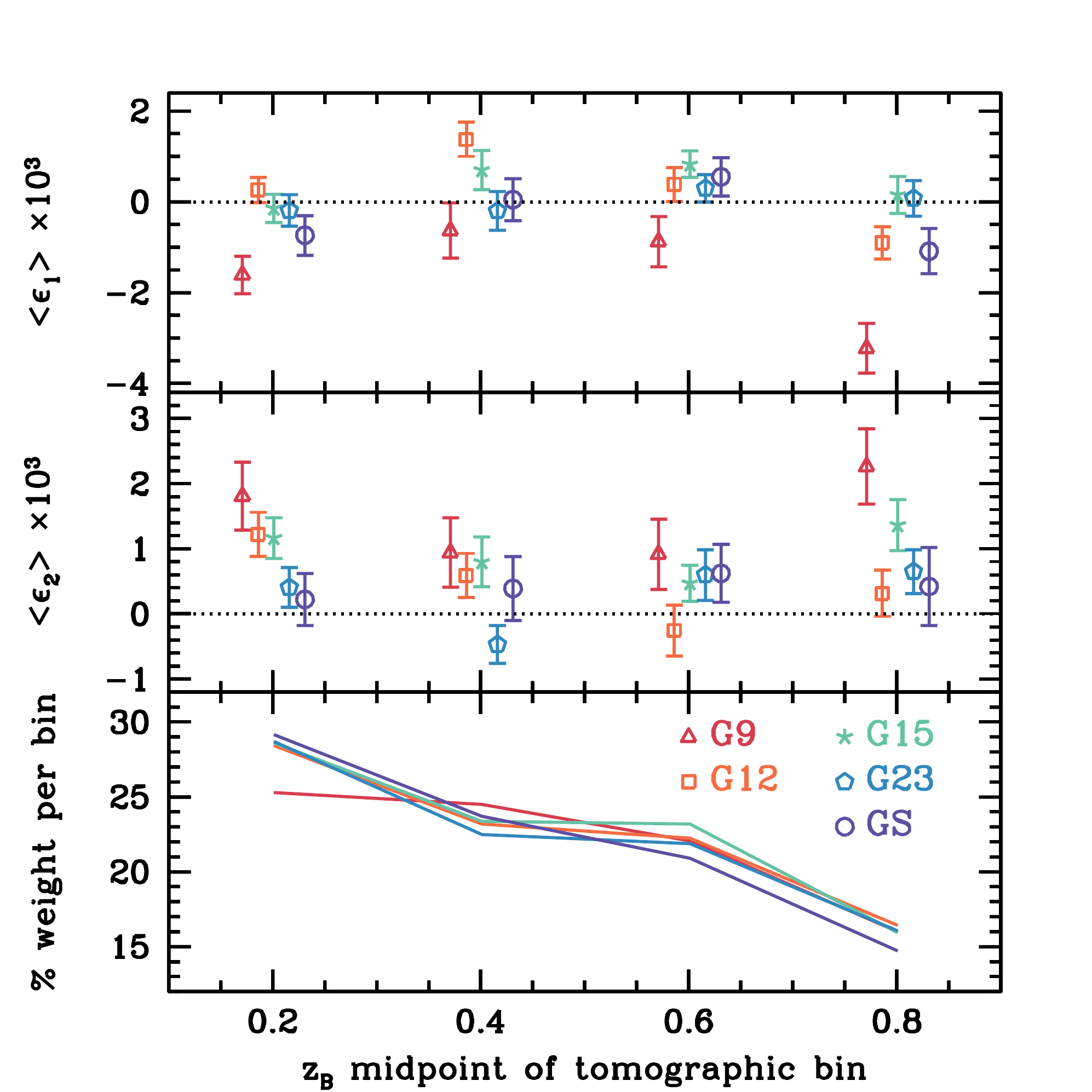}
  \caption{\label{fig:evsZB}Empirical additive calibration: the average weighted ellipticity $\epsilon_1$ (upper) and $\epsilon_2$ (middle panel) for the four tomographic slices.  The amplitude of the bias is shown for each of the five KiDS-450 patches; G9, G12, G15, G23 and GS, with points slightly offset on the horizontal access for clarity.  The lower panel shows the total weight of the galaxies in each tomographic slice as a fraction of the full survey weight.  About 10 per cent of the effective weight in KiDS-450 lies outside the tomographic limits used in this analysis.}
\end{figure}

In Section~\ref{sec:imsim} we review the KiDS image simulation analysis that finds these terms to be very small \citep[see][for further details]{fenechconti/etal:2016}.   This suite of image simulations is certainly the most sophisticated test of \emph{lens}fit to date, but there are a number of steps in the data acquisition, processing and analysis that are not simulated.   For example
\begin{enumerate}
\item{There are no image artefacts present, such as cosmic rays, asteroids and binary stars.}
\item{No astrometric shear is applied, which in the data analysis is derived using stars and subsequently corrected for in the model-fitting analysis.}
\item{The PSF model is determined from a set of known simulated stars, whereas the PSF model for the data derives from a stellar sample selected using the star-galaxy separation technique described in \cite{kuijken/etal:2015}.}
\item{Galaxy morphology is limited to two-component bulge-plus-disk models.}
\item{The simulated CCDs have a linear response.}   
\end{enumerate}
These higher-level effects are not expected to significantly change our conclusions about the multiplicative calibration $m$ from the image simulation analysis.  They do, however, impact on the additive term $c$ as demonstrated for example in Appendix~\ref{sec:artefacts}, where image artefacts from asteroids are shown to change the measured additive bias by an order of magnitude. It is therefore important to determine a robust empirical calibration scheme in order to correct for any sources of bias in the data that are not present in the image simulations.    Previous calibration schemes provide a $c$ correction per galaxy, by measuring the average ellipticity for samples of galaxies binned by their observed size, signal-to-noise and the PSF Strehl ratio at the location of the galaxy \citep{heymans/etal:2012,kuijken/etal:2015,hildebrandt/etal:2016}.  We do not use this methodology for KiDS-450 because \citet{fenechconti/etal:2016} show that separating galaxies by their measured size and signal-to-noise introduces a significant selection bias that changes the distribution of ellipticities which then impacts the shear deduced from the average ellipticity. The $c$-term measured by this method is therefore predominantly correcting for the selection bias introduced in its initial determination.   Instead we choose to determine an additive calibration correction $c$ from the average weighted ellipticity of all galaxies per tomographic bin, and per KiDS-450 patch as shown in Fig.~\ref{fig:evsZB}.  We find that the $c$-term is of the order $ \sim 10^{-3}$, differing between each tomographic bin and patch, with G9 showing the strongest bias. Errors are estimated from a bootstrap analysis.

\begin{figure}
  \includegraphics[width=\hsize]{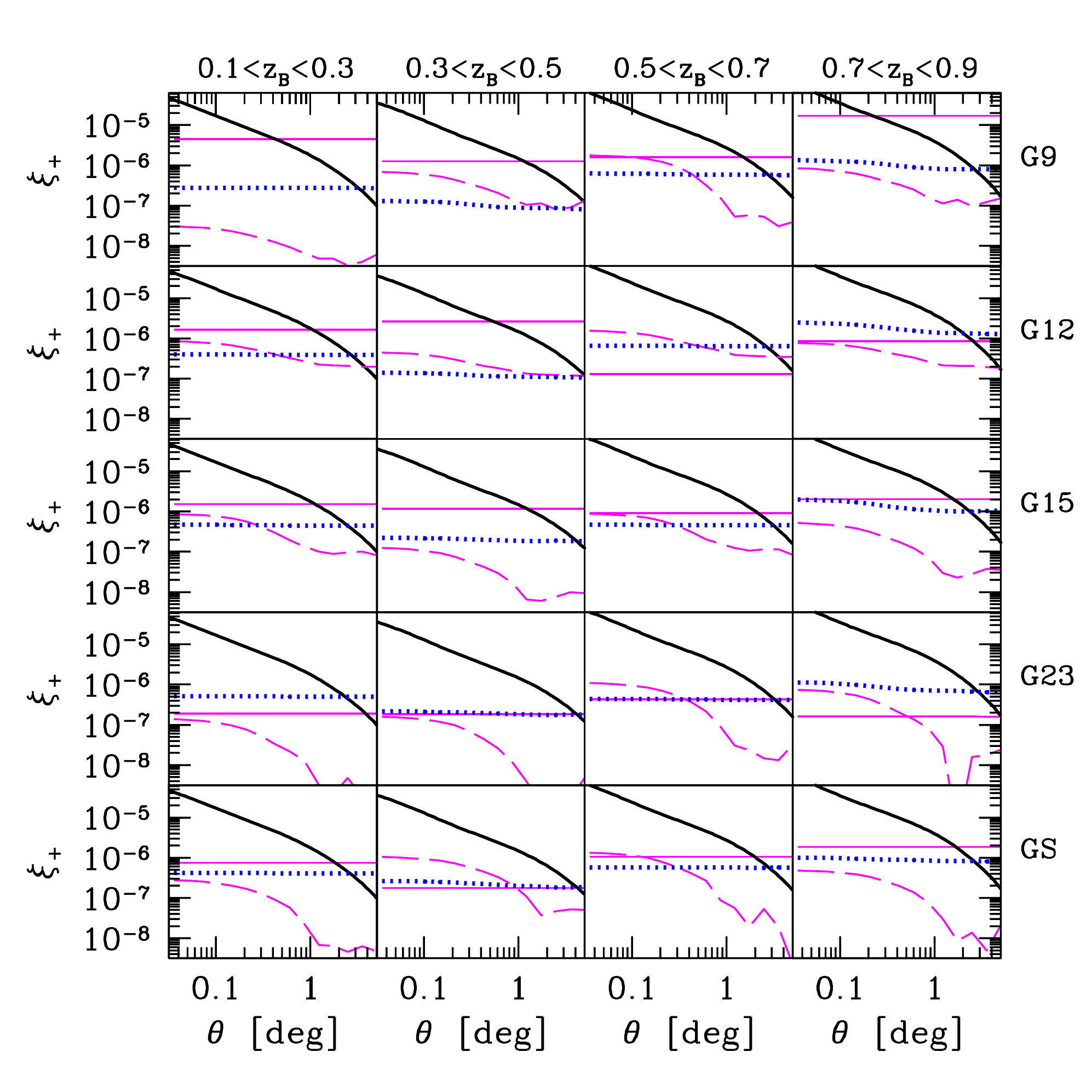}
  \caption{\label{fig:alpha_c}The systematic contribution to the two-point correlation function $\xi_+$ from the best-fitting $\alpha$ and $c$ values, determined from each tomographic slice (increasing in redshift from left to right) and KiDS-450 patch (labelled from top to bottom).  The dashed line in each panel shows $ \langle \alpha \epsilon^* \alpha \epsilon^* \rangle$ and the thin solid line shows $c_1^2 +c_2^2$, as determined from the data.  This can be compared to the best-fit cosmic shear signal from our primary analysis (thick solid) and to the systematic contribution from the $\alpha$ and $c$ predicted by the image simulation analysis of \citet{fenechconti/etal:2016} (dotted).}
\end{figure}

Owing to the strong ellipticity selection biases that arise when binning the data by measurements of size and signal-to-noise \citep{fenechconti/etal:2016} it is not possible to test whether the source of this additive bias derives from, for example, a population of small faint galaxies.  We can however test the dependence of the additive bias on observed position within the field of view and the number density of stars used to determine the PSF model \citep[see for example][]{vanuitert/schneider:2016} in addition to testing the dependence on PSF size and ellipticity.  We find no significant dependence of the additive bias with these quantities, which is consistent with the findings of the star-galaxy cross-correlation analysis presented in Section~\ref{sec:systematics}.  We further explore the possibility of PSF contamination when the data are split into tomographic slices by fitting the following model to the data, 
\be
\epsilon_j = \gamma_j + \epsilon_j^{\rm int} + \alpha_j \epsilon_j^* + c_j
\ee
where $\alpha_i$ and $c_i$ are free parameters, with $i=1,2$,  and $\epsilon^*$ is the measured PSF ellipticity.  The fit is made to each tomographic slice and patch, assuming $\langle \gamma_i + \epsilon_i^{\rm int} \rangle = 0$ in each sample.  We find measurements of $\alpha$ to be uncorrelated with $c$, such that including or excluding the $\alpha$ term from our systematics model has little impact on our conclusions about $c$.  We find $\alpha_1 \simeq \alpha_2$ with the average $\alpha$ across the patches ranging from $-0.03 < \alpha < 0.02$ across the four different tomographic bins with an error $\sim 0.01$.  Fig.~\ref{fig:alpha_c} shows the systematic contribution to the two-point correlation function $\xi_+$ from the best-fitting $\alpha$ and $c$ values, determined from each tomographic slice and patch.  The dashed line shows the PSF-dependent component of the systematics model\footnote{We use the shorthand notation of \citet{heymans/etal:2012} where $\langle ab \rangle$ indicates which two ellipticity measurements $a$ and $b$ are being correlated using the weighted data estimator in Eq.~\ref{eqn:xipm_est}.} $ \langle \alpha \epsilon^* \alpha \epsilon^* \rangle$.  We find that it is more than an order of magnitude lower than the best-fit cosmic shear signal (shown thick solid).  As $\alpha$ is found to be consistent with zero for the majority of tomographic slices and patches, we do not correct for this term in our analysis.  In contrast, the purely additive $c_1^2 + c_2^2$ contribution to $\xi_+$ (shown thin solid) is significant, exceeding the expected cosmic shear signal on large scales: therefore we do correct for this term.  For reference, the systematic contribution from the $\alpha$ and $c$ determined from the image simulation analysis of \citet{fenechconti/etal:2016} is shown dotted and found to be similar to the measured systematics, with the exception of the G9 and G12 patches and the second tomographic bin in the G15 patch.

\begin{figure}
  \includegraphics[width=\hsize]{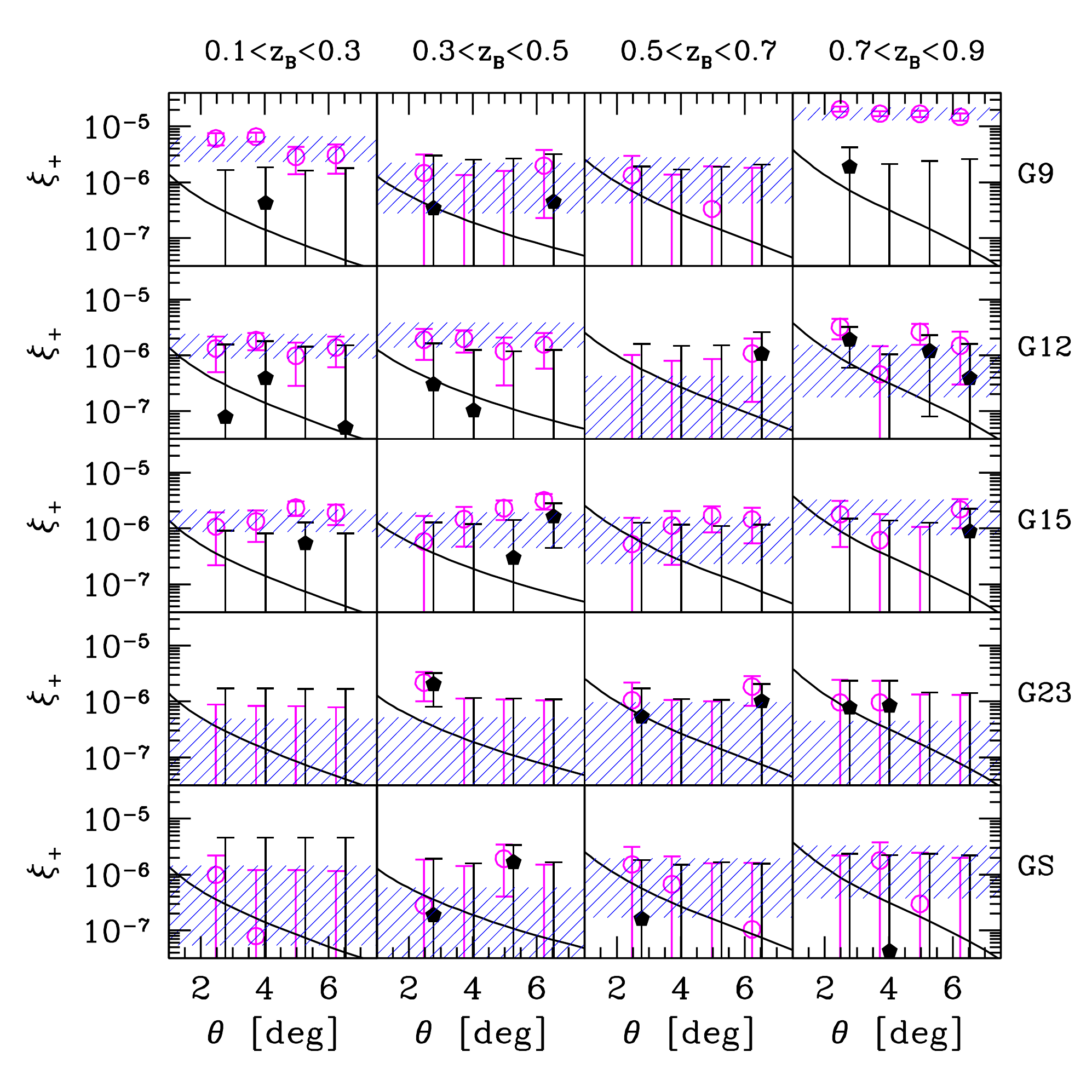}
  \caption{\label{fig:large-scales} The large-scale two-point correlation function with $\theta> $ \SI{2}{\degree} before (open symbols) and after (closed symbols) correcting each tomographic slice for the additive bias shown in Fig.~\ref{fig:evsZB}.    Each tomographic slice (increasing in redshift from left to right) and KiDS-450 patch (labelled from top to bottom) is shown.  The hashed region shows the amplitude of the correction and associated error.  The cosmological signal (shown solid) is expected to be consistent with zero on these scales.}
\end{figure}

In galaxy-galaxy lensing studies, it has become standard practice to measure the tangential shear around random positions, and subtract the measured `random signal' from the galaxy-galaxy lensing signal in order to remove additive systematic biases in the measurement.  We seek a similar method for cosmic shear to validate our additive calibration strategy.  There is no null signal, as such, but the cosmic shear signal on very large scales is expected to be consistent with zero within the statistical noise of KiDS-450.   Fig.~\ref{fig:large-scales} shows the measured $\xi_+$ on angular scales $\theta>$ \SI{2}{\degree}, before (open symbols) and after (closed symbols) correcting each tomographic slice for the additive bias shown in Fig.~\ref{fig:evsZB}.    The hashed region shows the amplitude of that correction and associated error.  After correction the large-scale signal is consistent with zero, and with the best-fit cosmological signal.   This verifies the calibration correction.  It also sets an upper limit on the angular scales that can be safely analysed for $\xi_+$.  We set this limit at \SI{1}{\degree}, where the measured amplitude of $\xi_+$ is more than an order of magnitude larger than the large-scale cosmic-shear signal that is also subtracted when employing this empirical calibration strategy.  On smaller scales of $\theta \sim$ \SI{5}{\arcmin}, where the cosmic shear signal-to-noise peaks (see Fig.~\ref{fig:cov_comp}), this large-scale cosmic-shear subtraction is completely negligible.  Note that there is no equivalent upper $\theta$ limit for $\xi_-$.  Additive terms do not typically contribute to the $\xi_-$ statistic as for square geometries $\langle c_t c_t \rangle = \langle c_\times c_\times \rangle$.  

%D4
\subsection{Star-galaxy cross correlation}
\label{sec:systematics}
We measure the correlation between star and galaxy ellipticities to determine if any tiles exist in our sample with a significant residual contamination resulting from an imperfect correction for the PSF.  We use the method described in \citet{heymans/etal:2012} to assess the significance of galaxy-PSF shape correlations in order to identify problematic tiles.  Previous surveys have used this strategy to flag and remove significant fractions of the data: 25 per cent \citep[CFHTLenS;][]{heymans/etal:2012}, 9 per cent \citep[RCSLenS;][]{hildebrandt/etal:2016} and 4 per cent \citep[KiDS-DR1/2;][]{kuijken/etal:2015}.   

Briefly, the method uses the fact that most galaxies in a tile have been observed in five different  sub-exposures, with different PSFs. It assumes that intrinsic galaxy ellipticities average to zero, and uses the degree of shape correlation between the corrected galaxies and the PSF models in the different sub-exposures to estimate the amount of PSF print-through in the tile's measured shear field. This PSF contamination is then cast in the form of a non-negative contamination, $\Delta\xi_\rmn{obs}$, to the 2-point galaxy ellipticity correlation function in that tile \citep[see eq.~10 in][]{heymans/etal:2012}. Mock shear maps with realistic noise properties are used to generate distributions of this statistic in the absence of systematic errors, in order to assess the significance of the measured values.

\begin{figure}
\includegraphics[width=\hsize, clip=true, trim=0cm 14.2cm 0cm 0cm]{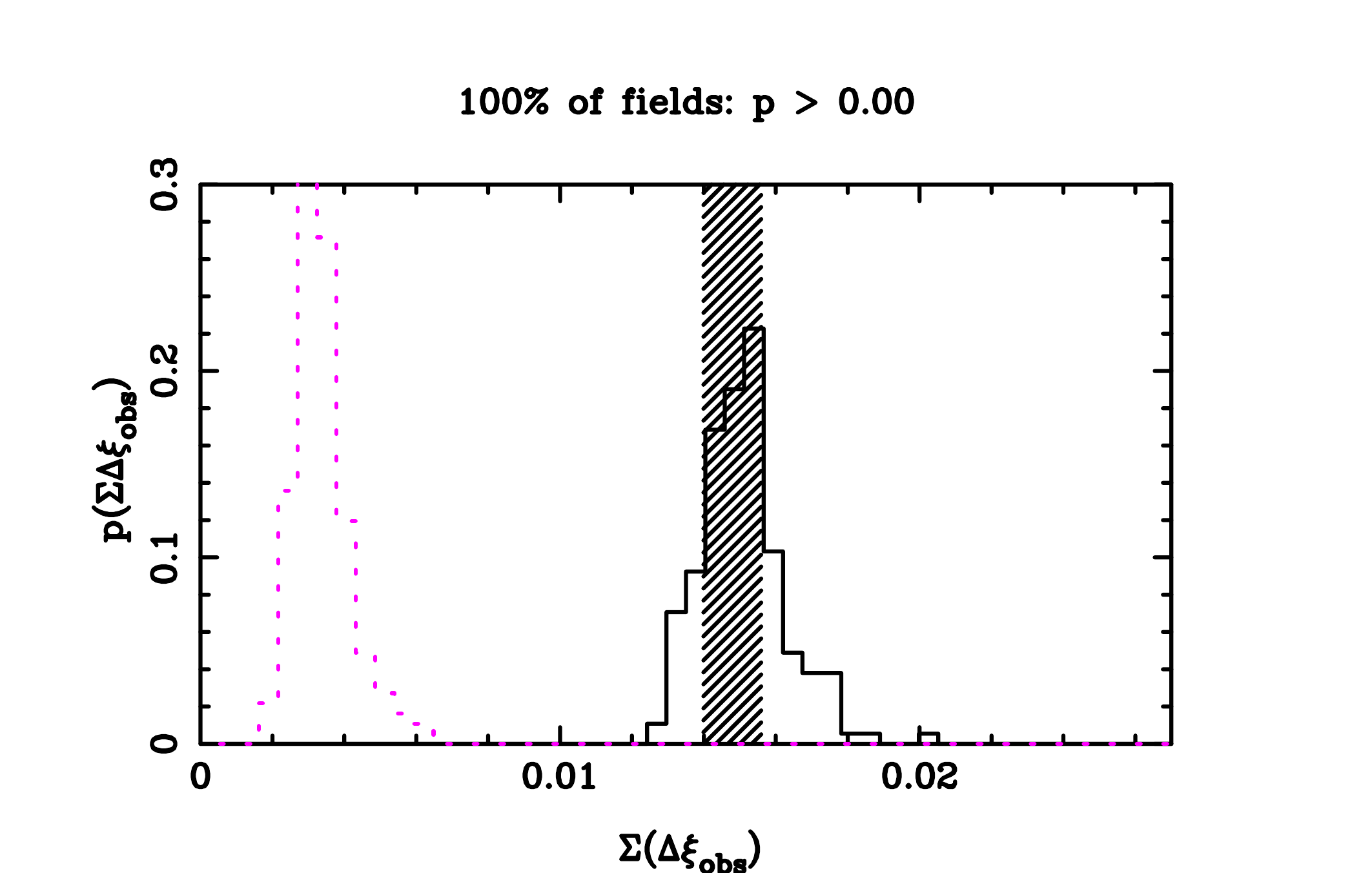}
\caption{\label{fig:delta_xi_sys}Amplitude of the star-galaxy shear cross-correlation statistic $\Delta\xi_\rmn{obs}$ summed over all KiDS-450 data tiles (hashed) and mock tiles (solid).  The contribution from cosmic shear to this statistic is shown by the dashed histogram. In KiDS-450 we do not reject any tiles based on this test since the value of $\Sigma \Delta \xi_\rmn{obs}$ is fully consistent with the expectation from simulations which model chance alignments between galaxies and the PSF due to cosmic shear, shape noise and shot noise.}
\end{figure}

The hashed bar in Fig.~\ref{fig:delta_xi_sys} shows the value of the $\Delta \xi_\rmn{obs}$ statistic, measured in each \SI{1}{\square\degree} tile, and then summed over the full KiDS-450 sample. For comparison, the histogram in Fig.~\ref{fig:delta_xi_sys} shows the distribution of $\Sigma \Delta \xi_\rmn{obs}$ measured for 184 systematic-free mock realisations of the KiDS-450 survey.  We find that the star-galaxy cross correlation measured in the data agrees well with the signal measured from our systematic-free mocks. 

\begin{figure}
\includegraphics[width=\hsize]{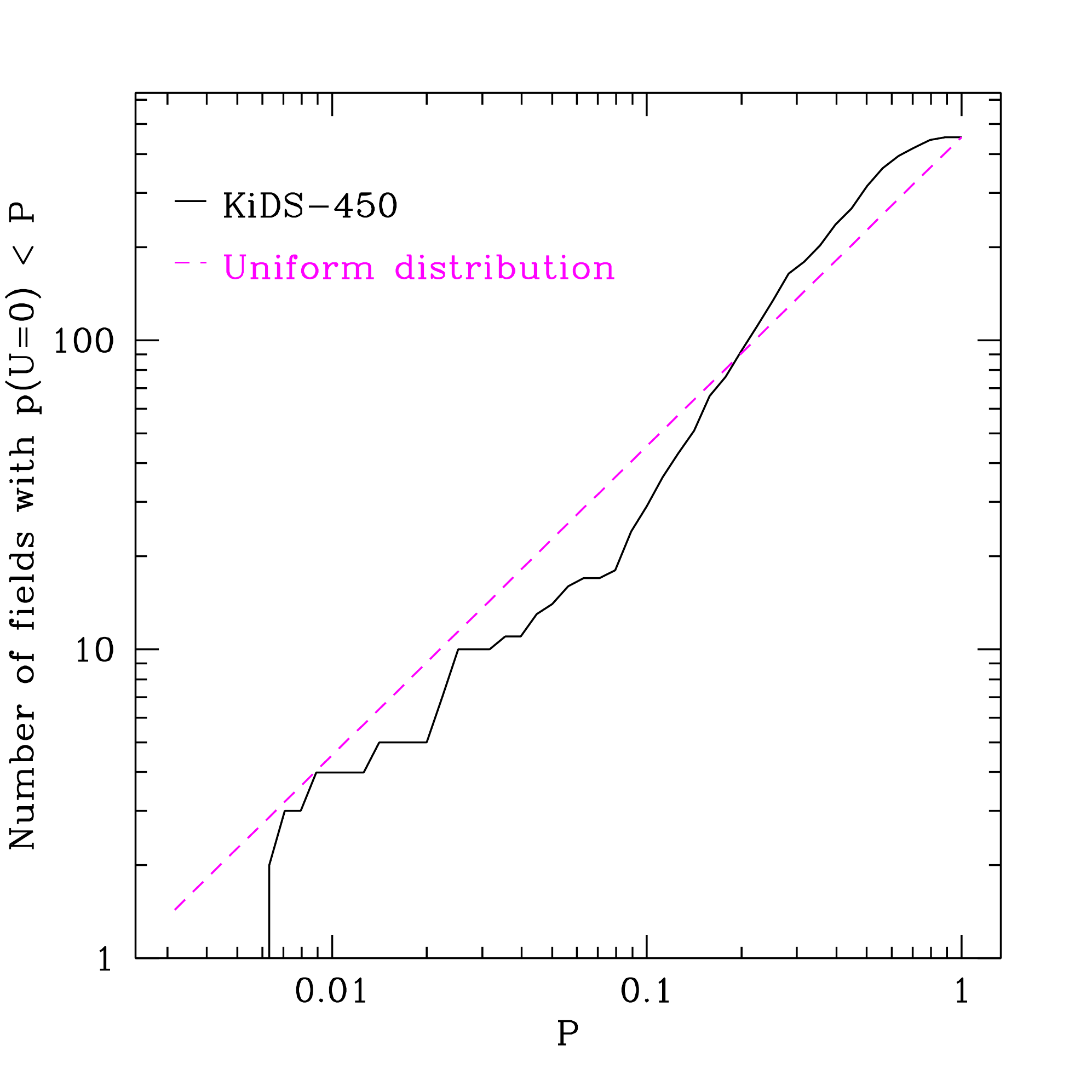}
\caption{\label{fig:p_u_cum}Cumulative probability distribution:  the number of tiles with a probability $p(U=0)<P$ where $p(U=0)$ determines how likely it is that the measured $\Delta \xi_{\rm obs}$ in each tile is consistent with zero systematics.  The data agree well with the dashed line which shows the cumulative probability for a uniform distribution.}
\end{figure}

This agreement is further explored in Fig.~\ref{fig:p_u_cum}.   For each tile we determine the probability $p(U=0)$ that  determines how likely it is that its measured $\Delta \xi_{\rm obs}$ is consistent with zero systematics (see \citealt{heymans/etal:2012} for details).  Fig.~\ref{fig:p_u_cum} shows how the measured cumulative probability distribution for the KiDS-450 tiles agrees well with a uniform distribution.  As such even the small handful of tiles with low $p(U=0)$ are expected for a dataset of this size.  From this analysis we conclude that, unlike previous surveys, we do not need to reject tiles. Evidently the reduction in PSF-dependent noise bias achieved by the self-calibrating \emph{lens}fit, combined with a benign PSF pattern, is of sufficiently high quality that there is no significant contamination by the PSF within the full KiDS-450 dataset.  

%D5
\subsection{E/B mode decomposition} 
\label{sec:EB}
The main disadvantage of using the $\xi_\pm$ statistic is the fact that it mixes curl-free gradient distortions (E-mode) and curl distortions (B-mode).  The weak lensing contribution to the B mode only occurs at small angular scales, $\theta < $ \SI{1}{\arcmin}, mainly due to source redshift clustering \citep{schneider/etal:2002}.  Separating the weak lensing signal into E and B modes therefore provides a stringent null-test.  A non-zero B mode could arise from residual systematics in the shear measurement method, intrinsic ellipticity selection biases from the object detection stage or the photometric redshift selection, or from the intrinsic alignment of nearby galaxies \citep[see for example][and references therein]{troxel/ishak:2015,joachimi/etal:2015}.  
 
\citet{COSEBIS} derive a set of complete basis functions, called COSEBIs, that optimally combine different angular scales from the $\xi_{\pm}$ measurement to produce a pure E/B separation.  When applied to the CFHTLenS data the COSEBIs analysis revealed significant high-order B modes when the data were separated into different tomographic bins \citep{asgari/etal:2016}.  As these tomographic B modes were not seen when using alternative E/B decomposition methods \citep{kitching/etal:2014}, we consider COSEBIs to provide the most stringent B-mode null-test.  A significant measurement of a COSEBIs B mode, however, only reveals the presence of a non-cosmological signal in the data.  It does not inform us how, or indeed whether, that signal biases cosmological constraints from a $\xi_{\pm}$ analysis, nor does it inform on the origin of the non-cosmological signal.    For the purposes of this paper, we therefore investigate the E- and B-type correlators $\xi_\rmn{E/B}$ proposed by \citet{Crittenden/etal:2002}, which are closely related to our chosen $\xi_{\pm}$ statistic.  These are given by
\be
\xi_\rmn{E}(\theta)=\frac{\xi_+^{\rm obs}(\theta)+\xi'(\theta)}{2}
\qquad\hbox{and}\qquad
\xi_\rmn{B}(\theta)=\frac{\xi_+^{\rm obs}(\theta)-\xi'(\theta)}{2}
\, ,
\label{eqn:xieb}
\ee
where
\be
\xi'(\theta)=\xi_-^{\rm obs}(\theta)+4\int_\theta^\infty \frac{\d\vartheta}{\vartheta} \xi_-^{\rm obs}(\vartheta)
        -12\theta^2 \int_\theta^\infty \frac{\d\vartheta}{\vartheta^3}\xi_-^{\rm obs}(\vartheta)\, ,
\label{eqn:xipr}
\ee
and $\xi_\pm^{\rm obs}$ is the observed two-point shear correlation function.  In the absence of B mode distortions, $\xi'(\theta) = \xi_+^{\rm obs}(\theta)$ and $\xi_+^{\rm obs}(\theta)=\xi_\rmn{E}$, i.e the observed shear correlation function is pure E mode.

Consider the toy model where any systematic contribution to the ellipticity measurement is uncorrelated with the true sheared ellipticity $\epsilon^{\rm true} $, and adds linearly such that  $\epsilon^{\rm obs} = \epsilon^{\rm true} + \epsilon^{\rm sys}$.  In the case where the contaminating systematic signal contributes equally to the tangential and cross distortions such that $\langle \epsilon^{\rm sys}_\rmn{t} \epsilon^{\rm sys}_\rmn{t} \rangle = \langle \epsilon^{\rm sys}_\times \epsilon^{\rm sys}_\times \rangle$, the systematics only contribute to the observed $\xi_+^{\rm obs}(\theta)$ with
\be
\xi_+^{\rm obs}(\theta) = \xi_+^{\rm true}(\theta) + \xi_+^{\rm sys}(\theta) 
\qquad\hbox{and}\qquad
\xi_-^{\rm obs}(\theta) = \xi_-^{\rm true}(\theta) \, ,
\label{eqn:xiobs_xisys}
\ee
where $\xi_+^{\rm sys} =  \langle \epsilon^{\rm sys}_\rmn{t} \epsilon^{\rm sys}_\rmn{t} \rangle + \langle \epsilon^{\rm sys}_\times \epsilon^{\rm sys}_\times \rangle$.    In this case, the observed $\xi_-^{\rm obs}(\theta)$ is systematic-free such that $\xi'(\theta) = \xi_+^{\rm true}(\theta)$ and
\be
\xi_\rmn{E}(\theta)=\xi_+^{\rm true}(\theta) + \frac{\xi_+^{\rm sys}(\theta)}{2}
\qquad\hbox{and}\qquad
\xi_\rmn{B}(\theta)=\frac{\xi_+^{\rm sys}(\theta)}{2}
\, .
\label{eqn:xiebsys}
\ee
A measurement of a non-zero B mode could therefore be mitigated in the cosmological analysis as
\be
\xi_+^{\rm true}(\theta) = \xi_+^{\rm obs}(\theta) - 2 \xi_\rmn{B}(\theta) \, ,
\label{eqn:xiplus_cor}
\ee
but only if the originating systematic is thought to contribute equally to the tangential and cross distortions.    This type of distortion would arise from random errors in the astrometry that are coherent over small scales, such that objects imaged in each exposure are not precisely aligned relative to one another.  It is also typical of the KiDS-450 PSF distortion patterns, $\epsilon^{*}$, where we find  $\langle \epsilon^{*}_\rmn{t} \epsilon^{*}_\rmn{t} \rangle - \langle \epsilon^{*}_\times \epsilon^{*}_\times \rangle$ to be consistent with zero on all scales for both the PSF correlation and the PSF residual correlation function.     Note that this is not necessarily the case for all camera PSFs \citep{hoekstra:2004}.

If the B mode is thought to originate from small-scale random astrometric errors, or from KiDS-450 PSF distortions, then the mitigation strategy outlined above would be a reasonable approach to take.  If the B mode is thought to originate from intrinsic galaxy alignments however, this strategy would be ill-advised.  The intrinsic galaxy alignment models and mocks that find significant B modes measure them to be less than $\sim 1/5$th of the intrinsic alignment contribution to the E mode \citep{heymansIA/etal:2006,giahi-saravani/schafer:2014}.  In contrast, the commonly used linear-alignment theory that we adopt to model the impact of intrinsic galaxy alignments (see Section~\ref{sec:IAmodels}) predicts B modes that only contribute to the sub-dominant `II' term at roughly an order of magnitude lower than the `II' E-mode signal \citep{hirata/seljak:2004}.    Whichever intrinsic alignment model one chooses, mitigating its effects through Eq.~\ref{eqn:xiebsys} would therefore fail.   Our toy model also breaks down for systematics introduced through selection biases, where the systematic is likely to be correlated with the true sheared ellipticity $\epsilon$.   Selection biases could arise at the source extraction stage where, at the faintest limits, there is a preference to select galaxies oriented in the same direction as the PSF \citep{kaiser/2000} and galaxies that are correlated with the gravitational shear \citep{hirata/seljak:2003}.  These selection biases can however be mitigated using a realistic suite of image simulations to calibrate the effect \citep[see the discussion in][]{fenechconti/etal:2016}.  \citet{vale/etal:2004} show that non-equal E and B modes can be introduced where there is variation in the source density across the survey, for example from changes in seeing.  The amplitude of this effect is, however, significantly smaller than the statistical noise in KiDS-450.

In previous cosmic shear analyses, significant detections of B modes were mitigated using different strategies.  \citet{benjamin/etal:2007} detected significant B modes and proposed to increase the uncertainty on the E-mode measurement by adding the measured B modes in quadrature.  A similar strategy was used by \citet{jee/etal:2013}.  An alternative method removes angular scales from the analysis where the B modes are found to be non-zero \citep{massey/etal:2007,fu/etal:2008}.  This strategy requires that the systematic  distortion has the same angular dependence for the E and B modes.  \citet{becker/etal:2016} report zero B modes for the DES-SV data based on a Fourier band-power analysis.  This conclusion relies on the assumption that small-scale systematics behave similarly to large-scale systematics as the B-mode band-power measurement does not probe the same angular scales as the primary $\xi_{\pm}$ cosmic shear statistic that is then used for cosmological parameter estimation \citep{abbott/etal:2015}. %\citepalias{abbott/etal:2015}. 
In the RCS\-LenS data \citet{hildebrandt/etal:2016} find a significant high-amplitude B mode that extends to large scales.  They argue that the uncertainty in the origin of this B mode does not allow for a robust mitigation strategy such that, in the era of high-precision cosmology, the high-amplitude large-scale measured B mode currently disqualifies the use of this survey for cosmic shear studies.

\subsubsection{Measurements of $\xi_E$ and $\xi_B$}

\begin{figure*}
\includegraphics[width=\hsize]{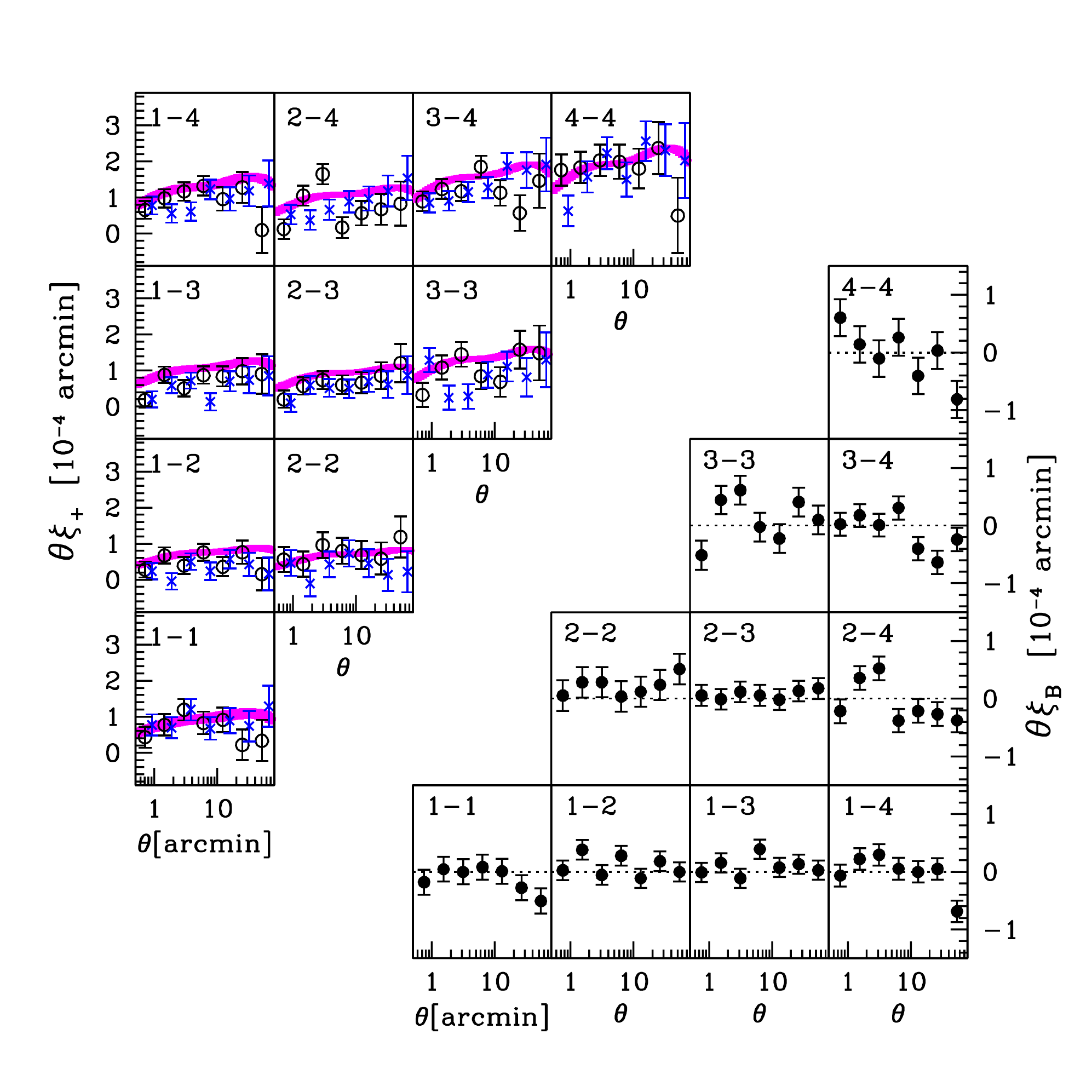}
\caption{\label{fig:Bmodes}Tomography measurements using $\xi_B$ for each redshift bin combination, as denoted in the caption inset in each panel. Lower right: the measured B mode $\xi_B$ (Eq.~\ref{eqn:xieb}).  Upper left: Measurements of $\xi_+^{\rm obs}$ (open symbols) are corrected using the measured B mode (crosses, Eq.~\ref{eqn:xiplus_cor}).  The data with and without B-mode correction can be compared to the best-fit cosmological model (solid) with its own error arising from uncertainty in the measured DIR redshift distribution.}
\end{figure*}

In Fig.~\ref{fig:Bmodes} we present measurements of $\xi_B$ using the same angular and tomographic binning as our primary measurement of $\xi_+$ shown in Fig.~\ref{fig:xipm}.    In practice, we measure $\xi_{\pm}$ in 3500 equally log-spaced $\theta$ bins from \SI{0.5}{\arcmin} to \SI{300}{\arcmin}.  In order to complete the integral in Eq.~\ref{eqn:xipr} with $\theta \rightarrow \infty$, we need a theoretical model to extend measurements of $\xi_{-}$ beyond $\theta=$ {300}{\arcmin}: our results are insensitive to a change between a Planck or CFHTLenS cosmology to model this large-scale signal \citep{planck/cosmo:2015, heymans/etal:2013}.  The small-scale measurements with $\theta<$ \SI{10}{\arcmin} are unchanged if we truncate the integral at $\theta=$ \SI{300}{\arcmin}.

We calculate the errors on $\xi_B$ analytically, as the error derives from shape-noise only, and find the B mode detected in Fig.~\ref{fig:Bmodes} to be significant.  The probability that $\xi_B$ is consistent with zero over all angular scales and tomographic bins is measured to be $p(\chi^2 | \nu) = 0.005$ .   The amplitude of the B mode is however small (note the different vertical axes between the B modes presented in the lower-right panels of Fig.~\ref{fig:Bmodes}, and the signal $\xi_+$ presented in the upper-left panels).  One approach to mitigate the impact of these systematics is to limit the angular scales used in the cosmological analysis.  We find that the B mode is consistent with zero, with a probability $p(\chi^2 | \nu) = 0.88$, when we restrict the measurement to $\theta> \SI{4.2}{\arcmin}$ (i.e removing the first three angular bins in each $\xi_+$ tomographic bin combination). We draw similar conclusions when the survey is analysed in 2D, finding significant B modes, but only on small scales $\theta \leq \SI{4.2}{\arcmin}$.  Results from a preliminary, non-tomographic power spectrum analysis \citep[see K\"ohlinger et al., in prep., and][]{kohlinger/etal:2016} find the B modes to be consistent with zero for $\ell \leq 3700$, thus supporting the conclusion that the origin of the systematic is on small angular scales.

In an attempt to isolate the origin of the detected B mode the following tests were carried out.  Unfortunately none aided our understanding of the source of the measured small-scale B mode:
\begin{enumerate}
\item{Lower quality data:} any tiles that had been flagged as potentially problematic through the data reduction validation checks, either for issues with scattered light or a lower-quality astrometric solution, were removed from the survey in a re-analysis.   The results were unchanged.
\item{Edge effects:} the survey was re-analysed removing a \SI{10}{\arcmin} width border around the outer edge of each pointing.  The results were unchanged.
\item{Pointing-to-pointing astrometry errors: the survey was re-analysed using only those galaxy pairs that were observed within the same pointing.  The small-scale signal was unchanged.}
\item{Quadrant errors: the survey was split into four, based on which quadrant the galaxy was located within each pointing.  The results from each quadrant were consistent.}
\item{PSF effects: In Section~\ref{sec:c_term} we argue that the PSF contamination to KiDS-450 is minimal with $|\alpha| < 0.03$, where $\alpha$ denotes the fractional PSF contamination to the shear measurement.   In order to re-produce the amplitude of the measured small-scale B mode from PSF effects alone we would require $\alpha=0.15$.  This PSF contribution would, however, also produce a significant large-scale B mode that we do not detect.}
\item{PSF residuals:  The B mode and E mode of the residual PSF ellipticity correlation function was found to be consistent with zero, where the residual is measured as the difference between the measured stellar ellipticity and the PSF model at the location of a star (see section~3.2.2 of \citealt{kuijken/etal:2015} for further details).}
\item{PSF errors:  We apply a comprehensive method in order to isolate any pointings that have a significant PSF residual, finding no evidence for this type of contamination in Section~\ref{sec:systematics}.  Nevertheless we undertake a simple further test where we measure the B mode of the statistic $C^{\rm sys} = \langle \epsilon \epsilon^*\rangle ^2 /  \langle \epsilon^*\epsilon^*\rangle$ \citep{bacon/etal:2003}, finding it to be more than an order of magnitude smaller than the measured B mode and consistent with zero.}
\item{Additive calibration correction:  We test the impact of the additive calibration correction detailed in Section~\ref{sec:c_term}.  Applying no correction to the survey increases the B modes measured on large scales,  but leaves the small-scale signal unchanged.} 
\item{Chip-position-dependent errors:  In Section~\ref{sec:c_term} we find no significant dependence on the amplitude of the additive bias $c$ on galaxy position within the field of view.  To further analyse this effect we bin the data by position $(x,y)$ within each pointing to construct a (noisy) map of $c(x,y)$.  We then measure the contribution to $\xi_B$ from a position dependent additive bias modelled using the mean measured $c(x,y)$ across the full survey and also per patch.  We find a small contribution to the measured small-scale B mode at the level of 10 per cent of the measured value.  If we correct for a position-dependent $c$-term using the noisy $c(x,y)$ map, however, we find that the noise serves only to increase the overall B mode by 10 per cent rather than decreasing the signal. For a more detailed analysis of position-dependent additive bias within CFHTLenS and KiDS-DR2 data see \citet{vanuitert/schneider:2016}.}
\item{Seeing and PSF variation:  We separate the pointings based on their measured seeing and PSF ellipticity, finding no significant difference in the resulting measured E and B modes.}
\item Astrometric errors:  We test whether random errors in the astrometric solution could cause B-modes.  Under the assumption that the VST OmegaCAM camera shear is relatively stable \citep[as motivated by][]{dejong/etal:2015}, we can estimate a spatially dependent astrometric error from the measured camera shear per pointing \citep[see for example][]{erben/etal:2013}.  Defining astrometric errors in this way produced E and B mode measurements that were insignificant in comparison to the measured $\xi_{E/B}$.
\end{enumerate}

\begin{figure}
\includegraphics[width=\hsize]{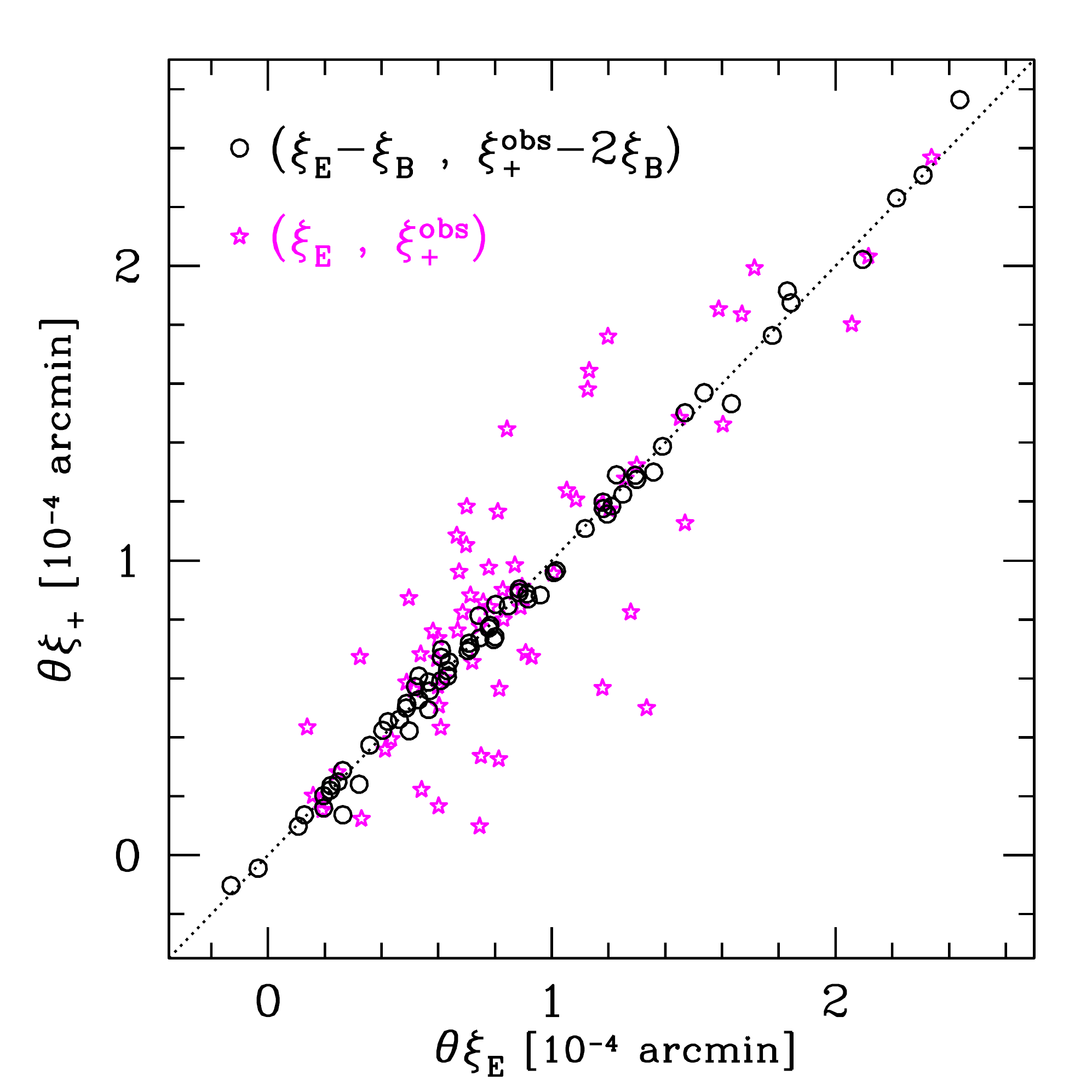}
\caption{\label{fig:xipcomp} Comparison of estimators for $\xi_+$ where in the absence of systematics $\xi_E = \xi_+^{\rm obs}$.  Each data point is one of the 70 components of the tomographic $\xi_+$ data vector, comparing the E-mode measurement $\xi_E$ with the measured two-point correlation function $\xi_+^{\rm obs}$ at the same $\theta$ scale in each tomographic combination of redshift bins (shown as stars).   Under the assumption that the contaminating systematic contributes equally to the tangential and cross distortions we can correct $\xi_E$ and $\xi_+^{\rm obs}$ with Eqs.~\ref{eqn:xiobs_xisys} and~\ref{eqn:xiebsys} (shown as circles).  After this correction has been applied the agreement between the two estimators for $\xi_+$ is significantly improved. }
\end{figure}

Unfortunately none of our tests were able to isolate the B mode or reveal its origin, but we can nevertheless proceed to test the assumption that whatever is the source of the B mode, the contaminating systematic contributes equally to the tangential and cross distortions. In Fig.~\ref{fig:xipcomp} we compare the measured E mode $\xi_E$ with the measured $\xi_+^{\rm obs}$ across all 70 points in the $\xi_+$ data vector (7 angular bins in 10 different tomographic bin combinations).    In the absence of systematics $\xi_E = \xi_+^{\rm obs}$, with some scatter in this relationship as the E-B decomposition method reduces the shot-noise variance on the measurement of $\xi_+$ by a factor of $\sqrt{2}$.    We correct $\xi_E$ and $\xi_+^{\rm obs}$ with Eqs.~\ref{eqn:xiebsys} and ~\ref{eqn:xiplus_cor}, finding that the agreement between the two estimators for $\xi_+$ is significantly improved when this correction is applied.   This comparison supports the assumption that the contaminating systematic contributes equally to the tangential and cross distortions: if that were not the case the agreement between $\xi_E$ and $\xi_+^{\rm obs}$ would deteriorate after the correction was applied.  This analysis therefore  provides an important consistency check and validates our approach of using the B mode to correct $\xi_+^{\rm obs}$ in Section~\ref{sec:MCMC_Bmode}.

%D6

\section[Pipeline redundancy: pixels to cosmological parameters]
{PIPELINE REDUNDANCY: PIXELS TO COSMOLOGICAL PARAMETERS}
\label{sec:redundancy}
This paper documents the primary pipeline analysis from raw pixels through to cosmological parameters.  In this section we record the cross-checks within the pipeline along with any redundancy, i.e., the extra components that are not always strictly necessary but are used to ensure that the primary analysis is robust.
\begin{enumerate}
\item{Data reduction:  The raw pixel data have been reduced using two independent pipelines, \textsc{theli} and \textsc{Astro-WISE}, and there is a good level of agreement between the two reductions.  The \textsc{theli} reduction, which is optimised for weak lensing analyses by including a global astrometric solution, is then used for the lensing image analysis.  The \textsc{Astro-WISE} reduction, which is optimised for the large-volume multi-colour data, is then used for the photometry analysis.}
\item{Catalogue Handling: There are numerous steps in the catalogue pipeline analysis from reduced images through to the final catalogue product.  The pipeline team used a central \textsc{git} repository, and major changes and upgrades to the scripts were verified by a second person within the team before they were committed to the master branch.}
\item{Photometric Redshifts:  As detailed in Section~\ref{sec:photoz_calibration} and Appendix~\ref{sec:app_z_tests} we investigate four different methods to determine the redshift distributions of our tomographic source samples.  The derived cosmological parameters for all four methods are in agreement with each other.}
\item{Shear Correlation function measurement:  We use the \textsc{athena} tree-code \citep{athena} to measure the tomographic shear correlation function $\xi_\pm$.  This was found to agree well with an alternative measurement using the \textsc{TreeCorr} software \citep{treecorr}.}
\item{Covariance Matrices:  As detailed in Section~\ref{sec:covariance} we investigate three different methods to determine the covariance matrix, finding consistent results for the derived cosmological parameters from both our numerical and analytical estimates.}
\item{Cosmological Parameter Estimates:  Our cosmological parameter likelihood analysis uses a modified version of \textsc{CosmoMC} \citep{lewis/bridle:2002} as described in \citet{joudaki/etal:2016}\footnote{\url{https://github.com/sjoudaki/cfhtlens_revisited}}. This pipeline underwent a series of cross-checks, verifying the cosmic shear and intrinsic alignment theoretical modeling through comparison to the \textsc{nicaea} software \citep{kilbinger/etal:2009} and private code within the team. In addition we used \textsc{MontePython} and \textsc{class} \citep{blas/lesgourgues/tram:2011,audren/etal:2013}, as described in \citet{kohlinger/etal:2016}, to provide another consistency check with a completely independent likelihood code.  The cosmological constraints from these two independent analyses are in good agreement.}
\item{$\xi_+$ vs $\xi_-$:  We found consistent cosmological constraints when using only $\xi_+$ or only $\xi_-$, with $\xi_-$ favouring slightly lower values of $S_8$ in comparison to $\xi_+$.  The constraints from $\xi_+$ alone are very similar to the constraints where the two statistics are analysed in combination, as $\xi_+$ carries the highest signal-to-noise as shown in Fig.~\ref{fig:covSN}.}
\item{2D vs tomography:  Our primary cosmological parameter likelihood analysis uses four tomographic bins.  We also determined the cosmological constraints from a `2D' case using all source galaxies selected with $0.1<z_B<0.9$ excluding systematics modelling.  The cosmological constraints were in excellent agreement with the tomographic `no-systematics' case} analysis.
\end{enumerate}

 %E

\section[Best-fit model parameters]
{BEST-FIT MODEL PARAMETERS}
\label{sec:detailed_results}

\begin{figure*}
\includegraphics[width=0.95\hsize]{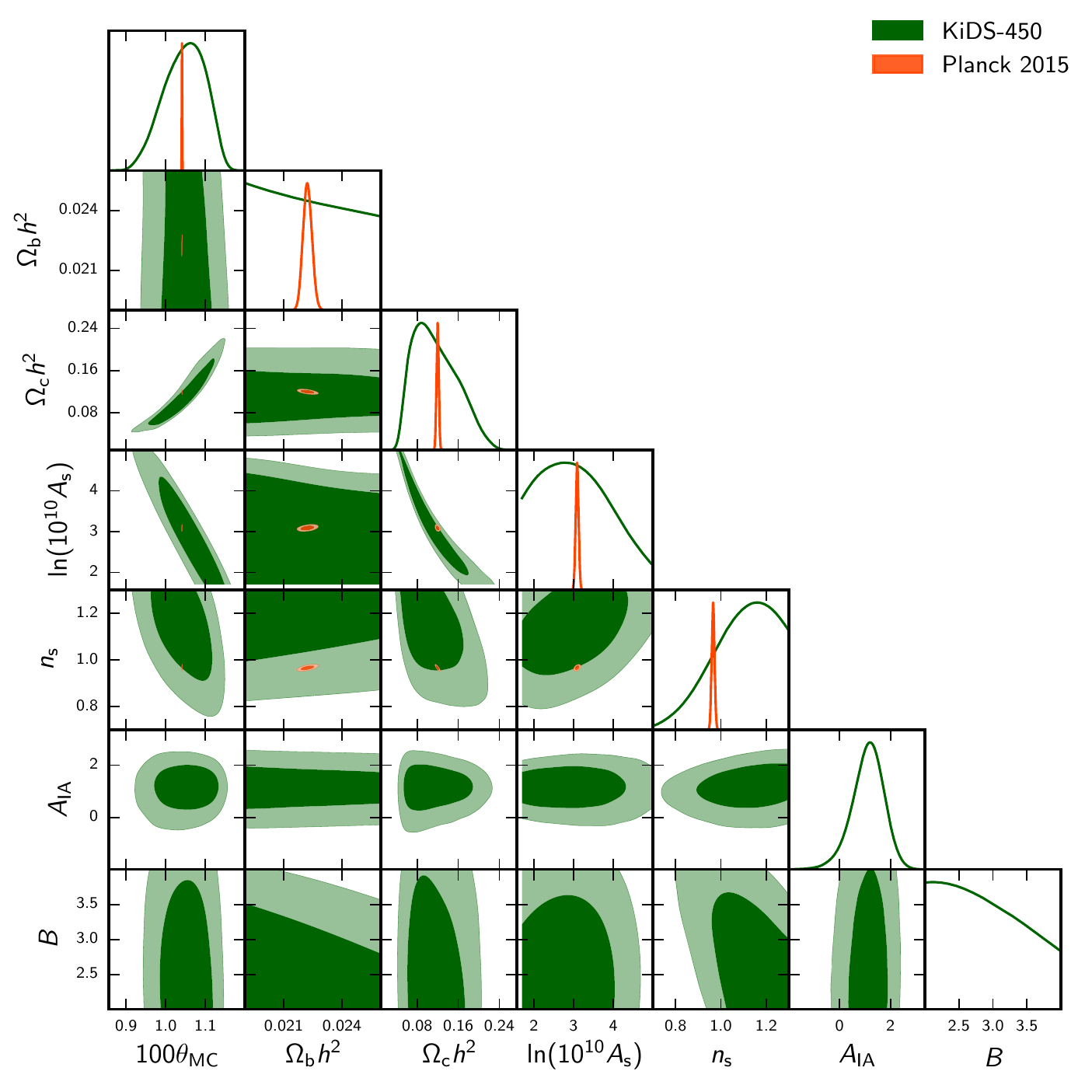}
\vspace{-1.3em}
\caption{\label{fig:triangle} Posterior distributions of the primary model parameters and their correlation.  Parameter definitions and priors are listed in Table~\ref{tab:priors}.}
\end{figure*}

\begin{figure*}
\begin{center}
\resizebox{13.0cm}{!}{{\includegraphics{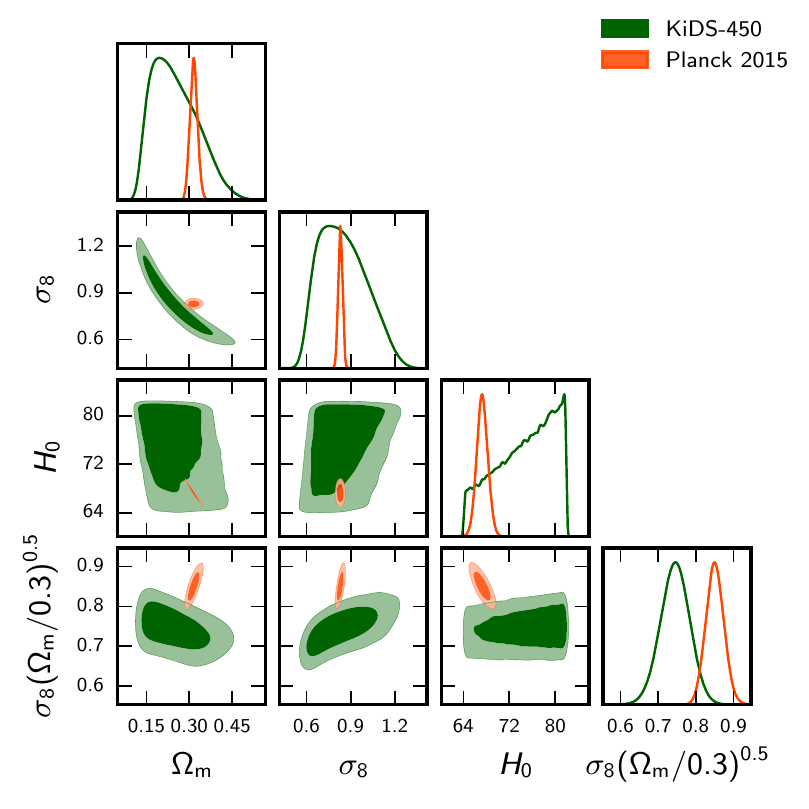}}}
\end{center}
\vspace{-2.3em}
\caption{\label{fig:triangle_derived} Posterior distributions of derived cosmological parameters and their correlation.}
\end{figure*}

In Tables~\ref{tab:primary_param} and \ref{tab:derived_param} we present detailed results for the best-fit model parameters for all MCMC runs described in Table~\ref{tab:MCMC_setups}. We include the seven primary parameters, along with a few derived parameters of interest. Note that the primary parameters $\Omega_{\rm b}h^2$, $\theta_{\rm MC}$, $n_s$ and $B$ are heavily constrained by the priors rather than by the KiDS-450 data. The same is true for the derived parameter $H_0$.
We report the $\chi^2_{\rm eff}$ of the best-fit model, the number of degrees of freedom, and the `Deviance Information Criterion' (DIC) \citep{joudaki/etal:2016}. Fig.~\ref{fig:triangle} shows 2D projections of the confidence region in 7-dimensional primary parameter space for all combinations of parameters in the primary KiDS-450 run. A similar plot for derived parameters can be found in Fig.~\ref{fig:triangle_derived}.

\begin{table*}
\begin{minipage}[t]{\textwidth}
\caption{\label{tab:primary_param}Mean and 68\%\ confidence intervals on the primary model parameters. Note that the parameters $\Omega_{\rm b}h^2$, $\theta_{\rm MC}$, $n_s$ and $B$ are heavily constrained by the priors rather than by the KiDS-450 data.}
\renewcommand{\footnoterule}{}  % to avoid a line before footnotes
\begin{tabular}{lrrrrrrr}
\hline
 & $ \Omega_{\rm b} h^2 [10^{-2}]$ & $ \Omega_{\rm c} h^2$ & $ 100\theta_{\rm MC}$ & $ {\rm{ln}}(10^{10} A_s)$ & $ n_s$ & $ A_{\rm IA}$ & $ B$ \\ 
\hline
KiDS-450 & $ 2.23^{+0.37}_{-0.33} $  & $ 0.116^{+0.029}_{-0.056} $  & $ 1.0472^{+0.0603}_{-0.0456} $  & $ 3.09^{+0.57}_{-1.21} $  & $ 1.09^{+0.20}_{-0.06} $  & $ 1.10^{+0.68}_{-0.54} $  & $ 2.88^{+0.30}_{-0.88} $ \\
DIR & $ 2.23^{+0.37}_{-0.33} $  & $ 0.119^{+0.031}_{-0.061} $  & $ 1.0504^{+0.0642}_{-0.0459} $  & $ 3.08^{+0.51}_{-1.26} $  & $ 1.02^{+0.12}_{-0.14} $  & $ 1.14^{+0.65}_{-0.55} $  & - \\
CC & $ 2.24^{+0.36}_{-0.34} $  & $ 0.143^{+0.047}_{-0.060} $  & $ 1.0711^{+0.0645}_{-0.0362} $  & $ 2.84^{+0.32}_{-1.14} $  & $ 1.05^{+0.16}_{-0.13} $  & $ 0.80^{+1.02}_{-0.96} $  & - \\ 
BOR & $ 2.22^{+0.11}_{-0.32} $  & $ 0.097^{+0.018}_{-0.052} $  & $ 1.0248^{+0.0534}_{-0.0622} $  & $ 3.44^{+1.21}_{-0.80} $  & $ 1.03^{+0.12}_{-0.12} $  & $ -0.92^{+0.99}_{-0.71} $  & - \\
BPZ & $ 2.22^{+0.38}_{-0.32} $  & $ 0.099^{+0.017}_{-0.054} $  & $ 1.0250^{+0.0538}_{-0.0674} $  & $ 3.49^{+1.51}_{-0.48} $  & $ 1.04^{+0.13}_{-0.13} $  & $ -1.10^{+0.96}_{-0.70} $  & - \\
DIR-no-error & $ 2.23^{+0.37}_{-0.33} $  & $ 0.120^{+0.031}_{-0.064} $  & $ 1.0495^{+0.0661}_{-0.0513} $  & $ 3.10^{+0.44}_{-1.40} $  & $ 1.02^{+0.13}_{-0.13} $  & $ 1.20^{+0.62}_{-0.52} $  & - \\
B mode & $ 2.23^{+0.37}_{-0.33} $  & $ 0.101^{+0.020}_{-0.055} $  & $ 1.0251^{+0.0568}_{-0.0638} $  & $ 3.43^{+1.57}_{-1.73} $  & $ 0.97^{+0.12}_{-0.14} $  & $ 1.11^{+0.67}_{-0.55} $  & - \\
$ \xi_+$ large scales & $ 2.24^{+0.36}_{-0.34} $  & $ 0.132^{+0.050}_{-0.052} $  & $ 1.0616^{+0.0666}_{-0.0355} $  & $ 2.79^{+0.28}_{-1.09} $  & $ 0.96^{+0.12}_{-0.16} $  & $ 1.07^{+0.79}_{-0.59} $  & - \\
no systematics & $ 2.22^{+0.38}_{-0.32} $  & $ 0.106^{+0.022}_{-0.058} $  & $ 1.0341^{+0.0610}_{-0.0614} $  & $ 3.32^{+1.68}_{-1.62} $  & $ 1.00^{+0.12}_{-0.12} $  & -  & - \\
\hline
\end{tabular}
\end{minipage}
\end{table*}

\begin{table*}
\begin{minipage}[t]{\textwidth}
\caption{\label{tab:derived_param}Mean and 68\%\ confidence intervals for certain derived cosmological parameters of interest. Note that the constraints on $H_0$ are dominated by the prior rather than the data. Further included are the best-fit $\chi^2_{\rm eff}$ values, number of degrees of freedom (dof), and DIC values for each setup.}
\renewcommand{\footnoterule}{}  % to avoid a line before footnotes
\begin{tabular}{l|rrrr|rrr}
\hline
 & $ \Omega_{\rm m}$ & $ \sigma_8$ & $ S_8=\sigma_8\times\sqrt{\Omega_{\rm m}/0.3}$ & $ H_0$ [km\,s$ ^{-1}$ \,Mpc$ ^{-1}$ ] & $ \chi^2_{\rm eff}$ & dof & DIC \\ 
\hline
KiDS-450 & $ 0.250^{+0.053}_{-0.103} $  & $ 0.849^{+0.120}_{-0.204} $  & $ 0.745^{+0.038}_{-0.038} $  & $ 74.7^{+7.2}_{-2.5} $  & 162.5 & 122 & 178.6 \\
DIR & $ 0.254^{+0.058}_{-0.109} $  & $ 0.826^{+0.115}_{-0.199} $  & $ 0.727^{+0.034}_{-0.030} $  & $ 75.0^{+6.9}_{-2.5} $  & 163.5 & 123 & 178.5 \\
CC & $ 0.304^{+0.086}_{-0.114} $  & $ 0.847^{+0.107}_{-0.234} $  & $ 0.816^{+0.070}_{-0.094} $  & $ 74.2^{+7.5}_{-3.5} $  & 159.0 & 123 & 191.8 \\
BOR & $ 0.213^{+0.035}_{-0.096} $  & $ 0.860^{+0.170}_{-0.174} $  & $ 0.693^{+0.037}_{-0.031} $  & $ 75.3^{+6.7}_{-2.3} $  & 172.1 & 123 & 182.5 \\
BPZ & $ 0.215^{+0.035}_{-0.098} $  & $ 0.884^{+0.186}_{-0.164} $  & $ 0.714^{+0.040}_{-0.034} $  & $ 75.2^{+6.7}_{-2.3} $  & 173.5 & 123 & 184.2 \\
DIR-no-error & $ 0.254^{+0.060}_{-0.114} $  & $ 0.830^{+0.116}_{-0.211} $  & $ 0.729^{+0.034}_{-0.029} $  & $ 75.0^{+6.9}_{-2.8} $  & 166.4 & 123 & 174.7 \\
B mode & $ 0.226^{+0.042}_{-0.105} $  & $ 0.850^{+0.179}_{-0.183} $  & $ 0.702^{+0.036}_{-0.029} $  & $ 74.2^{+7.5}_{-3.0} $  & 142.0 & 123 & 151.8 \\
$ \xi_+$ large scales & $ 0.282^{+0.093}_{-0.099} $  & $ 0.746^{+0.074}_{-0.187} $  & $ 0.692^{+0.040}_{-0.035} $  & $ 74.4^{+7.5}_{-3.5} $  & 133.7 & 93 & 141.2 \\
no systematics & $ 0.228^{+0.044}_{-0.106} $  & $ 0.846^{+0.145}_{-0.203} $  & $ 0.703^{+0.031}_{-0.025} $  & $ 75.1^{+6.8}_{-2.4} $  & 169.4 & 124 & 177.3 \\
\hline
\end{tabular}
\end{minipage}
\end{table*}

The full cosmological analysis was also carried out with the two other blindings. Those yielded a tension in the $S_8$ parameter with respect to Planck of 1.7$\,\sigma$ and 2.8$\,\sigma$, respectively.%F

%last page now done with lastpage package: works even if last page only contains a floating figure or table.
\end{document}